\documentclass[printer,usenatbib]{aa}
\usepackage{natbib}
\usepackage{times}
\usepackage[T1]{fontenc}
\usepackage{graphicx}

\usepackage[utf8]{inputenc}
\usepackage{tabularx}

\def\deg{\hbox{$^\circ$}}

\begin{document}

\title{Photometric survey, modelling, and scaling\break 
       of long-period and low-amplitude asteroids}

   \authorrunning{Marciniak et al.}
   \titlerunning{Long-period and low-amplitude asteroids}

\author{A. Marciniak \inst{1} 
 \and P. Bartczak \inst{1}
 \and T. M{\"u}ller \inst{2}
 \and J. J. Sanabria \inst{3}
 \and V. Al{\'i}-Lagoa \inst{2}
 \and P. Antonini \inst{4}
 \and R.~Behrend \inst{5}
 \and L. Bernasconi \inst{6}
 \and M. Bronikowska \inst{7}
 \and M. Butkiewicz - B\k{a}k \inst{1}
 \and A. Cikota \inst{8}
 \and R.~Crippa \inst{9}
 \and R. Ditteon \inst{10}
 \and G. Dudzi{\'n}ski \inst{1}
 \and R. Duffard \inst{11}
 \and K. Dziadura \inst{1}
 \and S. Fauvaud \inst{12}
 \and S.~Geier \inst{3, 13}
 \and R. Hirsch \inst{1}
 \and J. Horbowicz \inst{1}
 \and M. Hren \inst{8}
 \and L. Jerosimic \inst{8}
 \and K. Kami{\'n}ski \inst{1}
 \and P.~Kankiewicz \inst{14}
 \and I. Konstanciak \inst{1}
 \and P. Korlevic \inst{8}
 \and E. Kosturkiewicz \inst{1}
 \and V. Kudak \inst{15,16}
 \and F.~Manzini \inst{9}
 \and N. Morales \inst{11}
 \and M. Murawiecka \inst{17}
 \and W. Og{\l}oza \inst{18}
 \and D. Oszkiewicz \inst{1}
 \and F.~Pilcher \inst{19}
 \and T. Polakis \inst{20}
 \and R. Poncy \inst{21}
 \and T. Santana-Ros \inst{1}
 \and M. Siwak \inst{18}
 \and B. Skiff \inst{22}
 \and K.~Sobkowiak \inst{1}
 \and R. Stoss \inst{8}
 \and M. {\.Z}ejmo \inst{23}
 \and K.~{\.Z}ukowski \inst{1}
       }

    \institute{Astronomical Observatory Institute, Faculty of Physics, A. Mickiewicz University,
              S{\l}oneczna 36, 60-286 Pozna{\'n}, Poland. E-mail: am@amu.edu.pl 
      \and Max-Planck-Institut f{\"u}r Extraterrestrische Physik, Giessenbachstrasse 1, 85748 Garching, Germany 
      \and Instituto de Astrof{\'i}sica de Canarias, C/ V{\'i}a Lactea, s/n, 38205 La Laguna, Tenerife, Spain 
      \and Observatoire des Hauts Patys, F-84410 B\'edoin, France 
      \and Geneva Observatory, CH-1290 Sauverny, Switzerland 
      \and Les Engarouines Observatory, F-84570 Mallemort-du-Comtat, France 
      \and Institute of Geology, A. Mickiewicz University, Krygowskiego 12, 61-606 Pozna{\'n} 
      \and OAM - Mallorca, Cam{\'i} de l'Observatori s/n 07144 Costitx Mallorca, Illes Balears, Spain
      \and Stazione Astronomica di Sozzago, I-28060 Sozzago, Italy 
      \and Rose-Hulman Institute of Technology, CM 171 5500 Wabash Ave., Terre Haute, IN 47803, USA 
      \and Departamento de Sistema Solar, Instituto de Astrof{\'i}sica de Andaluc{\'i}a (CSIC),
     Glorieta de la Astronom{\'i}a s/n, 18008 Granada, Spain 
      \and Observato{\'i}re du Bois de Bardon, 16110 Taponnat, France 
      \and Gran Telescopio Canarias (GRANTECAN), Cuesta de San Jos{\'e} s/n, E-38712, Bre{\~n}a Baja, La Palma, Spain 
      \and Astrophysics Division, Institute of Physics, Jan Kochanowski University,
    {\'S}wi\k{e}tokrzyska 15, 25-406 Kielce, Poland 
      \and Institute of Physics, Faculty of Natural Sciences, University of P. J. \v{S}af{\'a}rik, Park Angelinum 9, 
      040 01 Ko\v{s}ice, Slovakia 
      \and Laboratory of Space Researches, Uzhhorod National University, Daleka st. 2a, 88000, Uzhhorod, Ukraine 
      \and NaXys, Department of Mathematics, University of Namur, 8 Rempart de la Vierge, 5000 Namur, Belgium 
      \and Mt. Suhora Observatory, Pedagogical University, Podchor\k{a}{\.z}ych 2, 30-084, Cracow, Poland 
      \and 4438 Organ Mesa Loop, Las Cruces, New Mexico 88011 USA 
      \and Command Module Observatory, 121 W. Alameda Dr., Tempe, AZ 85282 USA 
      \and Rue des Ecoles 2, F-34920 Le Cr{\`e}s, France 
      \and Lowell Observatory, 1400 West Mars Hill Road, Flagstaff, Arizona, 86001 USA 
      \and Kepler Institute of Astronomy, University of Zielona G{\'o}ra, Lubuska 2, 65-265 Zielona G{\'o}ra, Poland 
              }

   \date{Received 30 June 2017 / Accepted xx xx xx}

 \abstract
{The available set of spin and shape modelled asteroids is strongly biased against slowly rotating targets 
and those with low lightcurve amplitudes. This is due to the observing selection effects. 
As a consequence, the current picture of  asteroid spin axis distribution, rotation rates, 
radiometric properties, or aspects related to the object's internal structure might be affected too.}
{To counteract these selection effects, we are running a photometric campaign of a large sample of 
 main belt asteroids omitted in most  previous studies. 
Using least chi-squared fitting we determined synodic rotation periods and verified previous determinations.
When a dataset for a given target was sufficiently large and varied, we
performed spin and shape modelling with two different methods to compare their performance.}
{We used the convex inversion method and the non-convex SAGE algorithm, applied on the same datasets 
of dense lightcurves. Both methods search for the lowest deviations between observed 
and modelled lightcurves, though using different approaches. Unlike convex inversion, the SAGE method allows 
for the existence of valleys and indentations on the shapes based only on lightcurves.}
{We obtain detailed spin and shape models for the first five targets of our sample: 
(159) Aemilia, (227) Philosophia, (329) Svea, (478) Tergeste, and (487) Venetia.
When compared to stellar occultation chords, our models obtained an absolute size scale and  
major topographic features of the shape models were also confirmed. When applied to thermophysical modelling, 
they provided a very good fit to the infrared data and  allowed  their size, albedo, 
and thermal inertia to be determined.}
{ Convex and non-convex shape models provide  comparable fits to lightcurves.  
However, some non-convex models fit notably better to stellar occultation chords and to infrared data 
in sophisticated thermophysical modelling (TPM).
In some cases TPM showed strong preference for one of the spin and shape solutions. 
Also, we confirmed that slowly rotating asteroids tend to have higher-than-average values of thermal inertia,   
which might be caused by properties of the surface layers underlying the skin depth.}


\maketitle

\section{Introduction}

Physical parameters of asteroids such as the period of rotation and orientation 
of the spin axis are related to various processes that these bodies undergo. 
The rotation of large asteroids probably reflects the primordial 
spin acquired during the accretion phase in the protoplanetary disc 
\citep{Johansen2010}, 
which for smaller objects was later modified by impacts, collisions,  
and thermal forces, which are  strongest for small asteroids \citep{Bottke2006}. 
Asteroid rotations can reveal both their internal cohesion and the degree of fragmentation 
\citep{Holsapple2007}. 
Numerical simulations by \cite{Takeda2009} suggest that 
bodies of a rubble-pile structure usually spin down as a result of impacting 
events. 
Also, the long-term evolution under the thermal reradiation force (YORP effect) can both spin up 
and spin down asteroids \citep{Rubincam2000}. However, so far only the spin-up of the 
rotation period has been directly detected \citep[e.g.][]{Lowry2007, Lowry2014, Kaasalainen2007, 
Durech2008}

The spatial distribution of asteroid spin axes suggests that the largest bodies generally preserved their 
primordial, prograde spin, while smaller ones, with diameters less than 30 km, seem to be strongly 
affected by the YORP effect that pushes these axes towards extreme values of obliquities 
\citep{Hanus2013}. The spins of prograde rotators under the YORP effect influence  
can be captured into spin-orbit resonances,
sometimes even forming spin clusters \citep{Slivan2002, Kryszczynska2012}.

However, what is now known about these physical properties of asteroids 
is based on statistically non-representative samples. Most of the well-studied asteroids 
(those with the spin and shape model) 
are targets of relatively fast spin and substantial elongation of  shape, possibly also coupled 
with extreme spin axis obliquity, which  results in fast and large brightness variations 
(Fig. \ref{Pie_chart}). 
The reason for this state are the observing selection effects  
discussed in our first paper on this subject \citep[][hereafter M2015]{Marciniak2015},  
and summarised in the next section.

Asteroid shape models created by lightcurve inversion methods are naturally most detailed when 
created basing on rich datasets of dense lightcurves. 
High-quality lightcurves from at least five apparitions gained 
over a wide range of aspect and phase angles are a necessary prerequisite to obtain unique spin and shape 
solutions with main topographic features (usually coming in pairs of two 
indistinguishable mirror solutions for the pole). 
The obtained models can be convex representations of real shapes \citep[in the convex inversion method by][]
{Kaasalainen1, Kaasalainen2},
but can also be non-convex, more closely reproducing real asteroid shapes when supported by auxiliary data 
\citep[in KOALA and ADAM algorithms,][]{Carry2012, Viikinkoski2015}, but also based on lightcurves alone
\citep[in the SAGE algorithm,][Bartczak \& Dudzi{\'n}ski, MNRAS, accepted]{Bartczak2014}.

Even after the Gaia Solar System catalogue is released, which is expected at the beginning of the next decade, 
the most reliable way to study 
spins (sidereal periods and spin axis positions) of a number of new bodies of low amplitudes and long periods is 
the traditional dense photometry performed on a network of small and medium-sized ground-based telescopes. 
The precise shape modelling  technique is, and will most probably remain, the only tool allowing  
a substantial number of such challenging targets to be studied in detail because Gaia and most of the other sky surveys will deliver 
only a few tens of sparse datapoints for each observed asteroid, only providing ellipsoidal approximations 
of the real shapes. 
However, the number of targets with precise shape models cannot be as large as when 
modelling on sparse data because of the high demand of observing time, which reaches hundreds of hours 
for each long-period target (see Table \ref{obs} in  Appendix A).

Detailed asteroid shape models with concavities are in high demand for precise density determinations \citep{Carry},  
modelling the thermal YORP and Yarkovsky effects \citep{Vokrouhlicky2015} --  including self-heating -- and 
accurate thermophysical modelling \citep{Delbo2015} from which one can infer their sizes, albedos, 
surface roughness, and thermal inertia values, allowing further studies of their composition 
and surface and subsurface properties. 
Apart from studying asteroid parameters for themselves, such research has other very practical applications.
Large asteroids are very good calibration standards for infrared observatories like 
Herschel, APEX, and ALMA, perfectly filling the gap in the flux levels of stellar 
and planetary calibration sources \citep{Muller2002, Muller2014a}. However, their flux changes have to be clearly  predictable, 
and should not vary much over short timescales. Slowly rotating asteroids of low 
lightcurve amplitudes are best for such applications.

In this work we perform spin and shape modelling using two lightcure inversion methods: 
the convex inversion method \citep{Kaasalainen1, Kaasalainen2} and the non-convex SAGE algorithm 
\citep[][Bartczak \& Dudzi{\'n}ski, MNRAS, accepted]{Bartczak2014}.
Later we validate and at the same time compare the resulting shapes by fitting them 
to data from other techniques:
multi-chord stellar occultations, and all available thermal infrared data.
This way our shape models also get absolute size scale, both radiometric and non-radiometric.

The next section discusses the selection effects in asteroid studies, and
briefly describes our observing campaign to counteract them. 
Section 3 describes spin and shape modelling methods, 
and brings a description of thermophysical modelling and occultation fitting procedures 
used primarily to scale our models.
Section 4 contains the observing campaign intermediate results, another set of targets with 
corrected period determinations. 
In Section 5 we present models for five targets of our sample that have enough data 
for full spin and shape modelling, scale them by thermophysical modelling, and where possible  also by occultations. 
The last section describes the conclusions and planned future work.
 Appendix A contains observation details and new lightcurves.

\section{Selection effects and the observing campaign}

\subsection{Observing and modelling biases in asteroid studies}

\begin{figure}[h]
\includegraphics[width=0.5\textwidth]{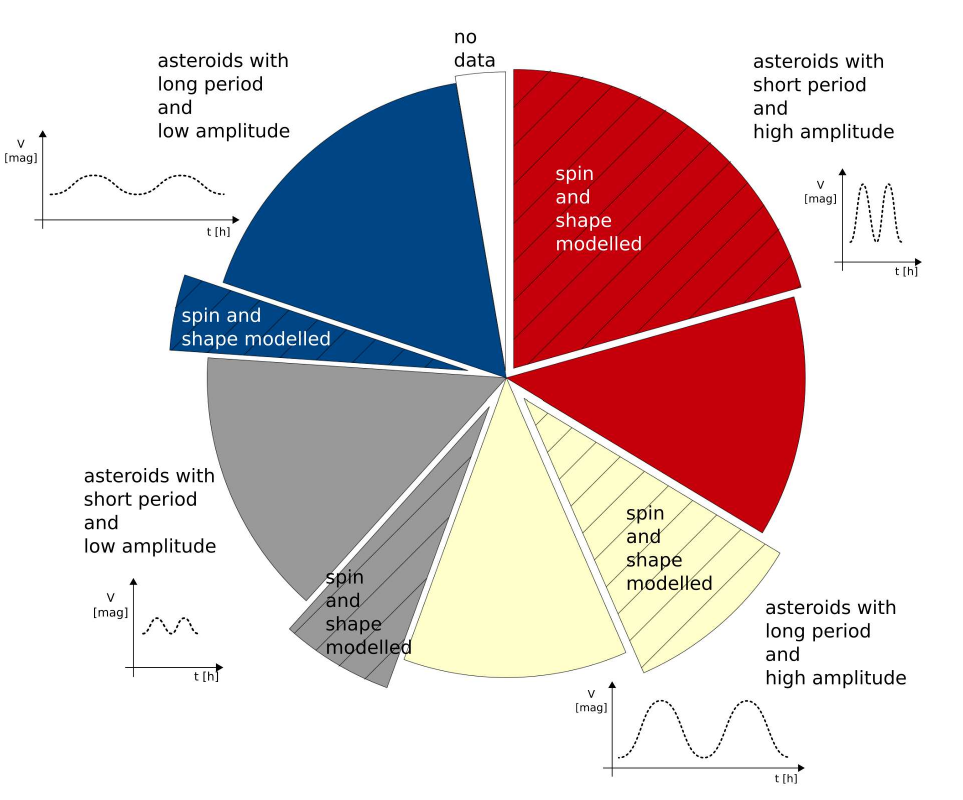}
\caption{Current distribution of known periods and maximum amplitudes among  the $\sim$1200 brightest 
         main belt asteroids \citep[based on LCDB,][updated 2016 September 5]{Warner2009}. 
         Division values are P=12~hours and a$_{max}$=0.25 mag. 
    The amount of spin and shape modelled targets is marked within each group. Asteroids with specific 
    features are over-represented, while others are largely omitted.
}
\label{Pie_chart}
\end{figure}

\begin{figure}[h]
\includegraphics[width=0.5\textwidth]{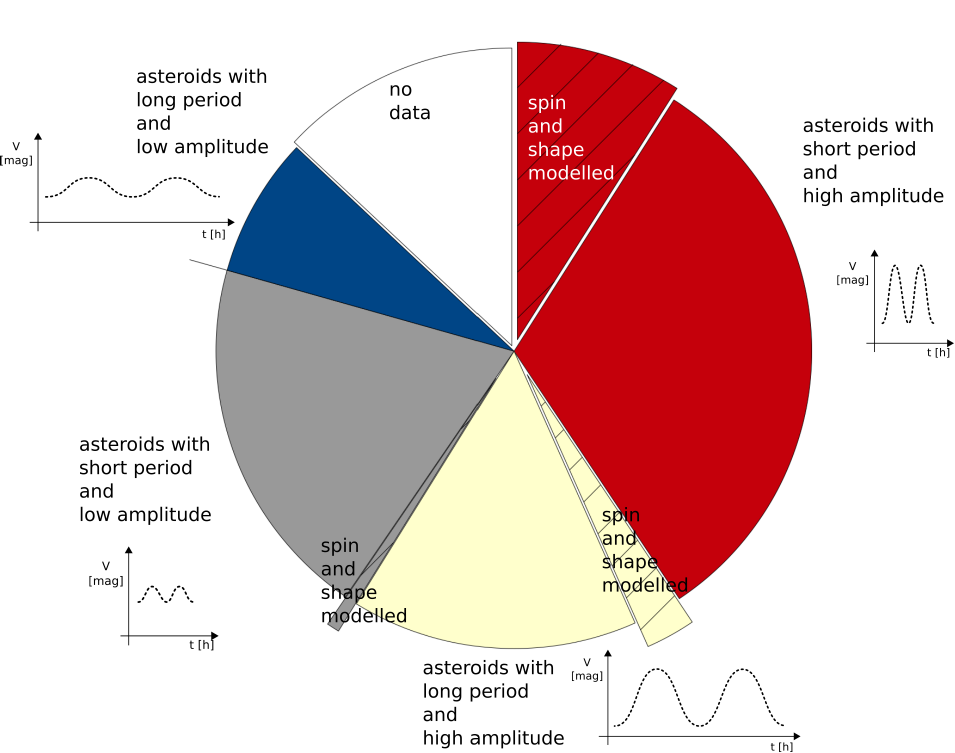}
\caption{Same as Fig. \ref{Pie_chart}, but for the $\sim$2270 fainter MB targets, with H between 11 and 13 mag  
    (source: LCDB). There are $\sim$270 large-amplitude targets 
    (from those on the right side of the chart) with available spin and shape model, 
    while only a few low-amplitude targets with a model (left side). 
    Judging from the sample of only those small asteroids that have available shape models, 
    and not taking into consideration the distribution of the amplitudes of all asteroids,  
    can create a false impression that almost all small asteroids are strongly elongated.
    }
\label{Pie_chart_faint}
\end{figure}

Statistical considerations in this section are based on the Minor Planer Center Lightcurve Database 
\citep[LCDB, ][updated 2016 September 5]{Warner2009}  
using a sample of the $\sim$1200 brightest 
main belt asteroids (those with absolute magnitudes H$\leq$11 mag, Fig. \ref{Pie_chart}),\footnote{The exact number of asteroids with certain H magnitude varies over time, due to 
updates in magnitude and albedo determinations gathered in LCDB.} 
which translates to diameters down to 12-37 km, depending on albedo (after MPC conversion table\footnote{{\tt http://www.minorplanetcenter.net/iau/lists/\break Sizes.html}}).
The rationale behind such a choice is that in this sample 97\% the main belt bodies have 
rotation period determined and available information on the lightcurve amplitude from at least one apparition. 
Among the fainter targets (H between 11 and 13 mag) there are many bodies with no information 
on the rotation parameters, so one cannot draw firm conclusions on the median period or amplitude. 
However, the selection effects discussed here are even more profound in the group of these fainter targets 
(equivalent diameters from 37 to 5 km, Fig. \ref{Pie_chart_faint}).

Because asteroid modelling using lightcurve inversion requires data from a wide variety of observing 
geometries, it is far more observationally demanding to gather a sufficient number of dense lightcurves over 
multiple apparitions for long-period targets (here those with P$\geq$12 hours) than for those 
with quicker rotation. However, not including them in spin and shape studies means omitting 
around half of the whole asteroid population in question (see the upper left and 
lower right part of Fig. \ref{Pie_chart}). 
Moreover, recent results from Kepler-K2 continuous observations spanning weeks show that there are substantially more 
slow-rotators among faint main belt asteroids and Jupiter Trojans  
than ground-based studies have shown 
\citep{Szabo2016, Szabo2017, Molnar2017}. 
Observations from the ground are naturally burdened with selection bias, 
absent when observing for long time spans from space.

Another problematic group of asteroids are those with low amplitudes of their brightness variations (here those with a$_{max}\leq$0.25 mag). 
They are almost as numerous as those with large amplitudes (greater than 0.25 mag); even so, they are 
spin and shape modelled very rarely (see the left part of Fig. \ref{Pie_chart}) because their study requires 
photometric data of very good accuracy, while data most often used for modelling asteroids nowadays 
come as a byproduct of large astrometric surveys. As such, these data are characterised by very low 
photometric accuracy \citep[0.1~-~0.2 mag on average,][]{Hanus2011}, so the modelling is missing 
most of the low-amplitude population \citep{Durech2016}. 

As a result there is a large `white spot' in the parameter space, where very little is known about 
large groups of asteroids (upper left part of Fig. \ref{Pie_chart}). 
We do not know their spin axis distribution, their shapes, or internal 
structure. 
Some of them may be tumbling, can be tidally despun by a large companion, or slowed down by the YORP effect.  
Also, their thermal inertia might be different than those rotating faster, as it seems 
to increase with the rotation period \citep{Harris2016} due to sampling of different depths 
that have different thermal properties. 
However, for now only 10\% of the asteroids observed 
in the infrared by IRAS and WISE space observatories have thermal inertia determined. 
It has been stressed that efforts should be made to carry out sophisticated thermophysical modelling 
of slowly rotating asteroids. Thermophysical modelling (TPM) techniques work best for objects with reliable shape and spin information.
  The existing multi-epoch, multi-wavelength thermal measurements can then be
  used to determine radiometric properties (effective size, geometric albedo,
  thermal inertia, surface roughness, emissivity) and to study if a given shape
  and spin solution can explain all measurements simultaneously  \citep[see e.g.][]{Muller2014b}.

\subsection{Observing campaign}

In order to counteract the above-mentioned selection effects, we are conducting an extensive and long-term observing 
campaign targeting  around a hundred bright (H$\leq$11mag) main belt asteroids that display both a long period of rotation (P>12 h) 
and a low lightcurve amplitude (a$_{max}\leq$0.25 mag), which are the objects that have been largely omitted in most of the previous 
spin and shape studies. 
We coordinate the multi-site campaign with about 20 observing stations placed around the world, from 
Europe through western US, to Korea and Japan. The detailed description of the campaign can be found in M2015.
Table \ref{sites} gives the information on the observing sites participating in this project. 
It also includes chosen sites of the group led by R. Behrend as we use some of the archival data gathered by this group, 
so far published only on the Observatoire de Geneve website\footnote{{\tt http://obswww.unige.ch/$\sim$behrend/page\_cou.html}}.

 \begin{table*}[h!]
   \begin{footnotesize}
 \begin{tabular}{|ccclc|}
 \hline
 \hline
  Site name & Abbreviation & IAU code & Location & Telescope\\
 \hline
  Borowiec Observatory (Poland) & Bor. & 187 & 52 N, 17 E & 0.4m \\
  Montsec Observatory (Catalonia, Spain) & OAdM & C65 & 42 N, 01 E & 0.8m\\
  Organ Mesa Observatory (NM, USA) & Organ M. & G50 & 32 N, 107 W & 0.35m\\
  Winer Observatory (AZ, USA) & Winer & 648 & 32 N, 111 W & 0.70m\\
  Bisei Spaceguard Center (Okayama, Japan) & Bisei & 300 & 35 N, 134 E & 0.5m and 1m \\
  Mt. Suhora Astronomical Observatory (Poland) & Suh. &   & 50 N, 20 E & 0.25m and 0.60m\\
  Le Bois de Bardon Observatory (France)  & Bardon &  & 45 N, 0 E & 0.28m\\ 
  Adiyaman Observatory (Turkey) & Adi. &  & 38 N, 38 E & 0.6m \\
  Derenivka Observatory (Ukraine) & Der.  & K99  & 48 N, 22 E & 0.4m \\
  JKU Astronomical Observatory, Kielce (Poland) & Kie. & B02 &  51 N, 21 E & 0.35m\\
  Pic du Midi Observatory (France) & Pic. & 586 & 43 N, 0 E & 0.6m\\
  Teide Observatory (Tenerife, Spain) & Teide & 954 & 28 N, 16 W & 0.8m \\
  Roque de los Muchachos (La Palma, Spain) & ORM   & 950 &  29 N,  18 W & 1m and 1.2m\\
  Kitt Peak National Observatory (AZ, USA) & KPNO  & G82 & 32 N, 112 W & 1m \\
  Lowell Observatory (AZ, USA)             & Lowell& 688  &  35 N, 112 W & 0.78m\\
  Command Module Observatory, Tempe (AZ, USA)& Tempe & V02  &  33 N, 112 W & 0.32 m\\
  Cerro Tololo Interamerican Observatory (Chile) & CTIO  & 807  &  30 S,  71 W & 0.6m \\
  La Sagra Observatory (Spain)             & La Sagra&   & 38 N, 3 W & 0.35m\\
  Piszkesteto Mountain Station (Hungary)   & Pisz. & 461 &  48 N, 20 E & 1m \\
  Sobaeksan Optical Astronomy Obs. (Korea) & Sobaek& 345 & 37 N, 128 E & 0.61m\\
  Flarestar Observatory (Malta)            & Flare.& 171 & 36 N, 14 E  & 0.25m\\
  Astronomy Observatory of Sertao de Itaparica (Brasil)  & OASI  & Y28 & 9 S, 39 W  & 1 m\\
 \hline
Observatoire des Engarouines  (France)& Engar. & A14 & 44 N, 5 E  & 0.21m\\
Le Cr{\`e}s (France) &  Le Cres & 177 & 44 N, 4 E  & 0.4m \\
Observatoire des Hauts Patys, B\'edoin (France) & Hauts Patys & 132 & 44 N, 5 E & 0.30m \\
OAM - Mallorca (Spain) & OAM & 620 & 40 N, 3 E & 0.3m \\
Stazione Astronomica di Sozzago (Italy) & Sozzago & A12 & 45 N, 9 E & 0.40m\\
 \hline
 \hline
 \end{tabular}
 \caption{Observing sites participating in this project}
 \label{sites}
 \end{footnotesize}
 \end{table*}

We perform unfiltered, or R-filter photometric observations of a given target until we get full rotation coverage and 
possibly also register notable phase angle effects. After that, the observations within one 
apparition are folded together in a composite lightcurve (Figs. \ref{Ortrud2016} - \ref{Venetia2015}) 
for synodic period determination. 
When the period is found to be in disagreement with the value in the MPC Lightcurve Database (LCDB), 
the observations concentrate on this target to confirm the new period value.
The observations are repeated in each apparition until data of good quality and quantity  from at least five well-spaced 
apparitions are gathered, including those already available in the literature.
In the course of the campaign the maximum amplitudes of some targets appeared to be larger than 0.25 mag, 
while periods of some others were shorter than 12 hours,
violating our initial selection criteria, nonetheless they remained on our target list.

Table \ref{obs} in  Appendix A summarises new observations for  11 targets studied in this paper 
(6 targets with corrected periods, and 5 with new models), 
presenting values important for spin and shape studies: 
mid-date of given lightcurve, sky ecliptic longitude of the target ($\lambda$), phase angle ($\alpha$), 
observing run duration, photometric error, and the observer's name with the observing site.

The best way to present the trustability of period determinations and the reliability of the obtained 
spin and shape models is to present the quality and quantity of supporting lightcurves and the model fit. 
Our data are presented in Appendix A in the form of composite lightcurves. 
Alongside lightcurves of modelled targets, we present the orientation on the zero phase of the best shape model, 
generated using the ISAM service\footnote{{\tt http://isam.astro.amu.edu.pl}}, described in \cite{Marciniak2012}. 
In Figs. \ref{Aemilia_fit}, \ref{Philosophia_fit}, \ref{Svea_fit}, \ref{Tergeste_fit}, and
\ref{Venetia_fit} we also present model example fits to lightcurves.

\section{Spin and shape modelling; scaling the models}

\subsection{Lightcurve inversion methods}
The Shaping Asteroids with Genetic Evolution (SAGE) modelling algorithm was 
developed at the Astronomical Observatory Institute of AMU Pozna{\'n}  \citep[][Bartczak \& Dudzi{\'n}ski, MNRAS, accepted]{Bartczak2014}. 
Thus, we utilise the local cluster with the SAGE code for 
the spin and shape modelling in parallel with the now classical convex inversion method 
by \cite{Kaasalainen1, Kaasalainen2}. 

SAGE  is a genetic algorithm that mutates the shape models 
to find the specimens that are best suited to lightcurve data. Although 
main belt asteroids can only be observed at relatively small phase angles (up to 30\deg at most), 
it has been shown that their lightcurves contain signatures of non-convex topographic features, so  
that these features can be successfully reproduced in the shape models 
(Bartczak \& Dudzi{\'n}ski, MNRAS, accepted). 
When modelling on lightcurves is a priori complemented by auxiliary data 
like adaptive optics or occultation contours in one multi-data inversion process, such non-concavities 
gain more support \citep[as in models created using ADAM algorithm,][]{Viikinkoski2015, Hanus2017}. 
However, when SAGE non-convex models based exclusively on lightcurves are a posteriori compared 
to  multi-chord occultations, their topographic features are confirmed, as has been shown in the case 
of binary asteroid (90) Antiope \citep{Bartczak2014}, but also in simulations and real-case studies 
performed recently by Bartczak \& Dudzi{\'n}ski (MNRAS, accepted).

The modelling here was performed independently using the convex inversion and SAGE methods, on the same datasets, 
taking as a starting value only the synodic period estimates from a set of composite lightcurves. 
The solutions for the poles and the shapes were searched over the whole possible range. 
From each method a set of internally consistent spin and shape 
solutions was obtained, and the uncertainty on the spin parameters was evaluated 
from the scatter of the best solutions for the pole (taking all the solutions with the best root mean square deviation (RMSD) 
enlarged by up to 10\%). 
The lightcurves produced by models from both methods 
fit the data around the noise level without big differences in the overall quality of the fit 
(measured by RMSD)
between the two methods, so it might seem that the  models fit the lightcurves in the same way. 
However, the overall sum of deviations does not reflect the subtle 
differences of the lightcurve fits between the two methods, like sometimes visible  better fitting 
of the SAGE models to critical features (e.g. deep minima or abrupt dimmings),  
where non-convex features most clearly manifest themselves. 
Such features, due to their short duration, usually contain far fewer datapoints than other lightcurve 
fragments, so their influence on the RMSD value is very small. However during the SAGE optimisation process 
the biggest weight is 
given to the worst fitting lightcurves, so in further iterations these fragments  
have a bigger influence on the shape model and are fitted better. Still, the final (unweighted) RMSD value 
might be the same, 
when other lightcurves have a  slightly worse fit, and the large number of points in them 
makes the small change more significant for RMSD. 
So, using only the RMSD of the fit, we have no means to tell which model best represents the real shape. 
Here we present one of possible solutions for the shape chosen from a family of very similar shape models; 
however, without a method to estimate shape uncertainties, it is hard to compare the performance of the two methods.

The shape models from the two methods were often  similar to each other, clearly indicating that convex models 
are the convex hulls of more complex shapes, successfully reproduced by the SAGE algorithm.  
However, in some cases the shapes looked distinctively different, and only the pole-on projections 
were similar.
The orientation of the two models in pairs of figures like \ref{AemiliaINVshape} and \ref{AemiliaSAGEshape} 
is the same, so these shape projections can be directly compared. 
Different positions of the x- and y-axes are caused by their different definitions: 
in SAGE models the rotation axis is the axis of biggest inertia, 
and the x-axis of the smallest inertia. In convex models, the z-axis should also correspond to the biggest inertia, but   
the x-axis is connected with the epoch of the first observation, so its orientation 
does not correspond to any specific feature of the shape model\footnote{ There is a different sequence of rotations in the reference frame definitions of the 
convex and non-convex models, 
so if both models were to be placed in the plane of sky, 
the rotation of -270$\deg$ around the z-axis 
would be necessary for the SAGE models to match the orientations of the convex models.}

\subsection{Thermophysical modelling} 

This  radiometric technique consists in the exploitation of thermal data in the mid- to far-infrared
and data in the visible. Thermophysical models allow the derivation of size, albedo, and thermal
properties for small bodies \citep[see][and references therein]{Delbo2015}.
There are different model implementations available, ranging from simple thermal models assuming
spherical shapes at opposition without heat conduction into the surface to more
sophisticated thermophysical model implementations which take complex shapes and rotational properties
into account;  at the same time heat conduction, shadowing effects, and self-heating
effects are calculated for a given illumination and observing geometry.
Here, we are interested in assigning reliable scales
to the obtained spin-shape solutions, deriving high-quality geometric albedos, estimating
the surface's thermal inertia, and finding indications for the levels of surface roughness.
For our analysis, we therefore used a TPM code developed by \citet{Lagerros1996, 
Lagerros1997, Lagerros1998} and extensively tested and validated \citep[e.g. by][]{Muller1998, Muller2002}.
The TPM allows the use of all kind of shape solutions (convex and non-convex). It considers the
true observing and illumination geometry to calculate the surface temperature distribution
for any given epoch. The 1D  heat conduction into the surface, shadowing, and self-heating
effects are calculated. Good examples for TPM applications to main belt asteroids can be found
in \citet{Muller2014a} for Ceres, Pallas,Vesta, and Lutetia, or in \citet{Marsset2017} for Hebe.

We applied the following procedure:
\begin{itemize}
\item We use a given convex or non-convex shape-spin solution (see previous section);
\item The small-scale surface roughness is approximated by hemispherical segment craters
  covering a smooth surface. We consider different levels of roughness ranging
  from 0.1 to 0.9 for the rms of the surface slopes;
\item The thermal inertia is considered as a free parameter, with values between zero
  (i.e. no heat conductivity, surface is in instantaneous equilibrium with the insolation)
  and 2000 Jm$^{-2}$K$^{-1}$s$^{-1/2}$ (bare rock surface with very high heat conductivity);
\item The characterisation of the reflected light is given by the H-G
  (or H-G1-G2) solutions; 
\item For each observed and calibrated infrared measurement we determine all possible
  size and albedo solutions for the full range of thermal inertias and roughness levels; 
\item We search for the lowest $\chi^2$ solution in size, albedo, and
    thermal inertia/roughness for all thermal IR measurements combined;
\item We calculate the 3-$\sigma$ solutions for the available set of
    thermal measurements: We consider 1/(N-$\nu$) where N is the
    number of (thermal) measurements and $\nu$ is the number of
    free parameters, here $\nu$ = 2 because we fit for diameter
    and thermal inertia. We also fit for albedo, but here we make use of another measurement (the H magnitude). We define
    the n-$\sigma$ confidence interval by accepting all solutions
    that have 
\begin{center}
\begin{equation}
     \chi^2$ < $\chi^{2}_{min}$ + $n^2,
\end{equation}
\end{center}
where $\chi^2$ is the actual
\begin{center}
\begin{equation}
    \chi^2 = \sum \left(\frac{obs - mod}{err}\right)^2;
\end{equation}
\end{center}
\item Solutions are only accepted if the reduced $\chi^2$ values are
    reasonably close to 1.0. In this case the `unreduced' $\chi^2$
    will have a minimum equal to N-2, and the 3-$\sigma$ limit for
    N observations is at N-2 + 3$^2$ = N - 7;
\item The minima for the reduced $\chi^2$ for each shape and spin solution
    are given in Table \ref{chi2}.
\end{itemize}

The results of this procedure are the following:
\begin{itemize}
\item We find the best radiometric size which corresponds to the size of an equal-volume
  sphere and can be used to scale the given shape-spin solution; 
\item We determine the geometric albedo (closely connected to the given H magnitude); 
\item We estimate the possible range of thermal inertias (higher  or lower values
  would introduce problems when comparing pre- and post-opposition  IR data); 
\item Assuming low roughness gives lower values for the thermal inertia, higher
  levels of roughness lead to slightly higher  thermal inertias. Our IR data are
  usually not good enough to break the degeneracy between thermal inertia and roughness, but we consider
  this aspect in the solutions in Table \ref{TPMresults};
\item In some cases the minimum $\chi^2$ values for the different shape-spin solutions
  for a given target are very different: in these cases we favour the solution
  with the best $\chi^2$ fit.
\end{itemize}

The radiometric technique is not very sensitive to the exact shape, and provides
sizes and albedos with around 5\% accuracy in the most favourable cases. It is
the most productive way of determining sizes and albedos for large samples of
asteroid IR measurements (as coming from IRAS, AKARI, WISE surveys), but it also allows
 spin properties to be constrained and  wide ranges of shape-spin solutions to be discarded.
The radiometric analysis uses thermal data from different epochs, phase angles,
wavelengths, and rotational phases. The resulting radiometric size is therefore
closely related to the full 3D body, while occultations are only representative
of the 2D cross section of the body.

\subsection{Stellar occultation fitting}

  Stellar occultations by main belt asteroids are being observed by a few active groups 
(like Noth American\footnote{\tt{ http://www.asteroidoccultation.com/observations/\break Results/}}, 
European\footnote{\tt{ http://www.euraster.net/results/index.html}}, 
or East Asian observers\footnote{\tt{ http://sendaiuchukan.jp/data/occult-e/\break occult-e.html}}), 
and published in the Planetary Data 
System\footnote{\tt{ http://sbn.psi.edu/pds/resource/occ.html}} \citep[PDS, see][]{Dunham2016}, 
providing great complementary data for asteroid physical studies. 
Occultation timing measurements of such events enable scaling of 
the otherwise scale-free shape models, 
and also  confirm their major and intermediate-size topographic features.
Very often they can also break the mirror-pole symmetry intrinsic to the lightcurve inversion models. 

When the occultation observation is successful and at least three well-spaced chords are obtained
with good accuracy, it is possible to overlay the occultation shadow chords and the photometric asteroid 
model \citep[as in e.g.][]{Timerson2009, Durech2011} 
with relatively small uncertainty regarding the exact position of the model contour.

Of the five targets modelled here, these multichord events were available for two of them  
and it allowed us to independently scale, compare, and verify their spin and shape models. 
The translation of the timings from PDS to chords on the Earth fundamental plane 
($\xi$, $\eta$) has been done using the method described in \cite{Durech2011}. 
Both convex and non-convex 3D shape models obtained here have been translated into scalable 2D contours, 
according to sky-plane shape orientation for a given moment, and then overlaid on the timing chords 
so as to minimise the overall rms deviations between the contour and the chords, taking into account the timing 
uncertainties. As a result, the models were scaled in kilometres with good accuracy; 
the maximum size of a given shape model was later  translated into the diameter 
of the equivalent volume sphere. Results are described and plotted in Section 4.
The list of all the observers of asteroid occultations that 
were utilised in this work can be found in Appendix B \citep{Dunham2016}.

\section{Corrected period determinations}

The first and rather unexpected result of our observing campaign was that as much as 25\% of the numerous bright  
main belt asteroids with both long period and small amplitude had a previously 
incorrectly determined synodic period of rotation (M2015). 
Their period quality codes in LCDB were 3, 2+, and 2. Although periods with code 2 and lower should be 
considered unreliable; usually, all period values with codes higher than 1+ are taken into account 
in the majority of spin state studies of asteroids.
The wrong period determination in the cases that we studied was due to previous incomplete or noisy lightcurve 
coverage, which often led the alias period to be incorrectly identified as the true rotation period.

As an example, in Figs. \ref{Ortrud2016} - \ref{Hooveria2016} we present a few more cases 
where we found rotation periods 
substantially different from the values accepted in LCDB \citep{Warner2009}. 
Below, we briefly review previous works on these targets and describe our findings. 
Their previous and new period values are presented in Table \ref{periods}. 
Together with targets for which we already had corrected period values  \citep[M2015, and][]{Marciniak2016},  
their overall number 
 (16) compared to the number of our targets for which we found secure period determinations (65) 
confirms our previous findings that around a quarter of bright long-period asteroids with low amplitudes 
had incorrectly determined rotation periods. 
More precisely, out of 16 targets with incorrect periods, four targets had period quality code 3, two had code 2+, 
and ten had code 2. So if only the reliable periods  (code 3 and 2+)  were considered, the percentage 
of incorrect values in the group of bright long-period, low-amplitude targets would be around 10\%.

\begin{table*}[h]
\begin{small}
\begin{tabular}{|l|c|c|c|l|}
\hline
& amplitude (LCDB && Period &\\
 asteroid name     & and \textit{this work})& Period (LCDB)& quality & Period (this work) \\
                   & \hspace{0.5cm}[mag]     &\hspace{0.5cm}[h]& code & \hspace{0.5cm} [h]  \\
\hline
Targets with new periods: &&&&\\
 (551) Ortrud       &  0.14 - \textit{0.19}&  13.05 & 2 & {\bf 17.416} $\pm$ 0.001\\
 (581) Tauntonia    &  0.07 - 0.20         &  16.54 & 2 & {\bf 24.987} $\pm$ 0.007\\
 (830) Petropolitana&  0.15 - \textit{0.42} & 39.0  & 2 & {\bf 169.52} $\pm$ 0.06\\
 (923) Herluga      &  0.16 - \textit{0.28} & 19.746& 2 & {\bf 29.71} $\pm$ 0.04\\
 (932) Hooveria     &  0.20 - \textit{0.24} & 39.1  & 2+ & {\bf 78.44} $\pm$ 0.01\\
 (995) Sternberga   &  0.06 - 0.20 & 14.612 & 2+ & {\bf 11.198} $\pm$ 0.002\\
\hline
Targets with models: &&&&\\
 (159) Aemilia      & 0.17 - 0.26          & 24.476 & 3  & 24.486 $\pm$ 0.002\\
 (227) Philosophia  & 0.06 - 0.20          & 52.98  & 2 A & {\bf 26.468} $\pm$ 0.003\\
 (329) Svea         & 0.09 - \textit{0.24} & 22.778 & 2+ & 22.777 $\pm$ 0.005\\
 (478) Tergeste     & 0.15 - 0.30          & 16.104 & 2+ & 16.105 $\pm$ 0.002\\
 (487) Venetia      & 0.03 - 0.30          & 13.34  & 3  & 13.342 $\pm$ 0.002\\
\hline
\end{tabular}
\caption{Synodic periods and amplitude values found within this project
compared to literature data gathered previously in LCDB. 
Boldface indicates period determinations substantially differing 
from previously accepted values.}
\label{periods}
\end{small}
\end{table*}

\subsection{(551) Ortrud}
The first report on lightcurve and period of (551) Ortrud was made by \cite{Robinson2002}, 
who determined a 13.05 h period based on an asymmetric, bimodal lightcurve from the year 2001. 
Although three consecutive works on this target, Behrend et al. (www) in 2003 and 2006, 
and \cite{Buchheim2007} in 2006 
reported a different period (17.59, 17.401, and 17.416 hours, respectively), the adopted value in LCDB remained unchanged 
due to the low quality code assigned to these determinations. 

During our observations, we found that only the period of $17.420 \pm 0.001$ hours can fit the data we gathered in 
2016 (Fig. \ref{Ortrud2016}), confirming the findings from the three latter works. 
So it turned out that the correct period has already been identified, but our data put it on firmer ground.
The amplitude was at the level of $0.19 \pm 0.01$ mag. The lightcurve, as in each observed apparition, 
is characterised by narrow minima and wide 
complex maxima.

\subsection{(581) Tauntonia}

Previously observed by group led by R. Behrend in 2005 and 2006, Tauntonia displayed very low amplitude lightcurves 
that seemed to fit a period of around 16.5 - 16.2 hours (Behrend et al., www).
\cite{Stephens2010} found instead that  the period was 24.90 hours, based on an asymmetric 
0.20 mag amplitude lightcurve from the year 2010. 

Our data from 2016 can be best folded with period $24.987 \pm 0.007$ hours, creating an unusual though consistent composite lightcurve 
 (Fig. \ref{Tauntonia2016}), and $0.18 \pm 0.02$ mag amplitude, 
confirming the determination by \cite{Stephens2010}.

\subsection{(830) Petropolitana}

The  only lightcurve observations of Petropolitana were reported by Behrend et al. (www), 
with a period estimated to 39.0 hours, based only on three separate fragments. 
In \cite{Hanus2016}, there is a model of this target based exclusively on sparse data from 
astrometric sky surveys, where the  sidereal period is 37.347 hours, found by scanning 
a standard period span of up to 100 hours. 

Our observations suggest a much longer period: $169.52 \pm 0.06$ hours, based on calibrated data 
with nightly zero point adjustments (Fig. \ref{Petropolitana2017}). The lightcurve behaviour is bimodal with a large amplitude ($0.42\pm 0.02$ mag).
So this is a very long-period target, but not  low-amplitude.

\subsection{(923) Herluga}

The only previous work on the lightcurve of (923) Herluga was published by \cite{Brinsfield2009}.
The period determined at that time, 19.746 h, was based on an imperfect composite lightcurve with 
some clearly misfitting fragments. 

Our observations of this target did not allow us to find a satisfactory fit to any period until 2016, 
when we gathered 11 long lightcurve fragments. The only period that fits the new 
data (and data from all the previous observations) is $29.71 \pm 0.04$ hours, which applied to the data from the year 2016 
reveals a complex, trimodal lightcurve where one of the minima is deeper than the others (Fig. \ref{Herluga2016}). 
The amplitude was unusually large for this target: $0.28 \pm 0.02$ mag.

\subsection{(932) Hooveria}
The first period determinations for Hooveria, 29.947 or 30.370 hours, were made by \cite{Sada2004}  
from a bimodal folded lightcurve behaviour. Another set of data was obtained by \cite{Warner2010} and a period of 
39.15 hours was found, producing a monomodal lightcurve of rather large amplitude for this type (0.22 mag).
In the same work, \cite{Warner2010} reanalysed the data obtained by \cite{Sada2004} and was also able to fit them 
with a 39.15-hour period, now making it  monomodal. 

Our extensive observations of Hooveria in late 2016 and careful nightly zero point adjustments using 
CMC15, APASS, and GAIA catalogue stars
have shown that the rotation period of Hooveria must be twice as long, being $78.44 \pm 0.01$ hours and
producing a bimodal lightcurve with clearly asymmetric extrema and $0.24 \pm 0.01$ mag amplitude (Fig. \ref{Hooveria2016}). 
Fitting these data with a 39-hour period would require large shifts in reduced magnitudes of steps  
bigger than 0.05 mag, much larger than the absolutisation errors.

\subsection{(995) Sternberga}

 All of the previous reports on the period of (995) Sternberga claimed different values:
\cite{Barucci1992} give 16.406 hours; Behrend et al. (www) estimated P > 12 h; 
\cite{Stephens2005} found 15.26 h, later corrected to 14.612 h in \cite{Stephens2013} based on new data 
of larger amplitude. 

Our analysis of this target since the beginning suggests that none of the previous values can be confirmed, 
and instead the period is either 22.404 hours or 11.202 hours \citep{Marciniak2014}. 
Finally, data from the apparition in 2016 confirmed the lower value providing a good fit 
to $11.198 \pm 0.002$ hours; this period was unambiguously found in spite of a very small amplitude of $0.06 \pm 0.01$ mag 
(Fig. \ref{Sternberga2016}). Also, it fits all the previously obtained data.

\vspace{0.5cm}

In summary, the substantial number of periods that needed a revision
was found among the brightest main belt targets (H$\leq$11) available to most 
 small telescopes. Among the fainter targets these effects can be expected to an even greater extent, due to 
more noise  in the photometric data. So one has to be careful when interpreting, for example a frequency-diameter plot, especially 
in the regions where fainter targets reside (diameters less than $\sim$ 30 km). 
Many  such targets might have incorrect period values, 
but a huge number of them are simply not present in the plot because their periods are unknown. 
Those that are present in the small diameter range of the frequency--diameter plot are strongly influenced by observing biases, 
favouring large amplitudes and short periods.

From our campaign, since the beginning of the project in 2013, we have gathered around 8000 hours of photometric data, 
resulting in a few tens of full composite lightcurves of our long-period, low-amplitude targets each year. 
This dataset enables spin and shape modelling of the first representatives of our sample.

\section{Individual models}

In the following we provide the description of previous works on given target and the new data obtained within 
this work, presented as composite lightcurves in Figures \ref{Aemilia2005} - \ref{Venetia2015} in  Appendix A. 
Next we describe the modelling process and the results of the spin and shape solutions presented in 
Table \ref{results} and pairs of figures (see Figs. \ref{AemiliaINVshape} and  \ref{AemiliaSAGEshape}). 
Table \ref{results} gives the spin solutions from both methods with uncertainty and RMSD 
(root mean square deviation) values. 
The first column gives the sidereal period value, the next four columns give two pairs of solutions for the 
north pole of the spin axis (J2000 ecliptic coordinates), all with uncertainty values. 
In the fifth column there is the observing span in years, number of apparitions ($N_{app}$), and individual lightcurves 
($N_{lc}$) used to create the models. The last column provides the code of the modelling method.
Tables \ref{occ_diameters159} and \ref{occ_diameters329} give the values for the diameters from the occultation fitting, 
and Table \ref{TPMresults}  the diameters from thermophysical modelling, both techniques described in the following sections. 
Additionally,  Table \ref{TPMresults} gives the best fitting albedo and thermal inertia values.
For  reference, the effective diameters from IRAS \citep{Tedesco2004}, AKARI \citep{Usui2011}, 
and WISE \citep{Mainzer2011, Masiero2011} surveys are given.

A model example fit to the lightcurves is presented in   Fig. \ref{Aemilia_fit}, \ref{Philosophia_fit}, and others. 
Additionally, to visualise what combination of aspect and shape can produce the given lightcurves, next to the composite 
lightcurves in  Appendix A we present shape models oriented at zero epoch using the ISAM 
service\footnote{{\tt http://isam.astro.amu.edu.pl}}. 
On the web page these plots can be set in motion, together with the rotating shape model.

\begin{table*}[h!]
\begin{small}
\begin{tabular}{rrrrcccccc}
\hline
Sidereal      & \multicolumn{2}{c}{Pole 1} & \multicolumn{2}{c}{Pole 2} & RMSD  & Observing span & $N_{app}$ & $N_{lc}$ & Method \\
period [hours]& $\lambda_p$ & $\beta_p$    & $\lambda_p$ & $\beta_p$    & [mag] &   (years)      &           &          &    \\
\hline
&&&&&&&&&\\
{\bf (159) Aemilia} & & & & & & & & & \\
$24.4787$   & $139\deg$     & $+68\deg$   & $348\deg$    & $+59\deg$    & 0.014 & 1981--2015 & 6 & 45 & convex LI \\  
$\pm 0.0001$& $\pm 18\deg$  & $\pm 8\deg$ & $\pm 18\deg$ & $\pm 6\deg$ &  &  &  &  & \\
$24.4787$   & $139\deg$     & $+66\deg$   & $349\deg$    & $+63\deg$   & 0.014  & " & " & " & SAGE \\  
$\pm 0.0001$& $\pm 7\deg$   & $\pm 5\deg$ & $\pm 7\deg$  & $\pm 6\deg$ &  &  &  &  & \\
&&&&&&&&&\\
{\bf (227) Philosophia} & & & & & & & & & \\
$26.4614$   & $ 95\deg$     & $ +19\deg$  & $272\deg$    & $-1\deg$    & 0.011  & 2006--2016 & 5 & 97 & convex LI \\  
$\pm 0.0001$& $\pm 5\deg$   & $\pm 4\deg$ & $\pm 6\deg$ & $\pm 2\deg$  &  &  &  &  & \\
$26.4612$   & $ 97\deg$     & $+16\deg$   & $271\deg$    & $ 0\deg$    & 0.009  & " & " & " & SAGE \\  
$\pm 0.0003$& $\pm 5\deg$   & $\pm 5\deg$ & $\pm 5\deg$  & $\pm 5\deg$ &  &  &  &  & \\
&&&&&&&&&\\
{\bf (329) Svea} & & & & & & & & & \\
$22.7670$   & $ 33\deg$   &   $+51\deg$   &  -    &  -   & 0.010  & 1986--2016 & 6 & 60 & convex LI \\  
$\pm 0.0001$& $\pm 15\deg$  & $\pm 10\deg$&  -    &  -   &  &  &  &  & \\
$22.7671$   &{\bf 21\deg}& {\bf +47\deg}  &  -    &  -   & 0.011    & " & " & " & SAGE \\  
$\pm 0.0002$& $\pm 7\deg$   & $\pm 5\deg$ &  -    &  -   &  &  &  &  & \\
&&&&&&&&&\\
{\bf (478) Tergeste} & & & & & & & & & \\
$16.10308$   & $ 2\deg$     & $-42\deg$   & $216\deg$    & $-56\deg$    & 0.011  & 1980--2016 & 6 & 48 & convex LI \\  
$\pm 0.00003$& $\pm 2\deg$  & $\pm 3\deg$ & $\pm 6\deg$  & $\pm 4\deg$  &  &  &  &  & \\
$16.10312$   & $ 4\deg$     & $-43\deg$   & {\bf 218\deg} & {\bf -56\deg}    & 0.011  & " & " & " & SAGE \\  
$\pm 0.00003$& $\pm 6\deg$  & $\pm 5\deg$ & $\pm 9\deg$  & $\pm 7\deg$  &  &  &  &  & \\
&&&&&&&&&\\
{\bf (487) Venetia} & & & & & & & & & \\
$13.34133$   & $ 78\deg$    & $+3\deg$    & $252\deg$    & $+3\deg$     & 0.012  & 1984--2015 & 8 & 34 & convex LI \\  
$\pm 0.00001$& $\pm 7\deg$  & $\pm 10\deg$& $\pm 8\deg$  & $\pm 12\deg$ &  &  &  &  & \\
$13.34133$   & $ 70\deg$    & $+8\deg$    & {\bf 255\deg} & {\bf +8\deg} & 0.011  & " & " & " & SAGE \\  
$\pm 0.00002$& $\pm 6\deg$  & $\pm 11\deg$ & $\pm 5\deg$  & $\pm 10\deg$  &  &  &  &  & \\
&&&&&&&&&\\
\hline
\end{tabular}
\caption{Parameters of the spin models of the five targets studied here, and  the uncertainty values. 
Column 1 gives the sidereal period of rotation; Cols. 2 -5 give  two sets of pole J2000.0 longitude and latitude; Col. 6  gives the rms deviations of the model lightcurves from the data; Cols. 7 - 9 give the photometric dataset 
parameters (observing span, number of apparitions, and individual lightcurve fragments). The last column contains 
the name of the lightcurve inversion (LI) method. The  preferred pole solutions are shown in bold. 
The second pole solution of (329) Svea, though possible in the lightcurve inversion, 
was clearly rejected by occultation fitting.}
\label{results}
\end{small}
\end{table*}

 \subsection{(159) Aemilia}
Lightcurves of (159) Aemilia have been previously obtained by 
\cite{Harris1989}, Behrend et~al.~(www), \cite{Ditteon2007}, and \cite{Pilcher2013}.
Initially there was controversy over whether the rotation period is close to 16 or  24 hours; this issue was resolved by \cite{Pilcher2013} based on multiple coverage from the year 2012 
folded with a period of $24.476$ hours. The lightcurve amplitudes varied from $0.17$ to $0.26$ mag.

We  observed Aemilia in  two other apparitions, in 2014 and 2015. Additionally, we present here  
unpublished lightcurves from 2005 obtained by the group led by Raoul Behrend and based on incomplete coverage. 
The morphology of the new lightcurves was similar to previously observed ones; there were  characteristic ``shelves'' 
after the maxima, one of which had a tendency  to evolve to a third maximum when observed 
at a larger phase angle (Figs. \ref{Aemilia2005} - \ref{Aemilia2015} in  Appendix A). The synodic periods 
of the composite lightcurves were around $24.49$ hours, with amplitudes from $0.24$ mag to $0.18$ mag.

\begin{figure}[h]
\includegraphics[width=0.5\textwidth]{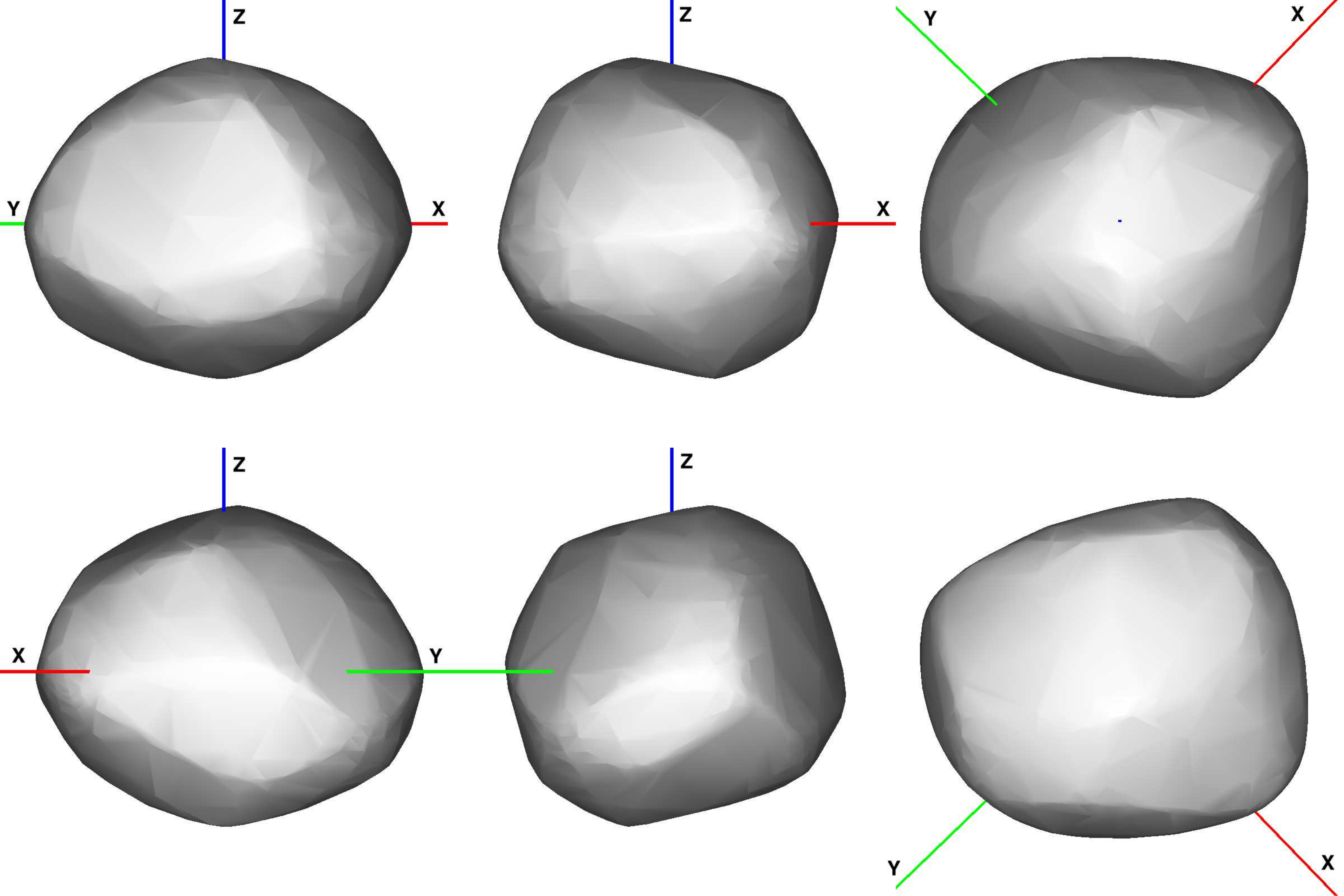}
\caption{Convex shape model of (159) Aemilia from the lightcurve inversion method shown in six projections.
The z-axis is the axis of rotation. Compare with Fig.\ref{AemiliaSAGEshape}.}
\label{AemiliaINVshape}
\end{figure}

\begin{figure}[h]
\includegraphics[width=0.5\textwidth]{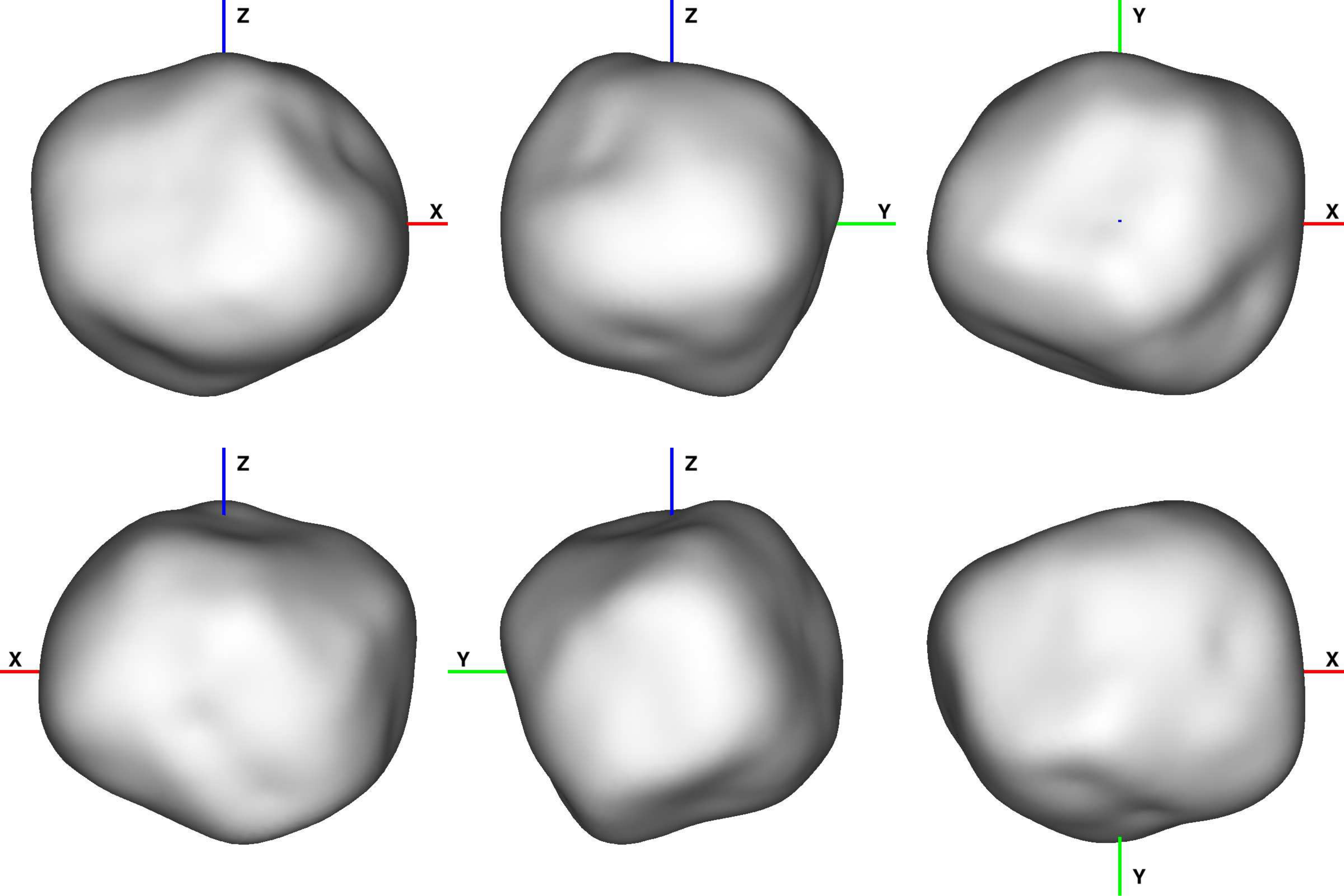}
\caption{Non-convex shape model of (159) Aemilia from the SAGE algorithm shown in six projections. The z-axis is the axis of rotation, while the x-axis is the longest axis of the shape model.}
\label{AemiliaSAGEshape}
\end{figure}

The dataset for the lightcurve inversions consisted of 45 individual lightcurve fragments 
from six apparitions (1981, 2005, 2006, 2012-2013, 2014, and 2015), 
well spread over the asteroid orbit and a range of phase angles (see Table \ref{obs}). 
We did not use the short and noisy fragment from 2008; all the other available data were used 
in the modelling process. The dataset consisted of around 200 hours of dense lightcurve observations.

In the convex inversion, the spherical harmonics expansion and convexity regularisation weight 
had to be increased in order to produce realistic physical shape models (Fig. \ref{AemiliaINVshape}). 
The sidereal period value 
and both solutions for the spin axis (Table \ref{results}) clearly stood out in the parameter space in terms of 
lowest RMSD (0.014 mag). The example fit to the lightcurves is shown in Fig. \ref{Aemilia_fit}. 
The last lightcurve from the apparition in 2014, and the first one from 2015, 
both obtained at large phase angles, had the worst fit  to the model lightcurves.
All the resulting shape solutions were roughly similar to each other.  Some shapes 
resembled a deltoid, while others were more ellipsoidal; there were small differences 
in the vertical dimensions. Here we present only one of the possible shapes for pole 1, which  
has been the standard practice in presenting lightcurve inversion solutions.

The non-convex model obtained with the SAGE algorithm fits the lightcurves similarly well (RMSD=0.014 mag, Fig. 
\ref{Aemilia_fit}) and  similar spin solutions  were found (Table \ref{results}), 
but the shape is more compact, with slight indentations and some large bulges (Fig. \ref{AemiliaSAGEshape}). 
The genetic evolution runs all led to the final shapes that were very similar to each other, 
and the only differences were in the depth of the largest `basins', which were still present 
on each final shape. The final solution had spin axis parameters 
close to the average of all the obtained solutions and had the lowest RMSD.
\begin{table}[h]
\begin{tabular}{ccc}
              &  Pole 1          &     Pole 2 \\
\hline
       CONVEX &  130 $\pm$ 7 km &  130 $\pm$ 8 km \\
       SAGE   &  135 $\pm$ 7 km &  138 $\pm$ 7 km \\      
\end{tabular}
\caption{Equivalent volume sphere diameters of (159) Aemilia models fitted to the occultation from 
2 May 2009. Compare with radiometric diameter from TPM in Table \ref{TPMresults}.}
\label{occ_diameters159}
\end{table}

The fitting to all four solutions (two mirror poles from the convex inversion and two from the SAGE algorithm) 
to the four-chord occultation from 2 May 2009 \citep{Dunham2016} does not provide a preferred solution for the pole or shape, 
but allows us to scale the model (see Fig. \ref{Aemilia_occult}).
The size of both convex and non-convex models fitted to this occultation yields equivalent volume sphere 
diameters from $130$ to $138$ km; the SAGE solutions are a few kilometres larger than the convex models 
(see Table \ref{occ_diameters159}). 
In Table \ref{TPMresults}, we present the radiometric size for the model solution that best fits in thermophysical modelling, 
i.e. $137$ km, in very good agreement with the  size from occultations.

The application of inversion models of (159) Aemilia in thermophysical modelling is a rare example 
of a remarkably good fit with no trend in the O-C plots (see Figs. \ref{Aemilia_TPM}). 
These O-C plots show nicely if a given model solution (size, shape,
thermal properties) can explain all the thermal measurements simultaneously.
Ratios close to 1.0 (solid line) indicate an excellent match between
observation and the corresponding model prediction; ratios in the
range 0.9 and 1.1 (dashed lines) reflect typical calibration
uncertainties of thermal measurements. As a rule of thumb, a 10\% flux
error roughly translates  into a 5\% error in the object's radiometric
size solution. Finding many data points outside the +/-10\% lines
usually indicates that the shape/spin solution has some problems.
Therefore, systematic offsets in the O-C plots indicate  a problem
with the radiometric size solution. Strong trends in the Obs/TPM
ratio with wavelength point towards problems with the thermal
surface properties (thermal inertia and roughness), an asymmetry
in the pre- and post-opposition  ratios are connected to an incorrect
thermal inertia, while outliers in the rotational-phase plot
point to shape-related issues. 
We used H=$8.100$ mag and G=$0.09$, 
after \cite{Pravec2012}, and infrared data from IRAS (6 x 4 band detections), AKARI (5 datapoints), 
and WISE W3/W4 bands (20 datapoints).
Both convex and non-convex models with both pole solutions fit the data similarly well, and substantially better 
than a spherical model  (see Table \ref{chi2}). 

The first model solution from the SAGE method ($\lambda$ = 139\deg, $\beta$ = 66\deg) seems to be the overall best solution 
(the reduced $\chi^2$ of 0.44)
and intermediate level of surface roughness, optimum thermal inertia around 50 SI units (higher for higher roughness, 
lower for lower roughness), effective size of around 137.0 km (around 10 km larger than in previous determinations), 
and geometric V-band albedo of 0.054. Uncertainty values can be found in Table \ref{TPMresults}.
The radiometric size is in agreement with lower values for the size from occultation fitting, 
but is still slightly higher than in all previous determinations that used a spherical model for the shape, 
also partly due to lower albedo than in previous works (see Table \ref{TPMresults}).


 \subsection{(227) Philosophia}
(227) Philosophia has been  observed by many authors, e.g. 
\citet{Bembrick2006}, 
\citet{Ditteon2007}, 
Behrend et al. (www), 
\citet{Alkema2013}, 
\citet{Pilcher2014a, Pilcher2014b},   
but the controversy regarding its 
rotation period remains (see our discussion on this target in M2015). In our previous work 
we considered a period of $26.46$ hours as the most probable, based on our monomodal lightcurve from 
the apparition on the verge of 2013 and 2014. Still, the currently accepted value in LCDB  
is twice as long, $52.98$ hours; however, it is annotated as not fully certain and ambiguous 
(code 2, and label A). The reported amplitudes ranged from $0.06$ to $0.20$ mag, but these values 
can be influenced by incorrect periods used for folding the lightcurves.

During the observing campaign within this work, we obtained extensive datasets from two more 
apparitions of Philosophia, in 2015 and 2016, in addition to the one from 2013-2014. 
In both of them a clearly bimodal behaviour 
over the shorter period timescale has been recorded, which resolves the problem of uncertain period, 
confirming our value of $26.46$ hours (see Figs. \ref{Philosophia2015} and \ref{Philosophia2016} in  Appendix A). 
This period fits all the available data from previous apparitions. In additional, we present here the 
data from apparition in 2006 from Behrend et al. (www) and \cite{Ditteon2007} folded together (Fig. \ref{Philosophia2016}).
Overall, the behaviour of the lightcurve variations changes from monomodal to bimodal with minima of unequal depth, 
and other irregularities. Curiously, monomodal lightcurves of this target do not display smaller amplitudes 
than bimodal ones, contrary to what is usually the case;  instead, the amplitude remains on a stable level 
of around $0.15$ mag in all apparitions.

\begin{figure}[h]
\includegraphics[width=0.5\textwidth]{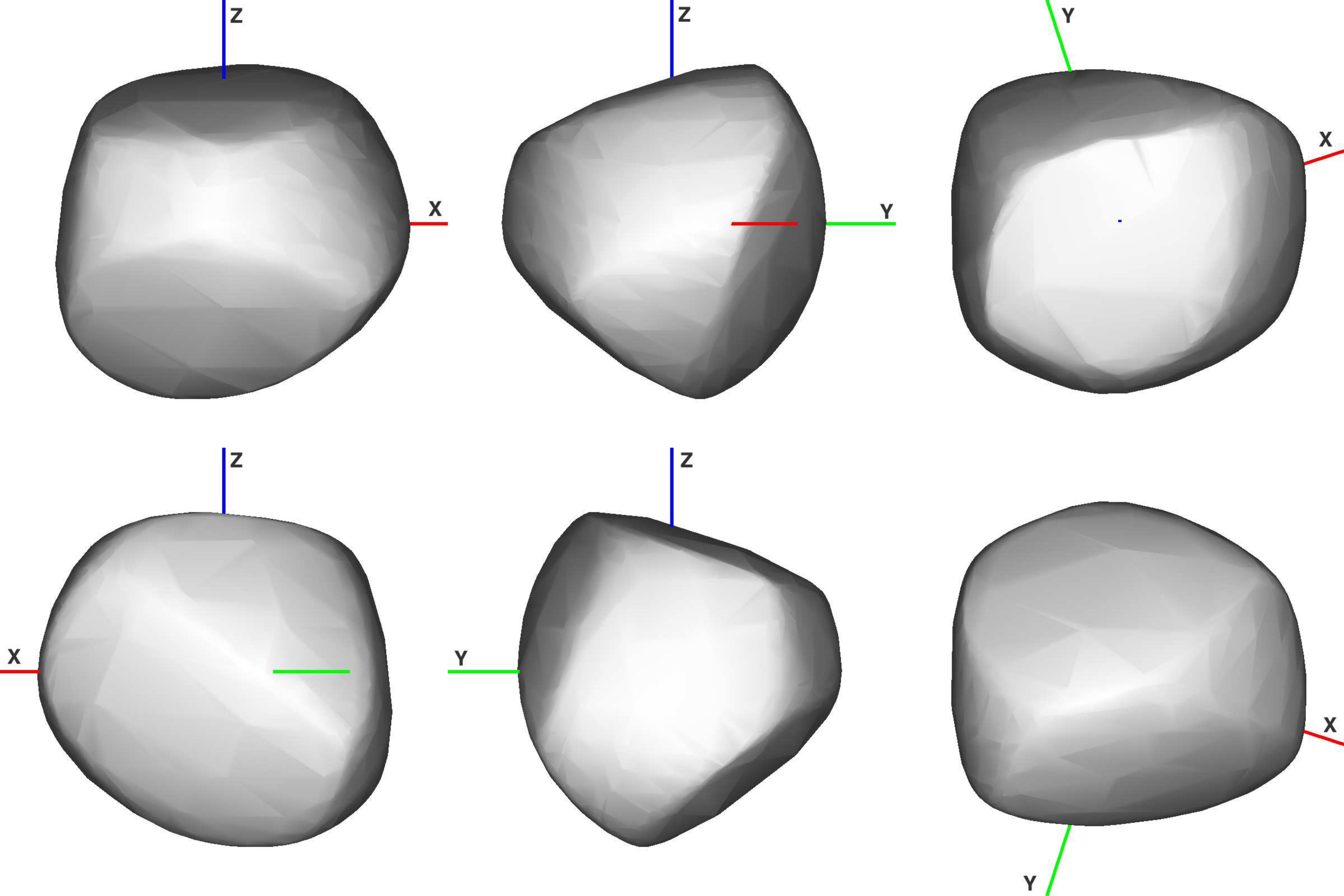}
\caption{Convex shape model of (227) Philosophia from the lightcurve inversion method shown in six projections}
\label{PhilosophiaINVshape}
\end{figure}

\begin{figure}[h]
\includegraphics[width=0.5\textwidth]{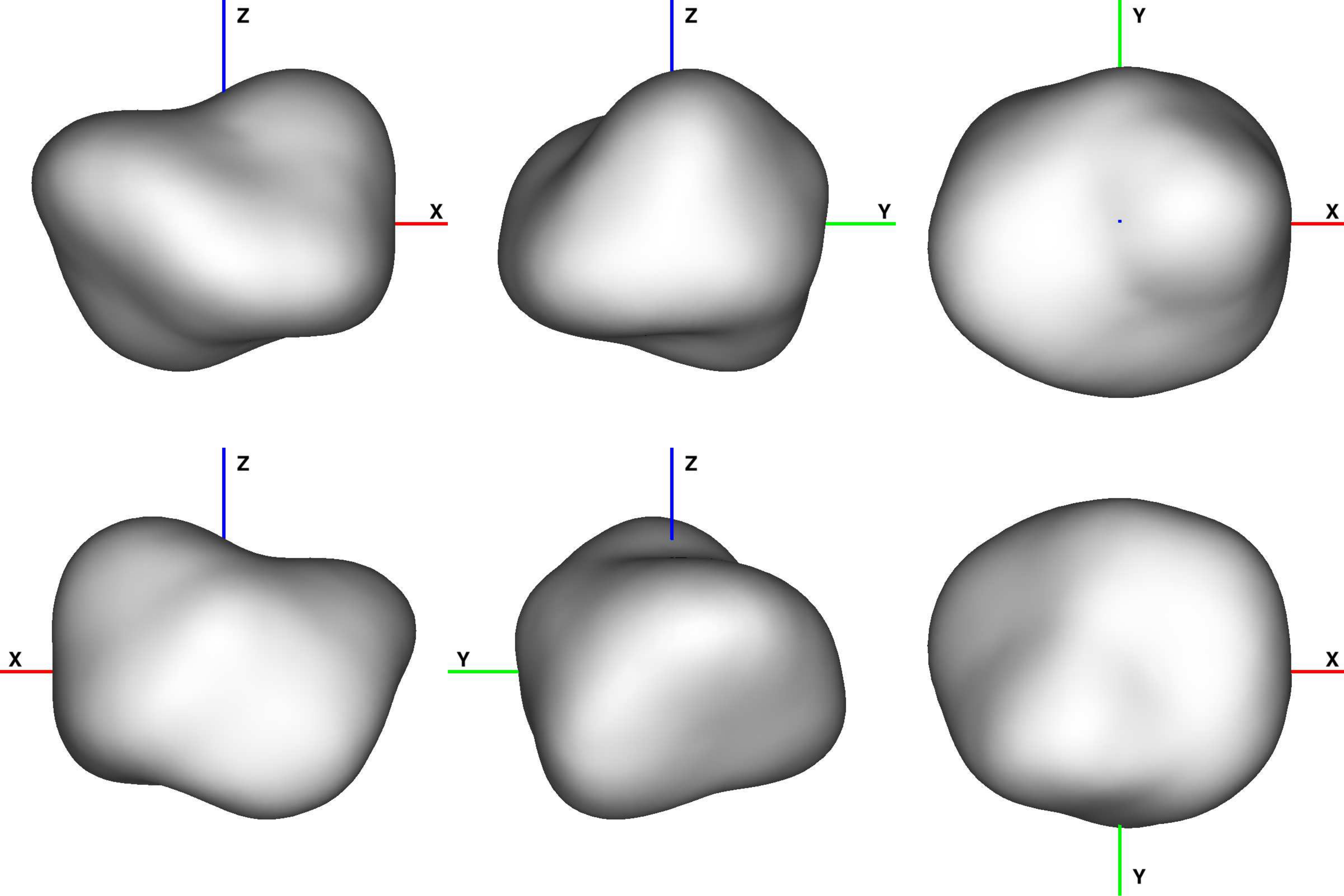}
\caption{Non-convex shape model of (227) Philosophia from the SAGE algorithm shown in six projections}
\label{PhilosophiaSAGEshape}
\end{figure}

Unfortunately, the data from the years 2004 and 2005 were  not available. For the modelling, we used all the other data 
from five distinct apparitions (2006, 2012-2013, 2013-2014, 2015, and 2016); there are  
as many as 97 separate lightcurve fragments, covering a total of around 500 hours. 
The modelling with the convex inversion method clearly pointed to two strong solutions for the spin axis, 
which appeared to have low inclination to the ecliptic (Table \ref{results}), as was expected from 
the lightcurve morphology changes. The shape model is quite atypical, 
with a triangular appearance when viewed from the equator (Fig. \ref{PhilosophiaINVshape}). This shape actually caused the 
most problems in the convex inversion as almost all the resulting shapes had an axis of greatest inertia tensor 
not coincident with the spin axis, regardless of the starting parameters. 
We present here two solutions where the difference between the rotation axis 
and the axis of greatest inertia were smallest. The fit to the lightcurves is satisfactory 
(RMSD = 0.011 mag, Fig. \ref{Philosophia_fit}) with the exception of the first two lightcurves from the year 2015. 

The SAGE algorithm also had problems with modelling this target. Some evolutionary paths were stuck in a blind track  
and finding a unique solution took much more CPU time than usual (one week compared to two days on the cluster 
consisting of ten 6-core 3GHz AMD processors and 2 GB RAM). Finally, two sets of solutions 
for the pole and shape were found (Table \ref{results}, Fig. \ref{PhilosophiaSAGEshape}); however,  
when starting the evolution around the expected mirror solution, the process often ended up 
near the other pole. Most probably, the 
mirror solution had the incorrect inertia tensor, thus was often rejected by the algorithm. Still, as the results 
from the convex inversion suggest, both pole solutions can fit the data on a similar level, so we consider the mirror pole 
solution equally possible. Here, the two above-mentioned lightcurves also fit worse than all the other fragments, 
and the overall RMSD value is 0.009 mag.
The non-convex shape model of Philosophia is even more specific:  one lobe is substantially  
larger than the other,  and there are many strongly non-convex features. However, its pole-on outline largely coincides 
with the corresponding solution from the convex inversion.

In  thermophysical modelling, Philosophia turned out to be  the worst constrained case of the five targets studied here.
Actually, the convex and SAGE models fit to thermal data was only slightly better than the corresponding spherical 
shape solution with the same spin parameters, indicating that inversion shape solutions are not yet perfect. 
We used an H value equal to 9.1 mag and a G value equal to 0.15,\footnote{after: {\tt https://mp3c.oca.eu}} and thermal data from 
IRAS (16 measurements), AKARI (6), and WISE W3/W4 (17). It seems that high-roughness solutions are favoured 
(Table \ref{chi2}).

The overall best fit in TPM is found for the first convex solution ($\lambda$=95\deg, $\beta$=+19\deg) with a $\chi^2$ 
of 1.2. The model fits best for a high level of surface roughness, optimum thermal inertia around 100-150 SI units, 
effective size in the range of 91-105 km (in agreement with previous determinations), and geometric V-band albedo 
of 0.038-0.044 (Table \ref{TPMresults}).

One explanation for this behaviour of the models is that the data are not well balanced with respect to phase angles:  there is only one data point at a negative phase angle (i.e. before the opposition). There is no clear trend with 
wavelength or with rotational 
phase (Fig. \ref{Philosophia_TPM}), but the data quality is not optimal. Also, the low pole of Philosophia might 
be the source of the problems;  in pole-on geometries for many months one of the hemispheres is heated constantly 
and that heat can penetrate to much deeper layers which have different thermal properties from the surface regolith. 
For a change, in geometries closer to equator-on, there are normal diurnal variations in the heat wave. 
Unfortunately, there is no multi-chord stellar occultation by Philosophia for comparison with the  
radiometric parameters or topographic features of the models obtained here.

 \subsection{(329) Svea}
Svea is one of the first targets from our survey for which we found substantially different period 
than that accepted in LCDB (see M2015). Observed previously by \cite{Weidenschilling1990}, \cite{Pray2006}, \cite{Menke2008}, 
and Behrend et al. (www), Svea displayed the ambiguous periods 15.201 hours or 22.778 hours. 
In M2015, we confirmed a 22.78-hour period 
based on data from the year 2013, and since that time we have gathered data from two more apparitions, in 2014, and 2016.
The lightcurve morphology of Svea is interesting and strongly variable; from clearly trimodal, through almost flat, to 
the more usual bimodal lightcurve of larger amplitude (see Figs. \ref{Svea2014} and \ref{Svea2015}). 
Available data from all apparitions fit the 22.78-hour period, and display amplitudes from $0.09$ to $0.24$ mag.

For the modelling, we were able to use our data from three apparitions coupled with data from 2005 provided by \cite{Menke2008}, 
from 2006 by Behrend et al. (www), and only one  of the four lightcurves from  1986 saved as a composit 
by Weidenschillig (1990). In total, there are 60 lightcurve fragments from six apparitions. 
\begin{figure}[h]
\includegraphics[width=0.5\textwidth]{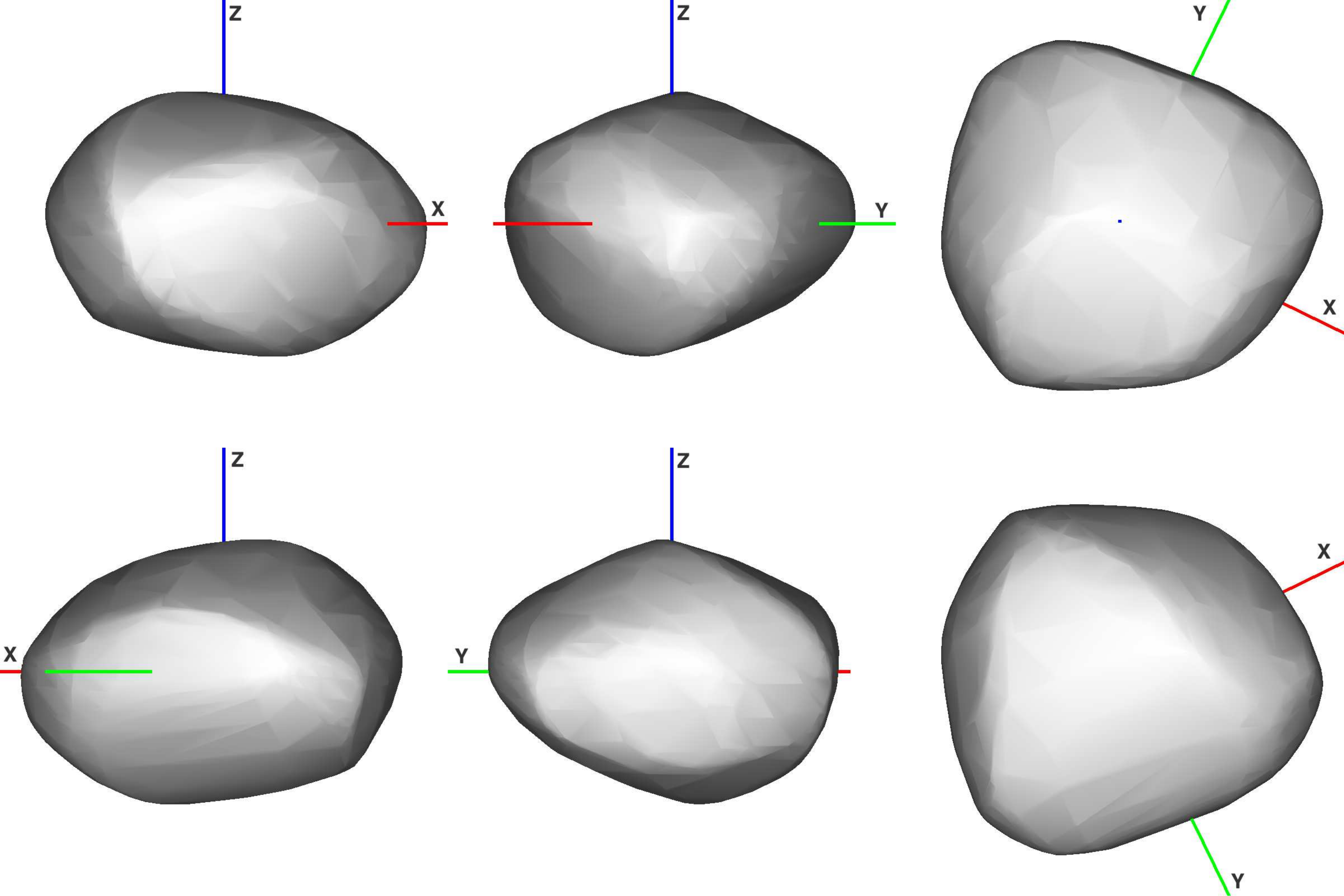}
\caption{Convex shape model of (329) Svea from the lightcurve inversion method shown in six projections}
\label{SveaINVshape}
\end{figure}

\begin{figure}[h]
\includegraphics[width=0.5\textwidth]{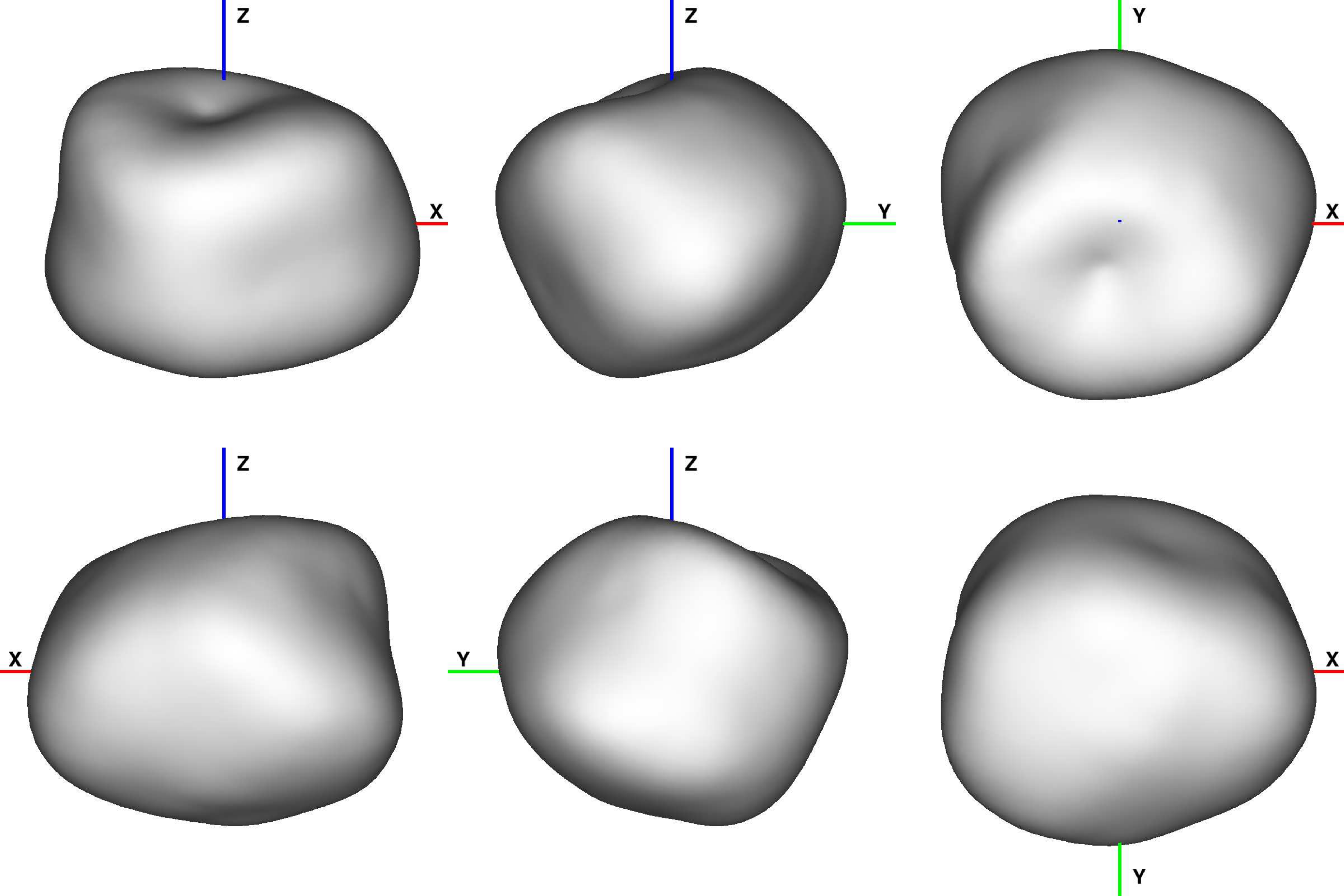}
\caption{Non-convex shape model of (329) Svea from the SAGE algorithm shown in six projections}
\label{SveaSAGEshape}
\end{figure}

In the modelling by the convex lightcurve inversion method, two resulting pole solutions were closer together than 
in the usual miror-pole symmetry, differing by only 124\deg\ in ecliptic longitude, with a similar values for 
pole latitude (Table \ref{results}). The shape model vertical dimensions were not well constrained, but the other 
features were stable (Fig. \ref{SveaINVshape}), providing a good fit to lightcurves at 0.010 mag level 
for both pole solutions (Fig. \ref{Svea_fit}). 

The SAGE spin solutions were 145\deg\ apart  (Table \ref{results}) and the corresponding shape models 
showed some large indentations near the equator and one of the poles (Fig. \ref{SveaSAGEshape}). 
The fit to the lightcurves shows 0.011 RMSD and is very similar to the fit by the convex models (Fig. \ref{Svea_fit}). 
\begin{table}[h]
\begin{tabular}{ccc}
                &   2011          &     2013 \\
\hline
       CONVEX  &  72 $\pm$ 4 km  &  74 $\pm$ 5 km \\
       SAGE    &  70 $\pm$ 4 km &   72 $\pm$ 3 km \\      
\end{tabular}
\caption{Equivalent volume sphere diameters of the (329) Svea models pole 1, fitted to two occultations: 
from 28 December 2011 and 7 March 2013. Compare with radiometric diameter from TPM in Table \ref{TPMresults}.}
\label{occ_diameters329}
\end{table}

In the case of Svea, there are two very good multi-chord occultations available \citep{Dunham2016} 
observed from Japan in 2011 (7 chords), and from Florida, USA, in 2013 (6 chords), 
giving a rare opportunity to test the shape models down to the medium-scale details. 
Additionally, in these events, a few negative results were recorded, allowing for better size constraints. 
Appendix B lists occultation observers and site names. 
Fitting our models of Svea to these occultations gave  remarkably good results, clearly allowing us 
to reject one of the mirror pole solutions (pole 2, shown in Fig. \ref{Svea_occult_mirror}), and 
confirming the first pole solution with indentations and other shape features of the SAGE model (Fig. \ref{Svea_occult}). 
The convex model for pole 1 also fits both occultations well, but the non-convex model fits markedly better.
This way the model gets unique validation and it shows that major topographic features present 
in the non-convex models made with SAGE are confirmed when auxiliary data are available.
  The fitting to two occultation events was done independently, but the results are internally consistent. 
Obtained size estimates range from $70$ to $74$ km for the effective diameter (see Table \ref{occ_diameters329}), 
which agrees with the radiometric size ($77.5$ km in Table \ref{TPMresults}) within the error bars.

Curiously, in thermophysical modelling it is the convex model (but also pole 1) that is slightly preferred. 
However, all the inversion solutions  clearly fit better to the thermal data than does the corresponding spherical shape solution 
with the same spin properties. In TPM the preference of pole 1 over pole 2 is stronger than the  preference of the best fitting 
convex model over the  non-convex solution; however, all the fits are at an acceptable level (see Table \ref{chi2}). 
Overall, the thermal data seem to point 
towards a spin axis close to $\lambda=33\deg$ and $\beta=+51\deg$. The convex inversion solution for this pole 
provides an excellent fit to all thermal data (reduced $\chi^2$ below 1.0) with an intermediate level of surface roughness, 
optimum thermal inertia around 75 SI units, effective size of around 77.5 km (confirming the value from occultations), 
and geometric V-band albedo of 0.055. The O-C plots for the best solution are shown in Fig. \ref{Svea_TPM}, and the uncertainties 
on the derived values are given in Table \ref{TPMresults}.
The infrared data that were used came from IRAS (20 measurements), AKARI (9), Wise W3/W4 (28), and MSX (8), 
and the adopted absolute magnitude and slope were 9.34 and 0.04, respectively.


 \subsection{(478) Tergeste}

Asteroid (478) Tergeste was observed previously for lightcurves in only two apparitions, by \cite{Harris1989} and 
Behrend et al. (www). In the latter, it displayed a 0.22 mag amplitude lightcurve of 16.104 hours period. 
We observed it in our project since 2013 through four consecutive apparitions, confirming the period around 16.104 hours 
and registering lightcurves of $0.15$ up to $0.30$ mag amplitudes. Those with larger amplitudes showed sharp minima and wide 
asymmetric maxima, while others were  smoother and more regular (see Figs. \ref{Tergeste2005} to \ref{Tergeste2016}).  

\begin{figure}[h]
\includegraphics[width=0.5\textwidth]{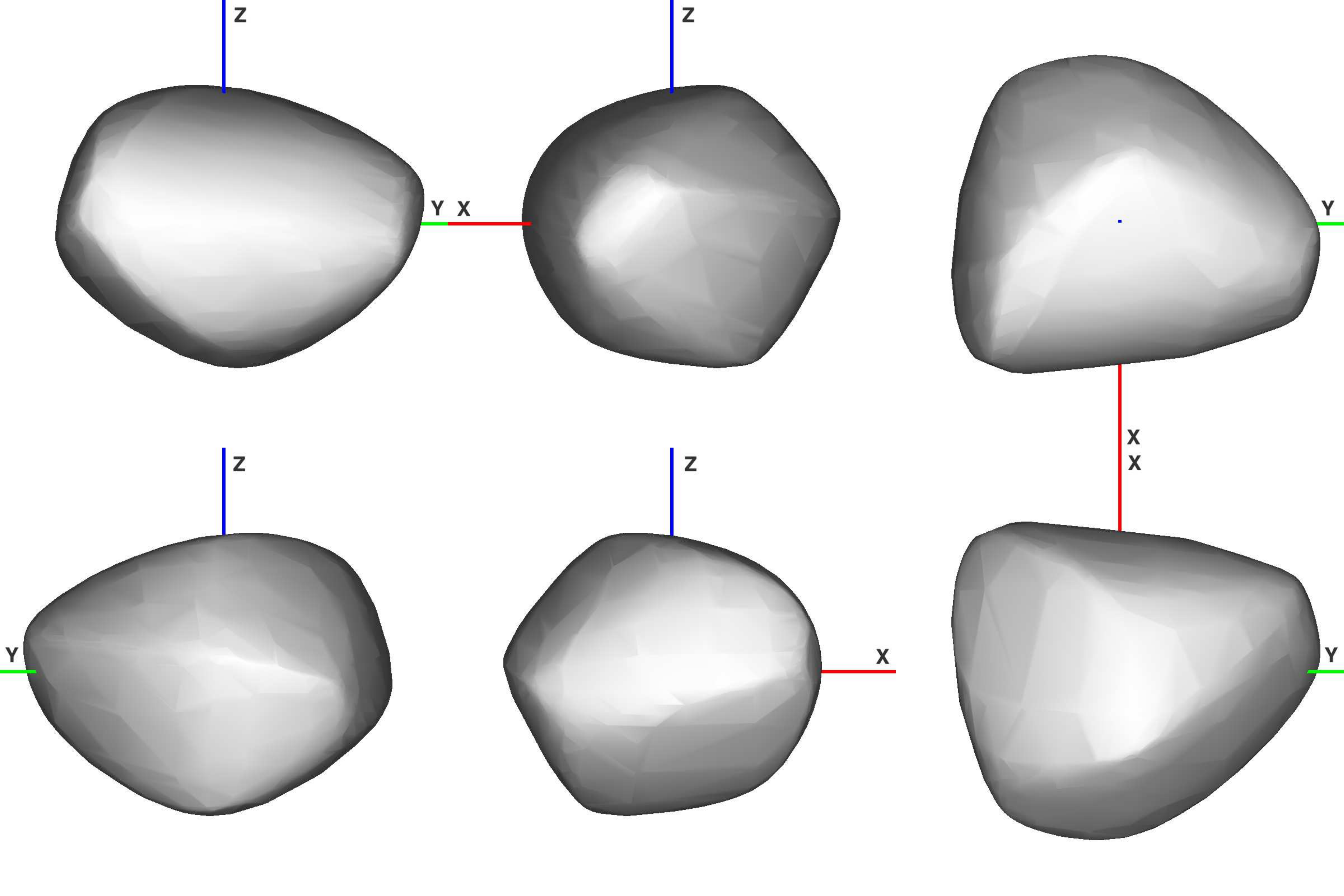}
\caption{Convex shape model of (478) Tergeste from the  lightcurve inversion method shown in six projections}
\label{TergesteINVshape}
\end{figure}

\begin{figure}[h]
\includegraphics[width=0.5\textwidth]{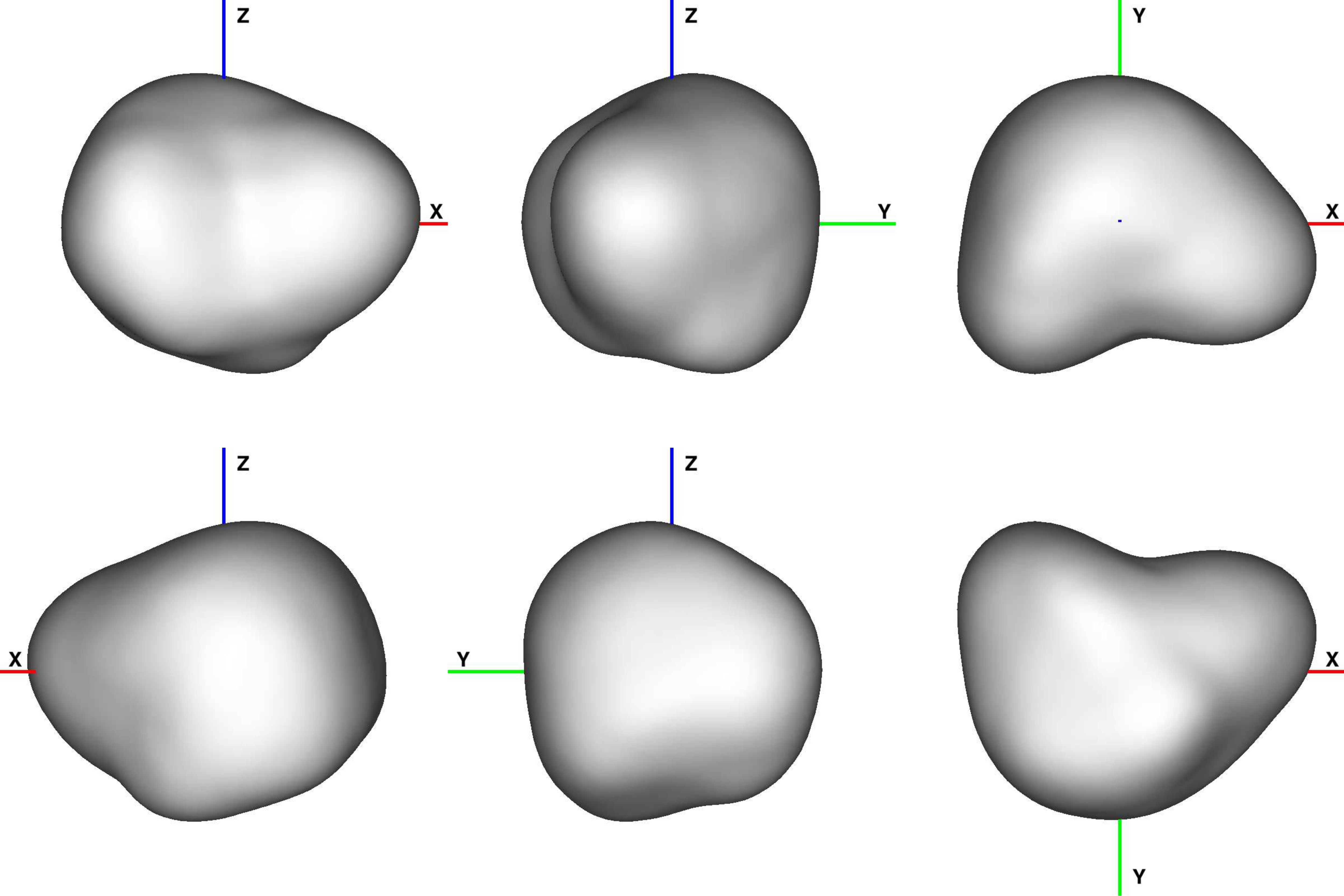}
\caption{Non-convex shape model of (478) Tergeste from the SAGE algorithm shown in six projections}
\label{TergesteSAGEshape}
\end{figure}

For the modelling, we used a dataset consisting of 48 lightcurves from six apparitions (in 1980, 2005, 2013, 2014, 2015, and 2016). 
In the convex inversion, a convexity 
regularisation weight had to be slightly increased in order to make some shape models physical (rotating around
the axis of greatest inertia tensor). There are two narrow solutions for the pole in the parameter space 
(Table \ref{results}), and the shape models are trapezoidal  (Fig. \ref{TergesteINVshape}). 
The fit to the lightcurves is on a 0.011 magnitude level (see Fig. \ref{Tergeste_fit}).

The non-convex SAGE models confirm these pole solutions within the small error bars (Table \ref{results}), but here the shapes 
are more complex, e.g. with a large valley visible from the pole-on view (Fig. \ref{TergesteSAGEshape}) in a place 
where the convex models showed a straight, planar area. Both spin solution models provide a similar fit to lightcurves (0.011 mag)  
\ref{Tergeste_fit}.
However, the Tergeste model fit (see Section 3.1) shows the  tendency of  
non-convex models to better fit deep and sharp lightcurve minima  (see middle plot of 
Fig. \ref{Tergeste_fit}). These local features, with only a few datapoints, cannot notably influence the overall RMSD value, 
but they clearly need some shadowing to be correctly reproduced (see the shape model projection in Fig. 
\ref{Tergeste2013}).

There is no multi-chord stellar occultation to discriminate between two equally possible pole solutions from lightcurve inversion 
for Tergeste, but surprisingly the thermophysical modelling shows a strong preference for one of the spin and shape solutions 
(Table \ref{chi2}). 
non-convex model 2 (at $\lambda$ = 218\deg, $\beta$ = -56\deg) provides a very good fit to the thermal data ($\chi^2$ around 1.0), 
while the other inversion spin and shape  solutions give fits that are  at least 1.5 times worse  (at the edge of being acceptable), 
and the spherical model gives a  fit that is  2.5 times worse. 
The preferred solution provides a very good fit to the thermal data (28 datapoints from IRAS, 8 from AKARI, 
and 18 from WISE W3/W4 bands,  adopting H=7.96 and G=0.15; Fig. \ref{Tergeste_TPM}) with an intermediate level of surface 
roughness, optimum thermal inertia around 75 SI units, effective size around 87.3 km, and geometric V-band albedo of 0.15 
(the last two  values are closest to AKARI determinations, see Table \ref{TPMresults}).

 \subsection{(487) Venetia}

Observed previously in as many as six apparitions, (487) Venetia displayed lightcurves of varying shape and amplitude.  
However, some of the observations only partially covered its 13.34-hour lightcurve  
 \citep[][Behrend et al., www]{Weidenschilling1990, 
 Shevchenko1992,        
 Neely1992,         
 Schober1994,        
 Ferrero2014}.

\cite{Erikson2000}, and \cite{Tungalag2002} published spin and shape solution for Venetia with similar 
spin axis coordinates, but a notable difference in sidereal period:\\
          \cite{Erikson2000}      $\lambda_p=268\deg$, $\beta_p=-24\deg$, $P=13.34153$ h\\
          \cite{Tungalag2002}     $\lambda_p=259\deg$, $\beta_p=-30\deg$, $P=13.33170$ h.\\
We  observed Venetia over three consecutive apparitions, registering full lightcurves that were often  almost featureless, 
while in other apparitions it showed a substantial amplitude of  $0.23$ mag  (Figs. \ref{Venetia2012} - \ref{Venetia2015}). 
This behaviour is a strong indication of an elongated object with low inclination of the spin axis. 

\begin{figure}[h]
\includegraphics[width=0.5\textwidth]{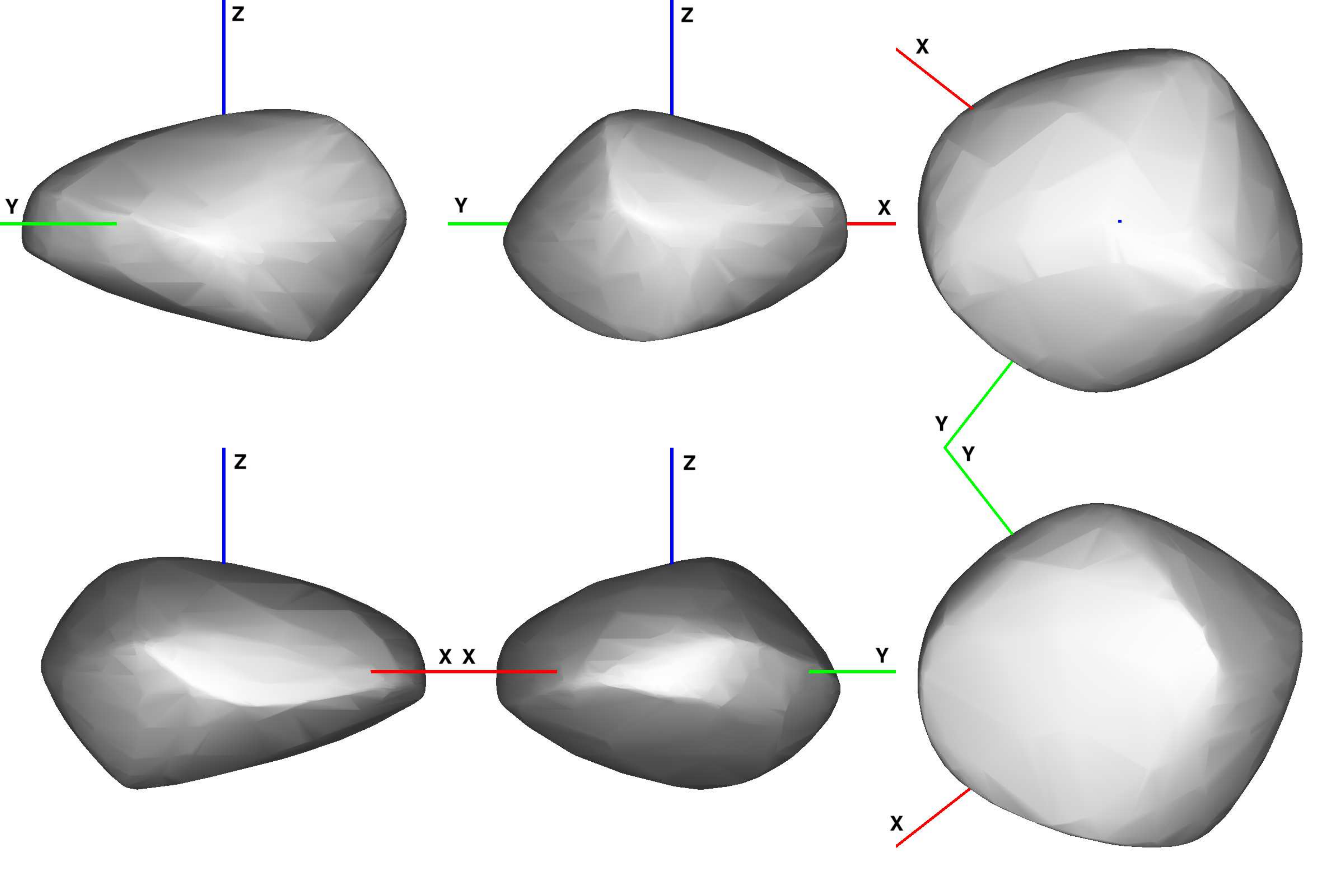}
\caption{Convex shape model of (487) Venetia from the lightcurve inversion method shown in six projections}
\label{VenetiaINVshape}
\end{figure}

\begin{figure}[h]
\includegraphics[width=0.5\textwidth]{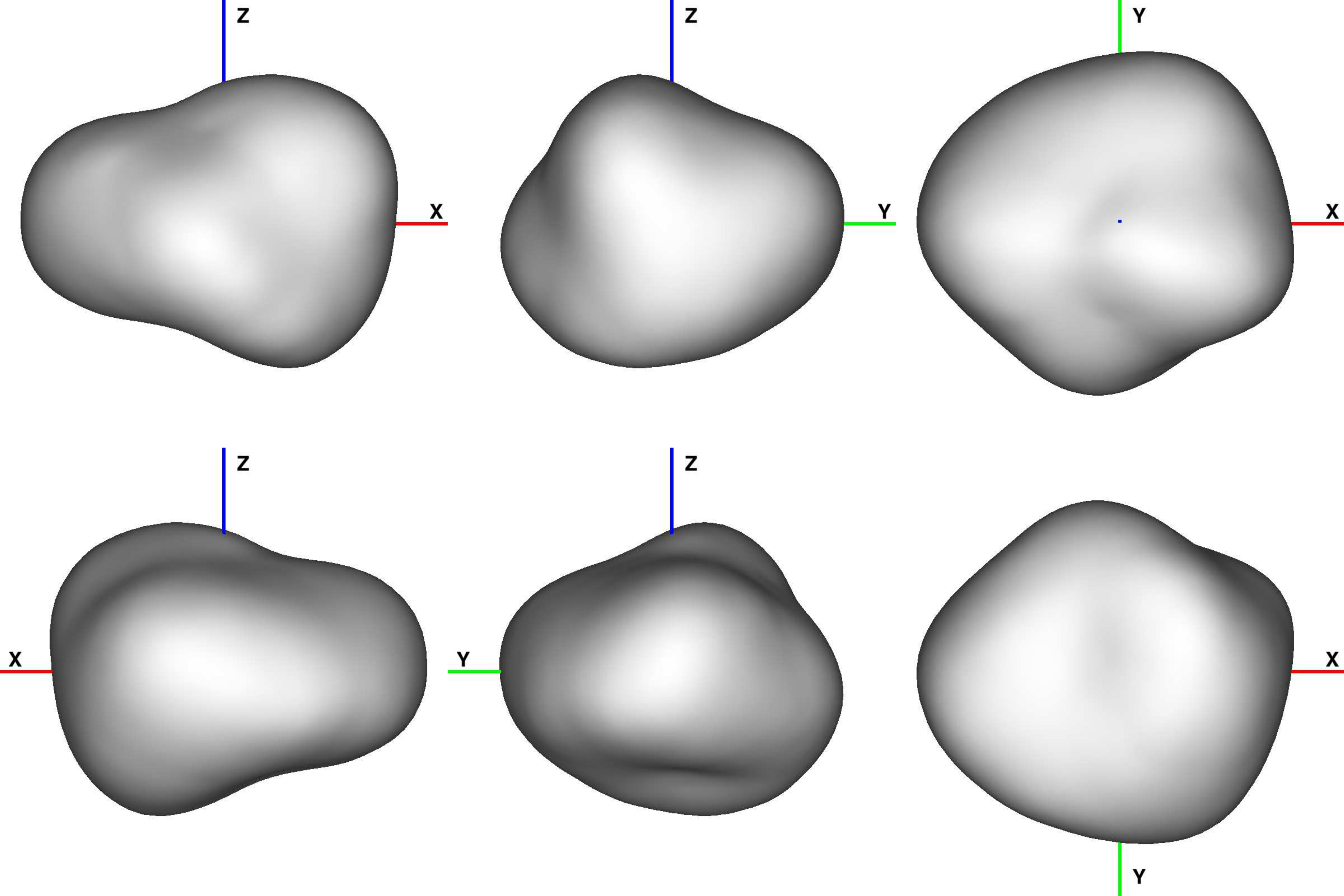}
\caption{Non-convex shape model of (487) Venetia from the SAGE algorithm shown in six projections}
\label{VenetiaSAGEshape}
\end{figure}

The lightcurve inversion indeed resulted in very small pole latitudes (see Table \ref{results}). The convex inversion model 
displays a somewhat angular flattened shape (Fig. \ref{VenetiaINVshape}), while the SAGE model has a smoother and more complex 
appearance (Fig. \ref{VenetiaSAGEshape}). 
Both model types  failed to reproduce tiny but complex brightness variations from pole-on geometries (Fig. \ref{Venetia_fit}) 
at the level of a few 0.01 mag, revealing the limits of lightcurve inversion. However, generally the fit was very good 
at the level of 0.011 mag RMSD in both methods. 
Our models are close in sidereal period to the value determined by \cite{Erikson2000}, and in pole longitude to both 
pole solutions published by \cite{Erikson2000} and \cite{Tungalag2002}; however, they disagree in pole latitude. Our slightly positive values are far from both of the previous determinations (Table \ref{results}).


Here too there are no available stellar occultations to verify or confirm one of the spin and shape solutions. However, 
thermophysical modelling shows a similarly strong preference for one of the spin and shape solutions, as in the previous case 
of (478) Tergeste. Best $\chi^2$=1.04 is as much as two times better for the (487) Venetia non-convex model 
at $\lambda$ = 255\deg, $\beta$ = +8\deg 
 than for any of its convex models, and 25\% better than its mirror non-convex counterpart (see Table \ref{chi2}). 
The thermal data came from IRAS (32 measurements), AKARI (7), and WISE W3/W4 (46), with adopted 
H and G values of 8.14, and 0.15, respectively.
The best fitting thermophysical parameters 
are an intermediate level of surface roughness, optimum thermal inertia of around 100 SI units, effective size $\sim$69.5 km, 
and geometric V-band albedo of 0.21 (see Table \ref{TPMresults}). The thermal data used here were well-balanced with  
 pre- and post-opposition geometries, also in the WISE data. One small issue is the small sinusoidal trend with rotational 
phase visible in the WISE data (squares in Fig. \ref{Venetia_TPM}).  
It might indicate some imperfections in the shape model or alternatively 
the increased infrared flux contribution  of surface layers underneath the skin depth 
from pole-on geometries. Some discrepancy can also be found for one WISE dataset 
compared to the model of (478) Tergeste (Fig. \ref{Tergeste_TPM}), also a possible indicator of some missing shape features.
Unfortunately, the WISE W1 data are too few and sparse to change the shape models when used as purely reflected light 
in parallel with all the lightcurve data.

\begin{table}[h]
\begin{tabular}{|rcc|}
\hline
 shape model ($\lambda$, $\beta$)    &  low roughness  &   high roughness \\
\hline
\multicolumn{3}{c}{(159) Aemilia}\\
\hline
 sphere (139\deg, +68\deg)  & 1.16 & 1.21\\
 sphere (348\deg, +59\deg)  & 1.15 & 1.21\\
 convex (139\deg, +68\deg)  & 0.61 & 0.53\\    
 convex (348\deg, +59\deg)  & 0.44 & 0.56\\
 SAGE   (139\deg, +66\deg)  & 0.44 & 0.47\\
 SAGE   (349\deg, +63\deg)  & 0.53 & 0.52\\
\hline
\multicolumn{3}{c}{(227) Philosophia}\\
\hline
 sphere ( 95\deg, +19\deg) & 2.67 & 1.34\\
 sphere (272\deg,  -1\deg) & 2.67 & 1.45\\
 convex ( 95\deg, +19\deg) & 2.37 & 1.22\\
 convex (272\deg,  -1\deg) & 2.49 & 1.34\\
 SAGE   ( 97\deg, +16\deg) & 1.93 & 1.28\\
 SAGE   (271\deg,   0\deg) & 1.93 & 1.40\\
\hline
\multicolumn{3}{c}{(329) Svea}\\
\hline
 sphere ( 33\deg, +51\deg) & 1.63 & 1.61\\
 sphere (157\deg, +47\deg) & 1.67 & 1.60\\
 convex ( 33\deg, +51\deg) & 0.98 & 0.97\\
 convex (157\deg, +47\deg) & 1.38 & 1.17\\
 SAGE   ( 21\deg, +47\deg) & 1.21 & 1.09\\
 SAGE   (166\deg, +39\deg) & 1.39 & 1.55\\
\hline
\multicolumn{3}{c}{(478) Tergeste}\\
\hline
 sphere (  2\deg, -42\deg) & 2.86 & 2.24\\
 sphere (216\deg, -56\deg) & 2.41 & 1.98\\
 convex (  2\deg, -42\deg) & 2.18 & 2.59\\
 convex (216\deg, -56\deg) & 1.53 & 1.81\\
 SAGE   (  4\deg, -43\deg) & 1.44 & 1.68\\
 SAGE   (218\deg, -56\deg) & 1.03 & 1.08\\
\hline
\multicolumn{3}{c}{(487) Venetia}\\
\hline
 sphere ( 78\deg,  +3\deg) & 2.39 & 1.62\\
 sphere (252\deg,  +3\deg) & 1.38 & 1.09\\
 convex ( 78\deg,  +3\deg) & 2.01 & 2.69\\
 convex (252\deg,  +3\deg) & 1.82 & 2.88\\
 SAGE   ( 70\deg,  +8\deg) & 1.30 & 1.79\\
 SAGE   (255\deg,  +8\deg) & 1.04 & 1.23\\
\hline
\end{tabular}
\caption{ Reduced $\chi^2$ minimum values of various models fit to infrared data in thermophysical modelling.  
The first column gives the shape model type and spin axis position.}
\label{chi2}
\end{table}

\begin{table*}[h]
\begin{small}
\begin{tabular}{ccclcrc}
\hline
                 &           &          &          &  \multicolumn{3}{c}{Radiometric solution for combined data}\\
Target           &$D_{AKARI}$&$D_{IRAS}$&$D_{WISE}$& Diameter & Albedo & Thermal inertia \\
                 &   [km]    &   [km]   &  [km]    &  [km]    &        & [Jm$^{-2}$s$^{-0.5}$K$^{-1}$]  \\
\hline                                                       
159 Aemilia      &  130.0   & 125.0   & 127.4  & $137$   & $0.054$     & $50$ \\                                
                 &          &         &        & $\pm 8$ & $\pm 0.015$ & $\pm 50$ \\
227 Philosophia  &   95.6   &  87.3   & 105.3  & $101$   & $0.041$     & $125$\\                             
                 &          &         &        & $\pm 5$ & $\pm 0.005$ & $\pm 90$ \\
329 Svea         &   70.4   &  77.8   & 69.2   & $78$    & $0.055$     & $75$ \\                                 
                 &          &         &        & $\pm 4$ & $\pm 0.015$ & $\pm 50$ \\
478 Tergeste     &   85.6   &  79.5   & 77.2   & $87$    & $0.15$      & $75$ \\
                 &          &         &        & $\pm 6$ & $\pm 0.02$  & $\pm 45$ \\                                 
487 Venetia      &   66.1   &  63.1   & 65.6   & $70$    & $0.21$      & $100$ \\                                 
                 &          &         &        & $\pm 4$ & $\pm 0.02$  & $\pm 75$ \\
\hline
\end{tabular}
\caption{Asteroid diameters from AKARI, IRAS, and WISE compared to values obtained here on combined data 
for the preferred pole solution (Col. 5) using TPM. 
The last two columns contain the derived albedo and thermal inertia values. Errors are full 3-$\sigma$ range.}
\label{TPMresults}
\end{small}
\end{table*}


\section{Summary and future work}

This  work is a first step towards actual debiasing the available set of spin and shape models for asteroids 
to include real targets of abundant group with long rotation periods and low amplitude lightcurves.
We determined here spin and scaled shape solutions with albedo and thermal inertia values for
the first five asteroids from our sample. The diameters  
are in most cases in good agreement with previous determinations from the IRAS, AKARI, and WISE surveys,
though  our values are usually a few kilometres larger. The reason for this small discrepancy 
might be that the cited sizes are usually based on single-epoch measurements, i.e. corresponding more
to the apparent cross-section, and on a simple thermal model. The radiometric results obtained here are based
on multiple wavelength, epoch, phase-angle, and rotational-phase data, 
and refer to the scaling size for a given 3D shape solution.
 
Spin and shape models, and thermal inertia values for these targets are determined here for the first time
(except for (487) Venetia).
When most of our sample is modelled and applied this way, the existing bias in these parameters 
will be largely diminished, at least for bright targets (i.e. for most of large and medium-sized 
main belt asteroids). We predict that we will complete the task over the course of the next three years.

Our results based on five test cases have shown that asteroid models obtained with both convex 
and non-convex lightcurve inversion are largely comparable. In some applications (ocultation fitting and 
thermophysical modelling); however, non-convex models often do somewhat better, sometimes even allowing a choice 
between two mirror pole solutions.
Thanks to the large amount and the high quality of the data used, both model types  are smooth and fit the data 
close to noise level. The  differences between the shape models do not manifest themselves in the RMSD value, 
but they do in the subtle details of the lightcurve fit.

On the contrary, models based on sparse data are usually characterised by 
low-resolution angular shapes that tend to be problematic in further applications like the above. 
Nonetheless, sparse data models are good for general statistical studies of spin properties, provided that 
the data are properly debiased, which is not a trivial task \citep[see e.g.][]{Cibulkova2016}.
As the Gaia mission is expected to provide absolute photometric data of much better accuracy than previously used 
sky surveys, some of the biases described in this work are expected to decrease, 
like those against long-period targets with large amplitudes. Still, to a large extent, 
low-amplitude targets are going to be problematic for the Gaia mission algorithm for asteroid modelling, 
as has been shown by \cite{Santana-Ros2015}. A substantial amount of low-amplitude asteroids
(even up to 80\% of targets with equivalent ellipsoid dimensions a/b$\leq$1.25, especially those 
with poles of low inclination to the ecliptic) 
will be either rejected or wrongly inverted by this algorithm. 
Thus, it is essential to focus ground-based photometric studies on these more demanding targets 
to make the well-studied population as complete and varied as possible, 
and also to start to alleviate biases expected in the future.

Some of our targets that should soon be modellable coincide with asteroids 
for which the Gaia mission is expected to provide reliable mass estimates, so  
after scaling them, e.g. by thermophysical modelling, 
it will be possible to calculate their densities. Practically all of our targets are characterised by complex 
lightcurves, i.e.  a certain signature of asymmetric, complex shapes. Approximating these shapes with simple 
ellipsoids \citep[as in the Gaia algorithm for asteroids,][]{Cellino2009}  
can lead to large errors in derived volumes, which would consequently propagate to large errors in densities
\citep[e.g.][]{Carry}. 
Our modelling is going to provide precise shape models 
that can be further validated and scaled using stellar occultations, adaptive optics imaging, or thermophysical modelling. 
This way the derived volumes and densities should be possibly closest to real values. 

Since most of our targets are bright, both in the visible and the infrared range, many of them have thermal data 
of good quality, and some even have continuous thermal lightcurves, which -- coupled with reliable shape models --  
are a good input for thermophysical modelling and further studies on their physical parameters (e.g.  
thermal inertias, albedoes, and sizes) and also on the development of the TPM method itself. Some of them may prove to be good 
candidates for secondary calibrators for infrared observatories like ALMA, APEX, or IRAM \citep{Muller2002} 
as their infrared flux is only weakly and slowly variable (although in a predictable way),
which are desirable features of calibrator asteroids.

Cases like 227, 478, and 487 add support to the suggestion of \cite{Harris2016} that slowly rotating asteroids 
have higher thermal inertia values, but a larger sample is still needed. 
Our modelled targets applied in careful thermophysical modelling show best fitting values from 50  to 125 SI units,  
which seems to fit the trend to higher values of thermal inertia for rotation periods longer than 10 hours 
\citep[see fig. 5 in][]{Harris2016}. 
With slower rotation, the heat penetrates deeper to more compact subregolith layers with 
substantially higher density and thermal conductivity, which both seem to rapidly grow with depth. 
Thermal inertia appears to grow by a factor of 10 (main belt asteroids) and 20 (near-Earth objects) 
with a depth of just 10 cm \citep{Harris2016}.
Alternatively, the growth observed here might also be related to the objects' sizes:
a low thermal inertia of 15 has been found for large (fine-grained regolith covered) asteroids with sizes much larger 
than 100 km, but we are looking here at objects below or close to 100 km. They might have less low-conductivity material 
on the surface, due to reduced gravity. 
Our future works are going to provide thermal inertia values for a larger sample of slow-rotators, 
a highly needed input for further studies of subsurface layers of asteroids.

\begin{acknowledgements}
       This work was supported by grant no. 2014/13/D/ST9/01818 from the National Science Centre, Poland. 
       The research leading to these results has received funding from the European Union's
       Horizon 2020 Research and Innovation Programme, under Grant Agreement no 687378.
       VK was supported by the grant from the Slovak Research and Development Agency with number APVV-15-0458. 
       The Joan Or{\'o} Telescope (TJO) of the Montsec Astronomical Observatory (OAdM) 
       is owned by the Catalan Government and is operated by the Institute for Space Studies of Catalonia (IEEC).
       The 0.82m IAC80 Telescope is operated on the island of Tenerife by the Instituto de
       Astrofisica de Canarias in the Spanish Observatorio del Teide.
       Based on observations obtained with the SARA Observatory 1.0m Jacobus Kapteyn Telescope at ORM, 
       and 0.6m telescope at CTIO,
       which are owned and operated by the Southeastern Association for Research in Astronomy (saraobservatory.org).
\end{acknowledgements}

\bibliographystyle{aa}
\bibliography{bibliography}

\clearpage

\begin{figure*}[h]
\includegraphics[width=0.33\textwidth]{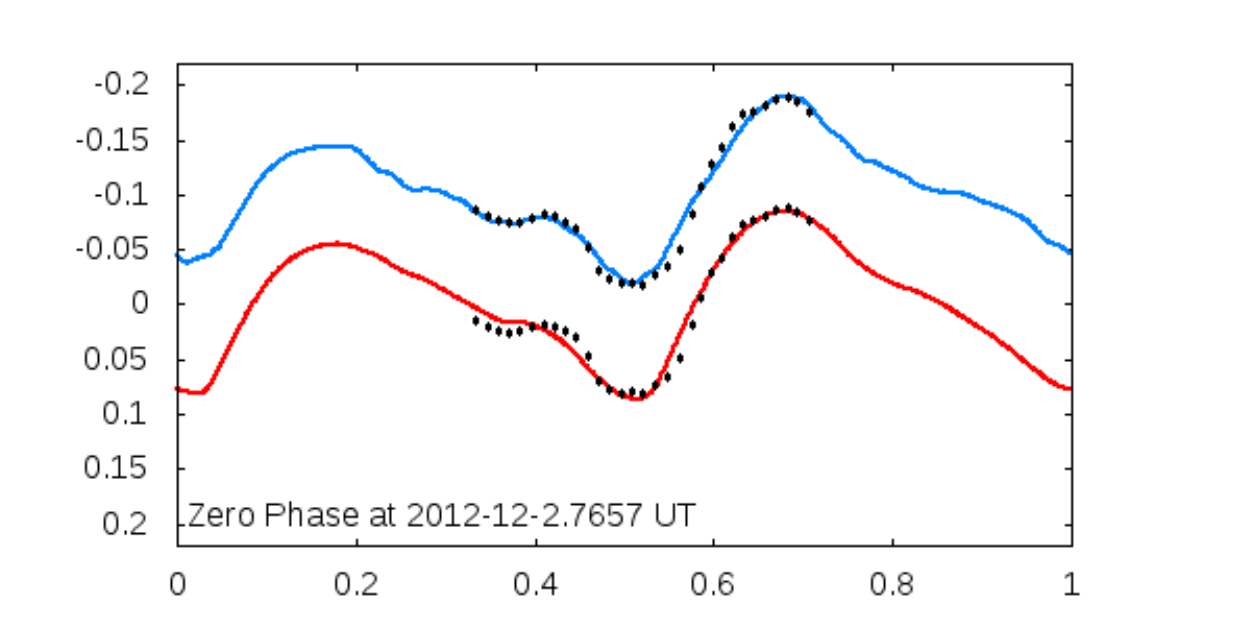}
\includegraphics[width=0.33\textwidth]{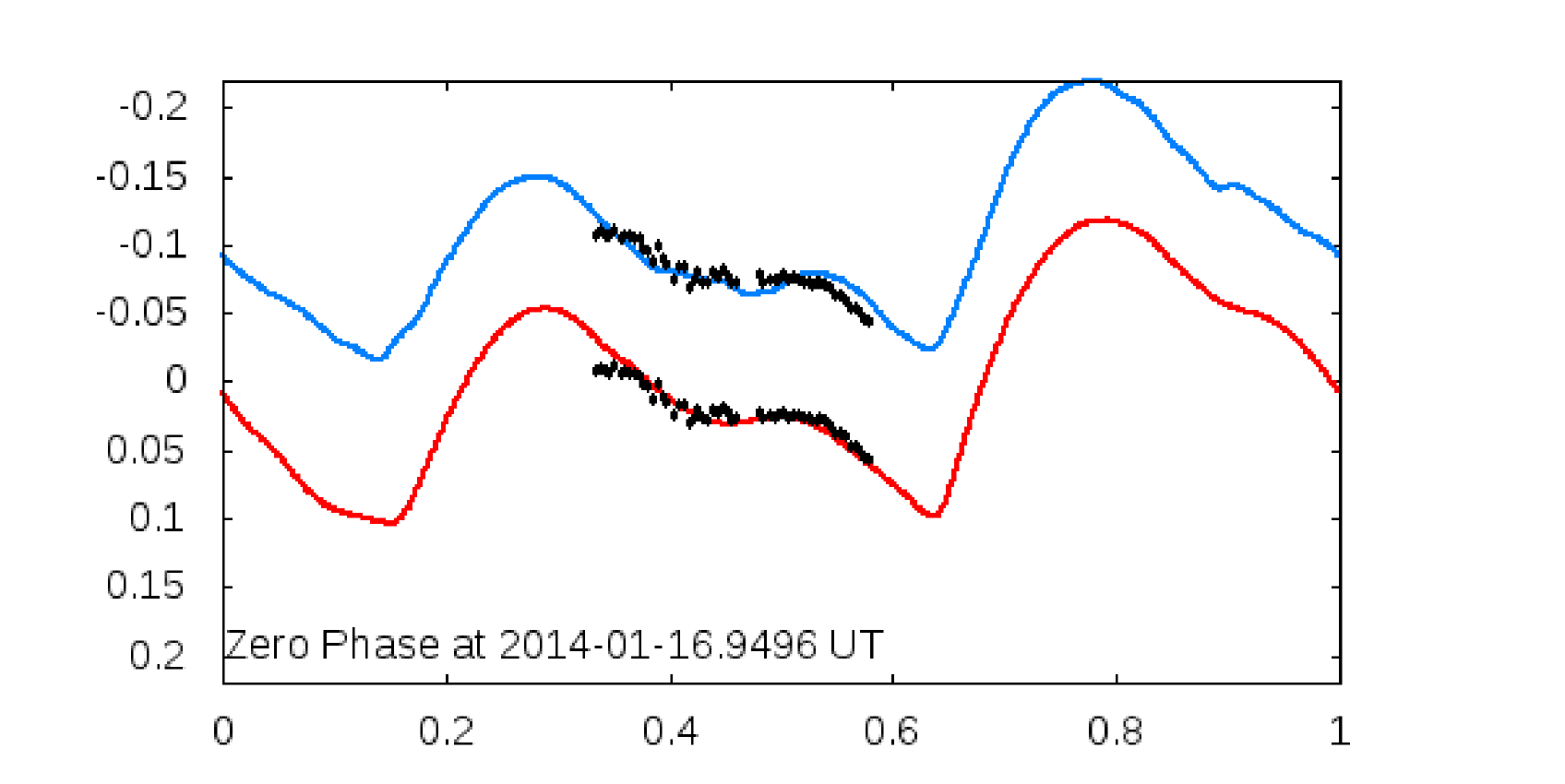}
\includegraphics[width=0.33\textwidth]{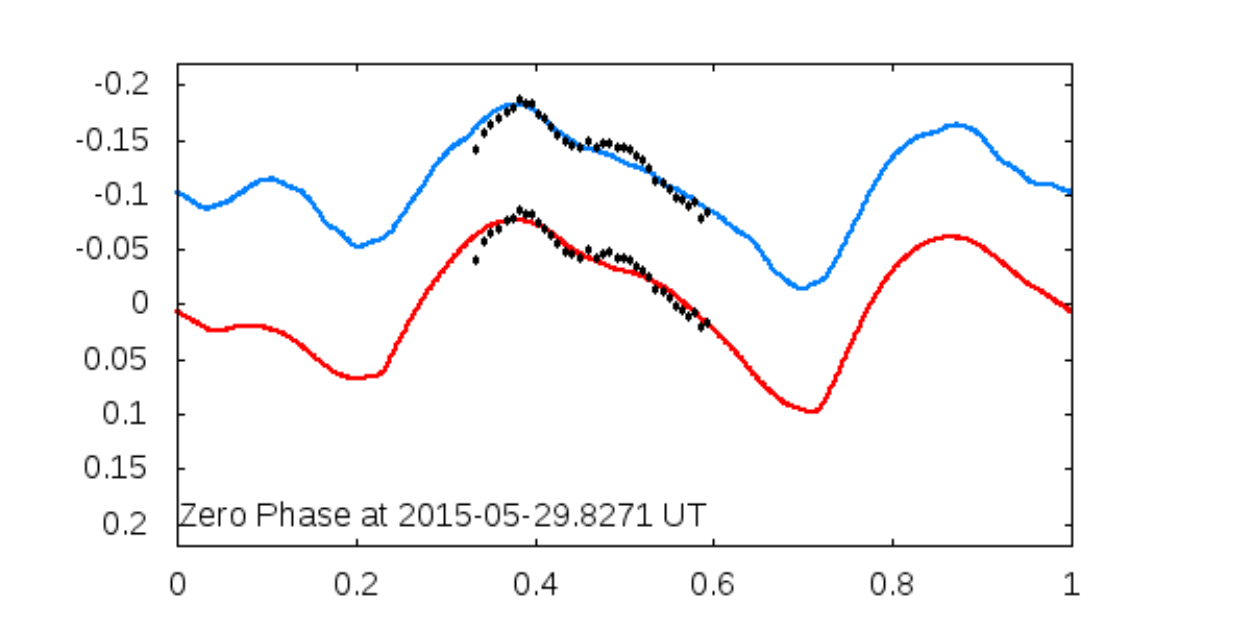}
\caption{Convex (upper curve) and non-convex (lower curve) model lightcurves of (159) Aemilia fitted to data 
from various apparitions (black points)}
\label{Aemilia_fit}
\end{figure*}

\begin{figure*}[h!]
\begin{tabular}{cc}
 \includegraphics[width=0.4\textwidth]{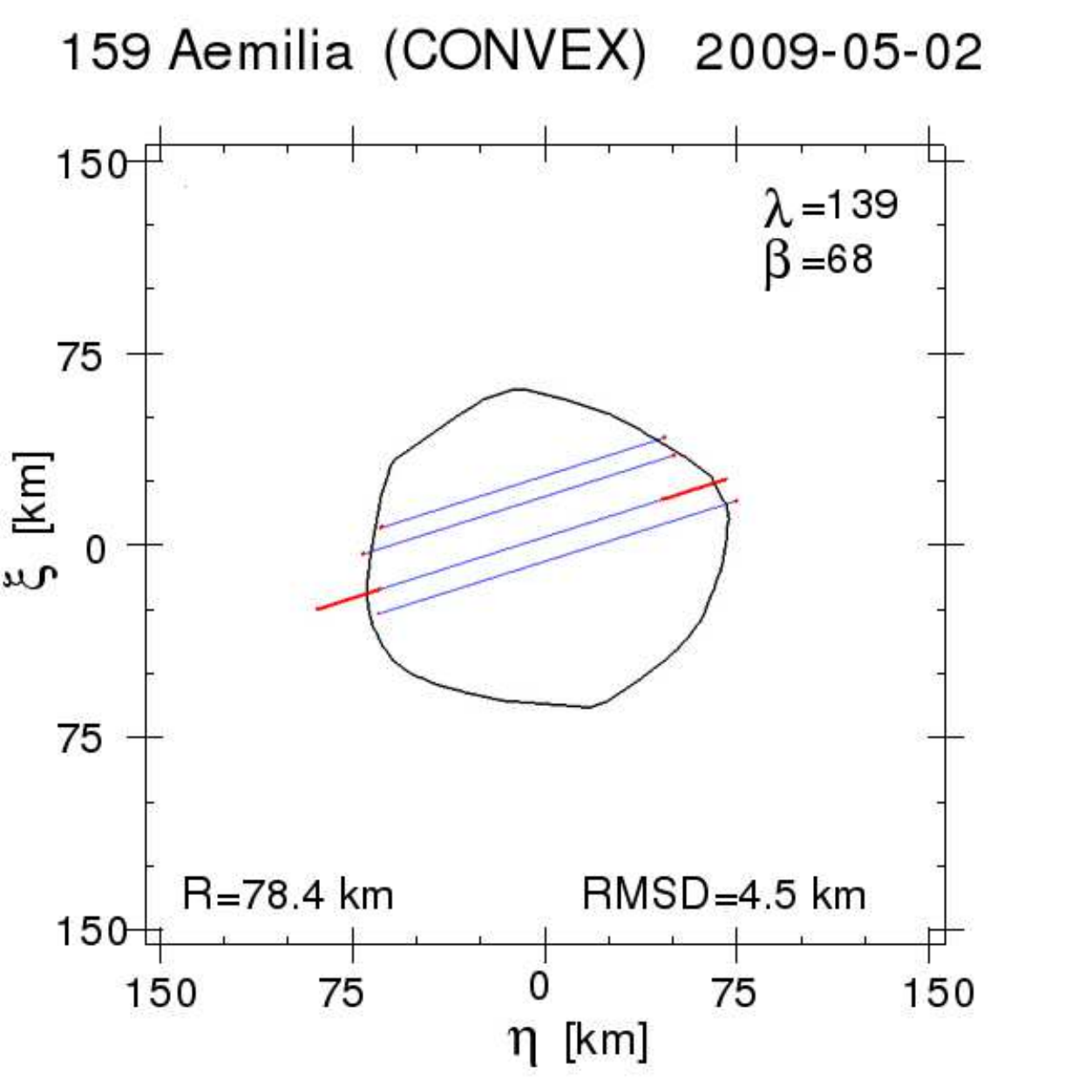}&
 \includegraphics[width=0.4\textwidth]{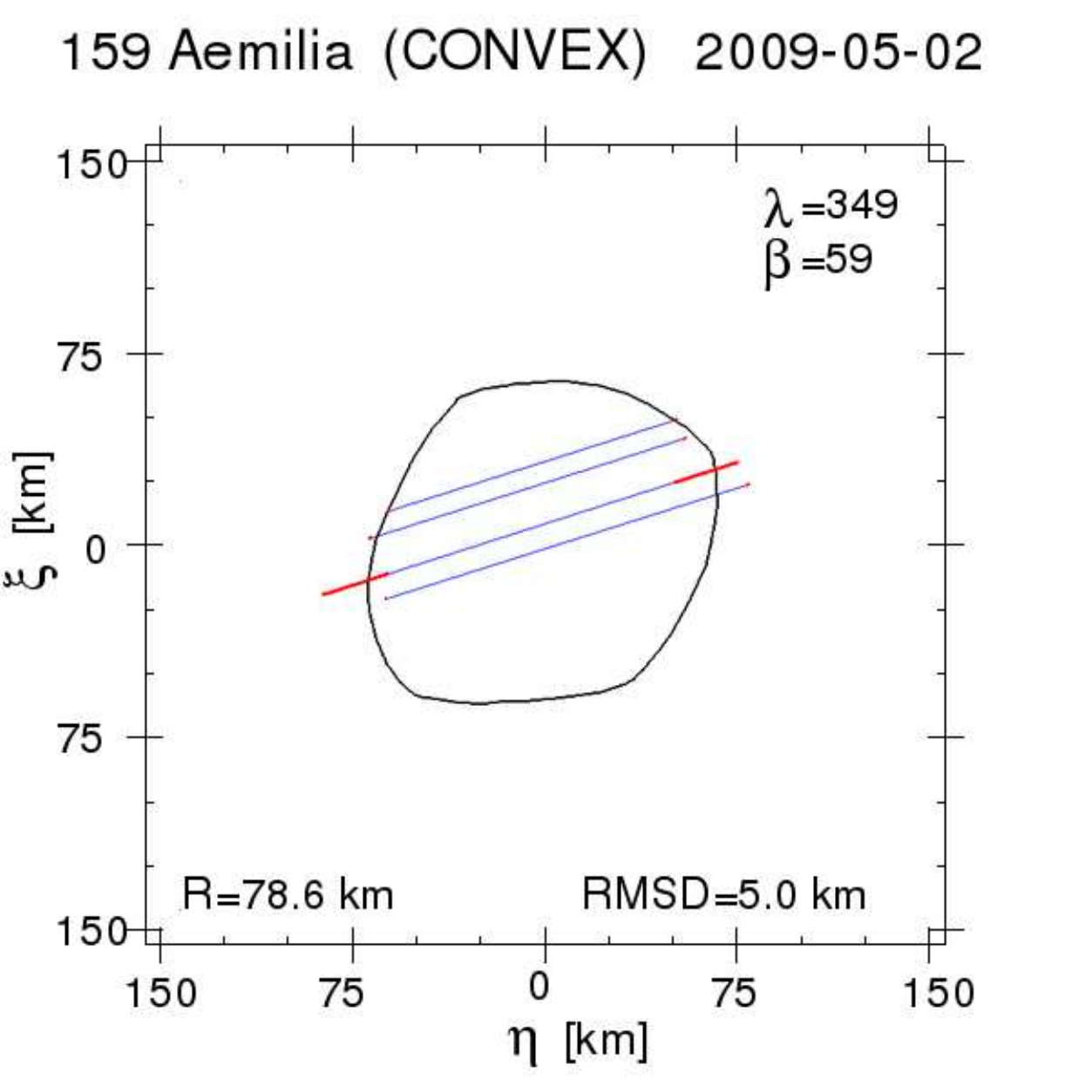}
\\
 \includegraphics[width=0.4\textwidth]{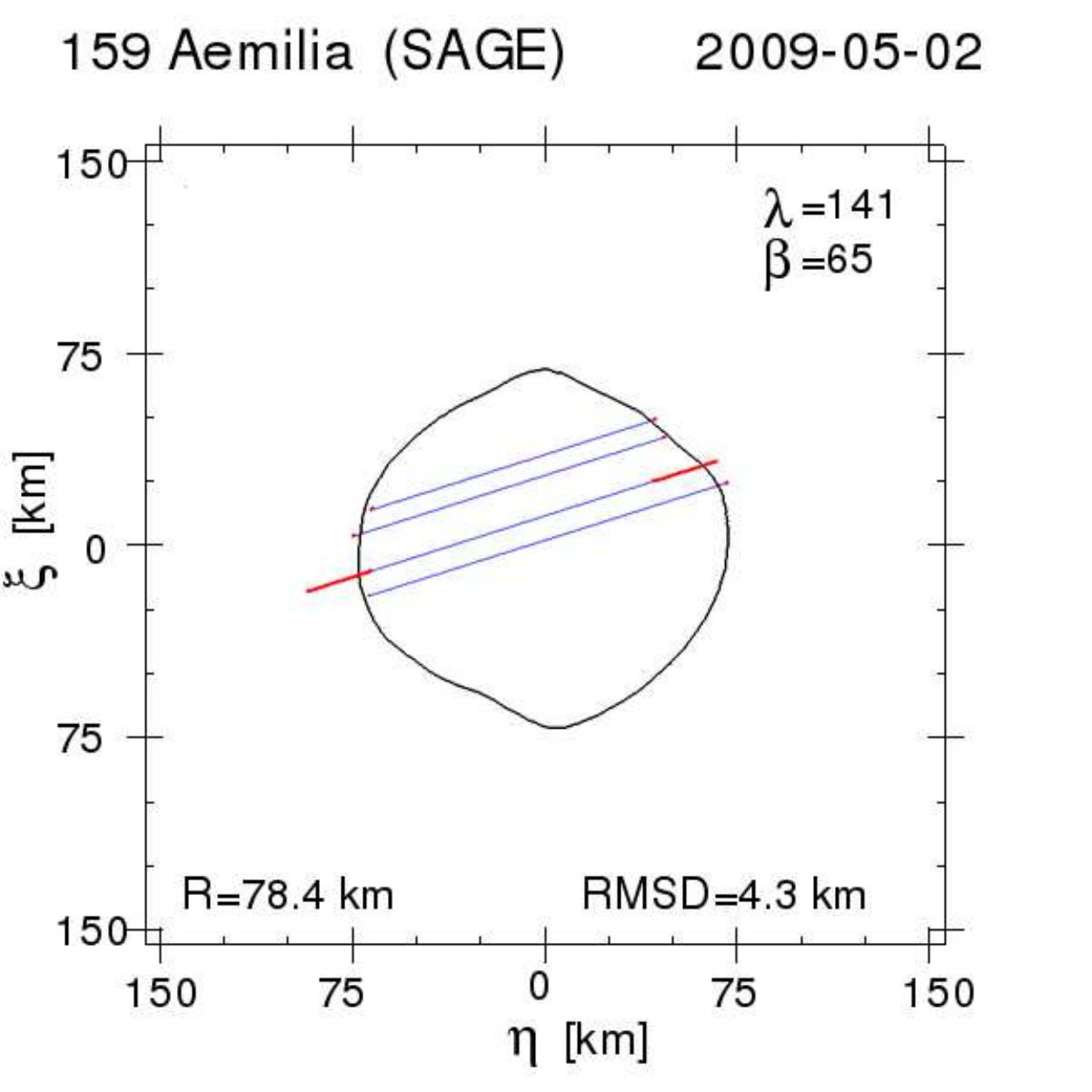}&
 \includegraphics[width=0.4\textwidth]{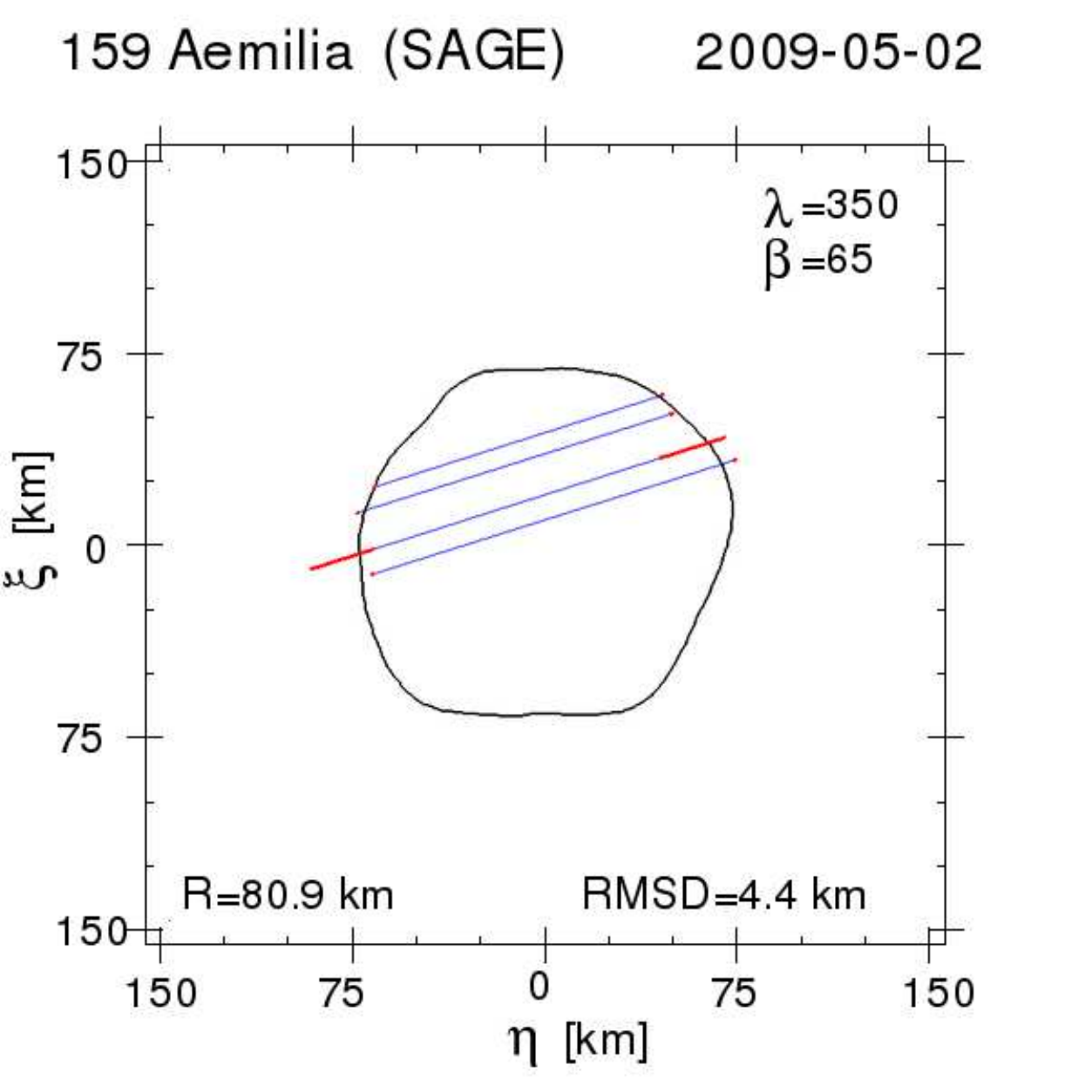}
\end{tabular}
\caption{Stellar occultation fits of convex (top) and non-convex (bottom)
    models of (159) Aemilia. At the end of each chord a timing uncertainty is marked.
    R is the radius of the largest model dimension.
    For equivalent volume sphere diameters see Table \ref{occ_diameters159}.}
\label{Aemilia_occult}
\end{figure*}

\clearpage

\begin{figure*}[h!]
\begin{tabular}{cc}
 \includegraphics[angle=90,width=0.4\textwidth]{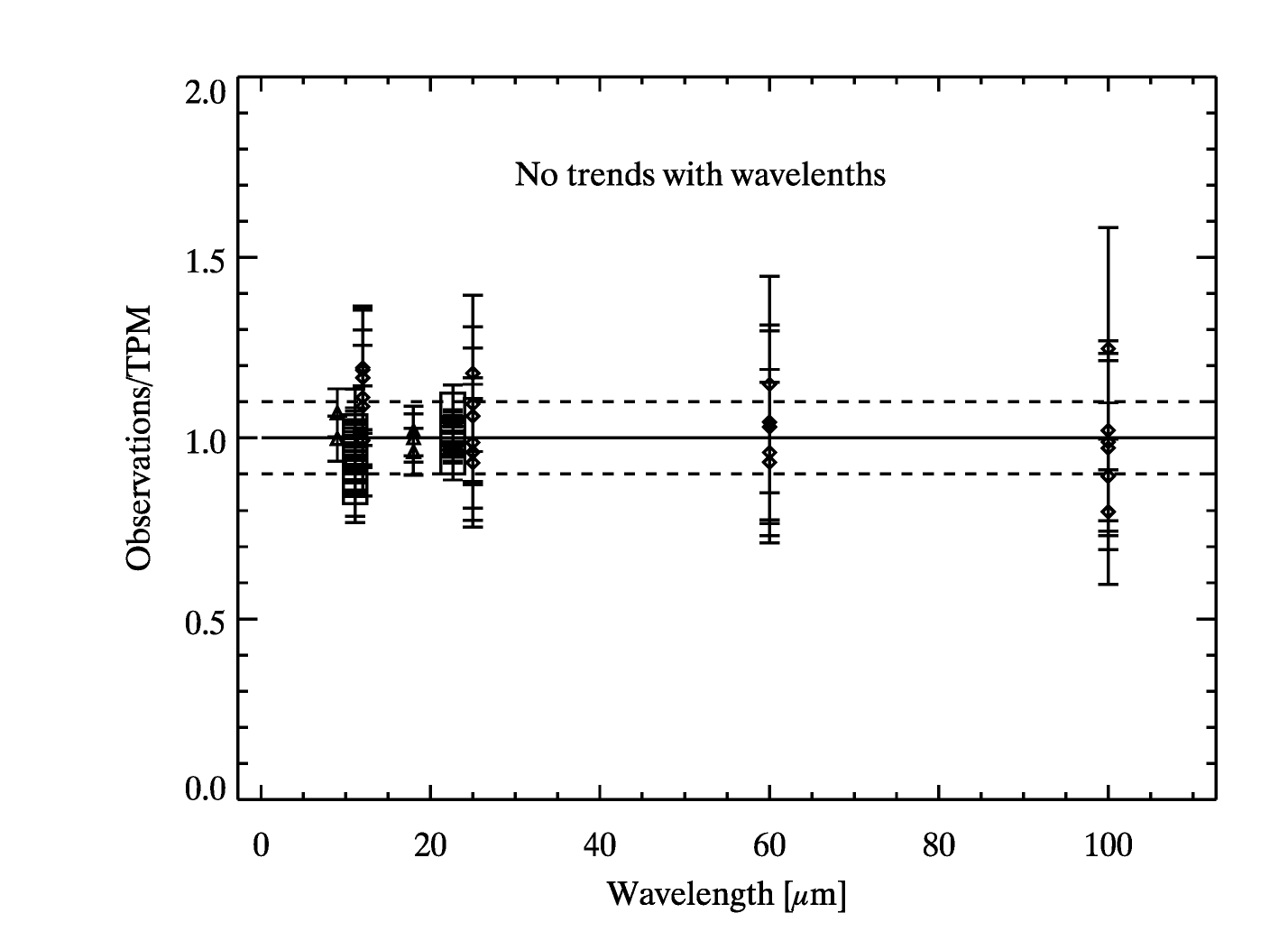}&
 \includegraphics[angle=90,width=0.4\textwidth]{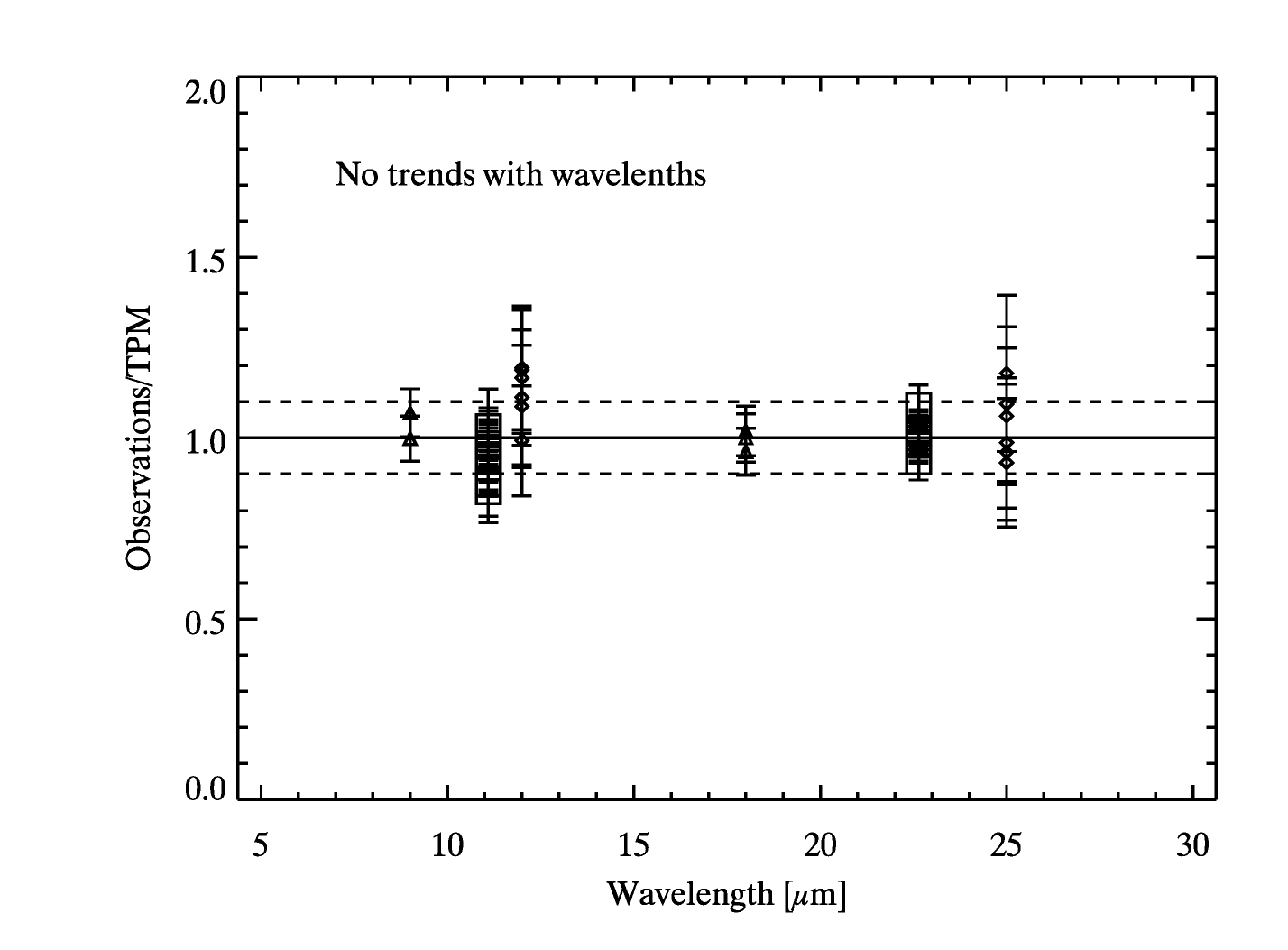}
\\
 \includegraphics[angle=90,width=0.4\textwidth]{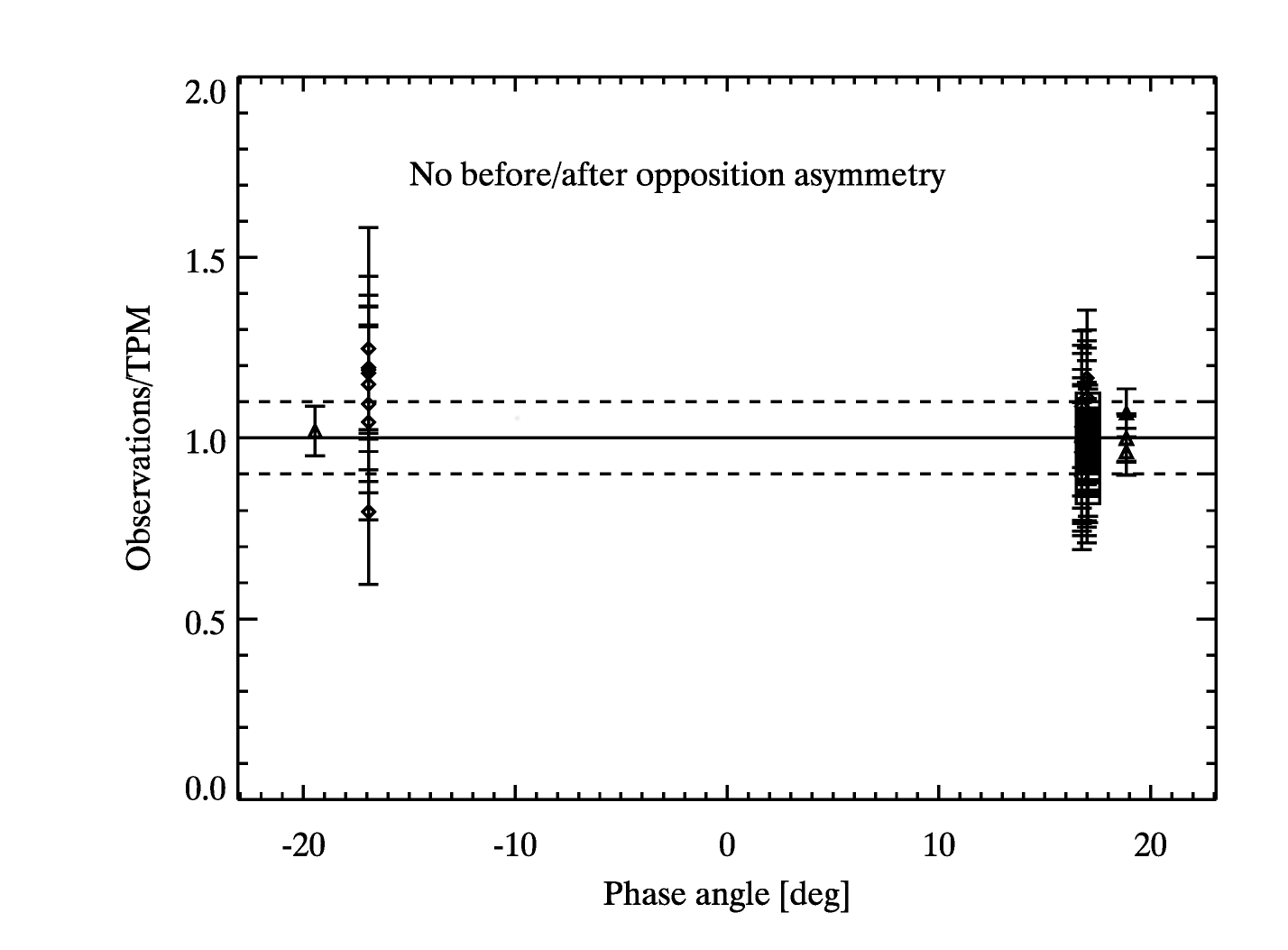}&
 \includegraphics[angle=90,width=0.4\textwidth]{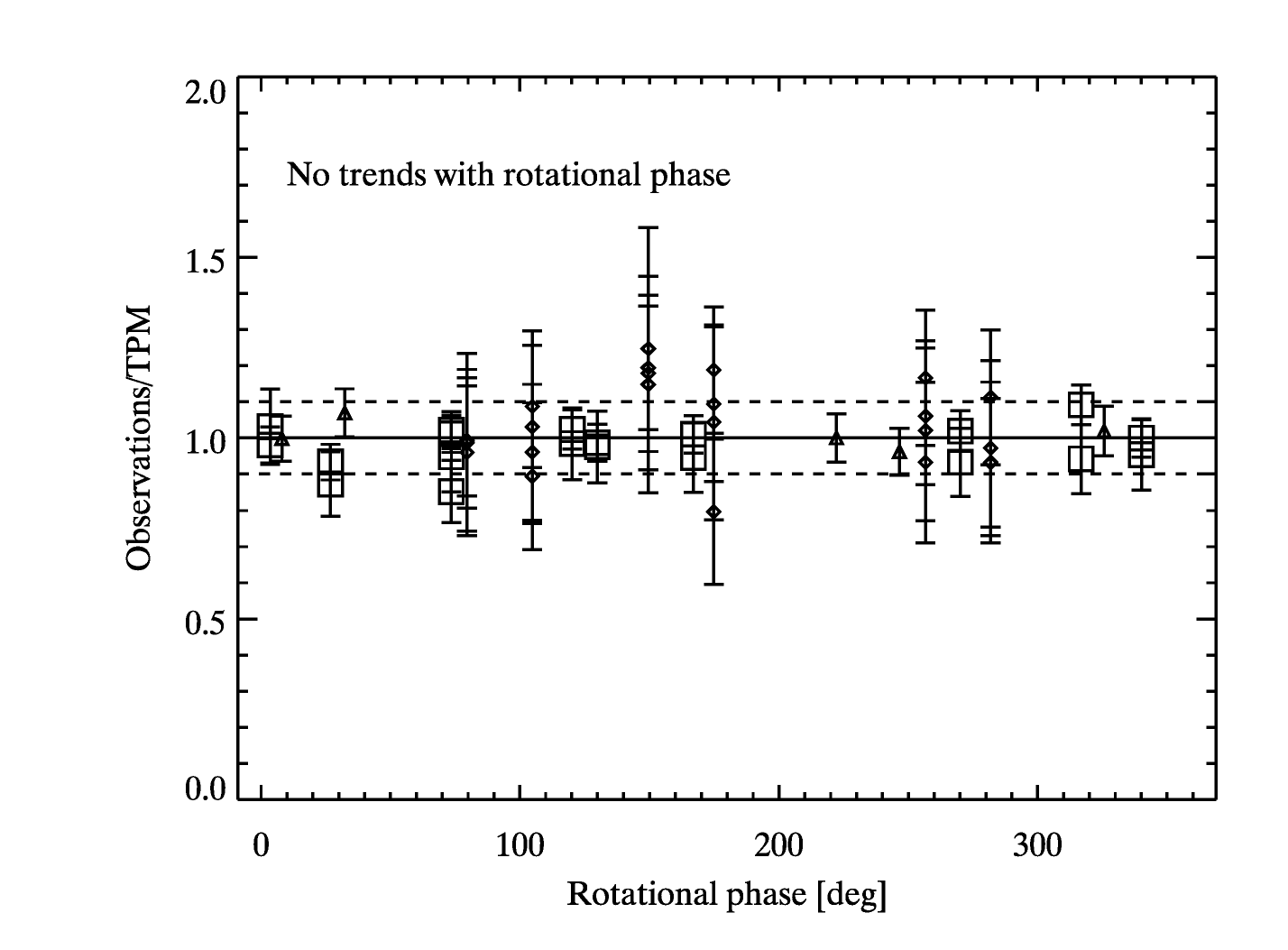}
\end{tabular}
\caption{O-C diagrams for the thermophysical model of (159) Aemilia using SAGE model 1. 
They illustrate how well the spin/shape model works against thermal infrared data. 
The dashed lines indicate +/-10\% in the observation-to-model ratio,  
which corresponds to typical flux errors of thermal measurements. 
There are no trends with wavelength, rotation, or pre- and post-opposition  asymmetry.
For the best fitting thermal parameters see Table \ref{TPMresults}. 
Triangles: data from AKARI, squares: WISE W3/W4, small diamonds: IRAS.}
\label{Aemilia_TPM}
\end{figure*}

\clearpage

\begin{figure*}[h!]
\includegraphics[width=0.33\textwidth]{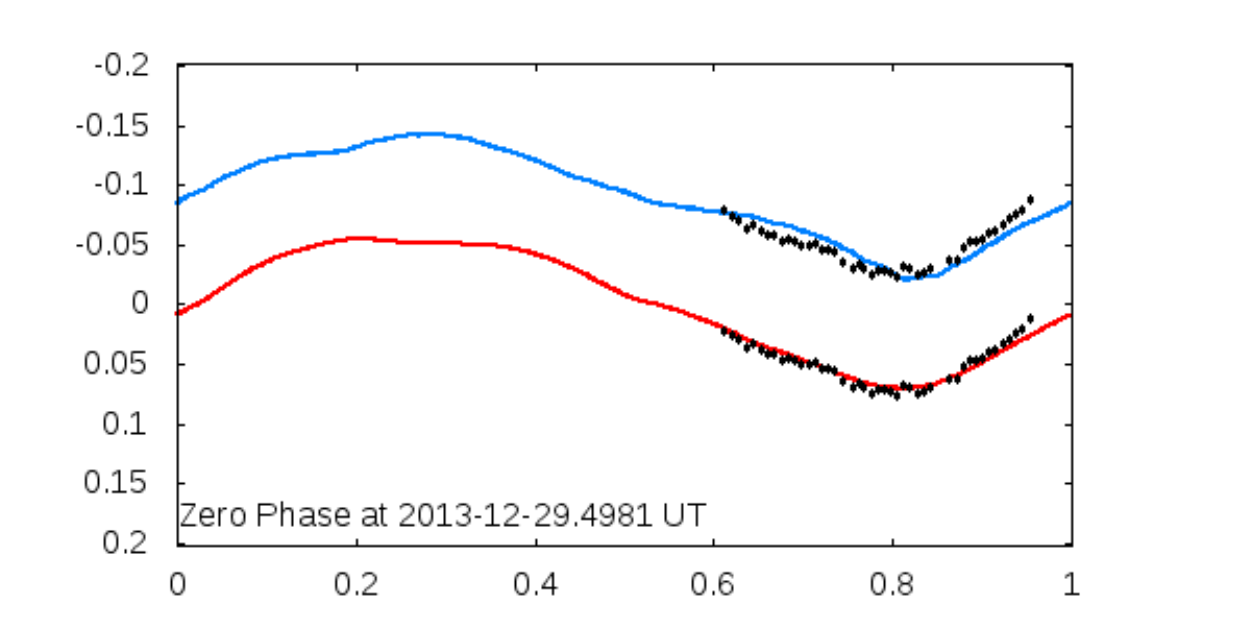}
\includegraphics[width=0.33\textwidth]{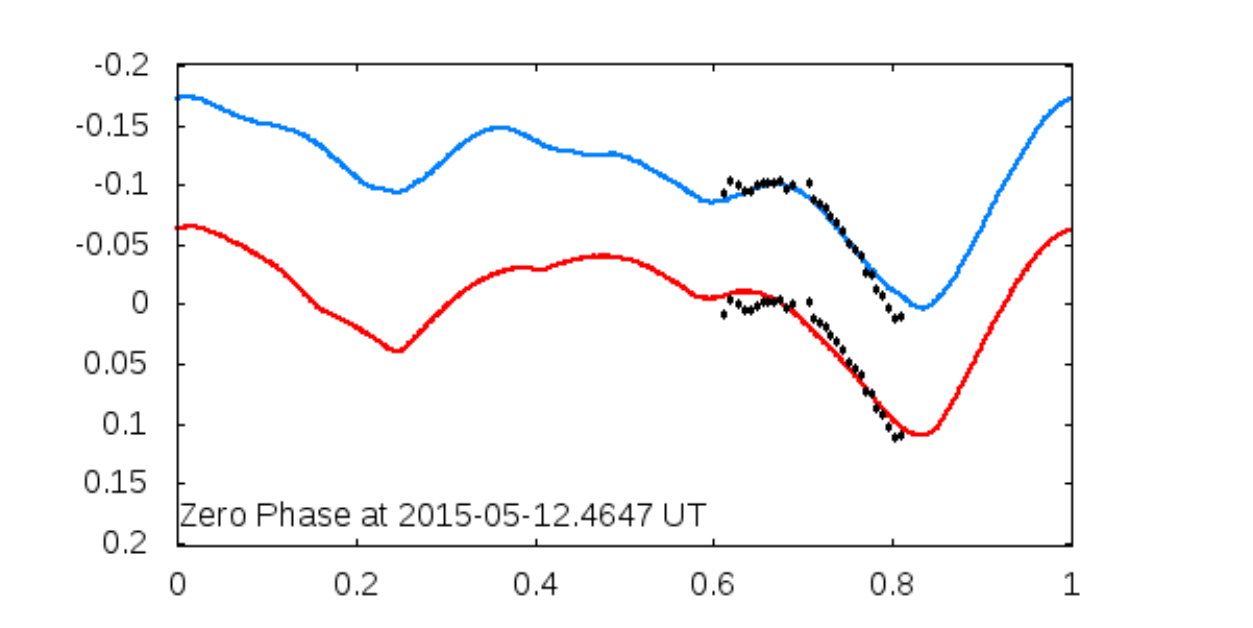}
\includegraphics[width=0.33\textwidth]{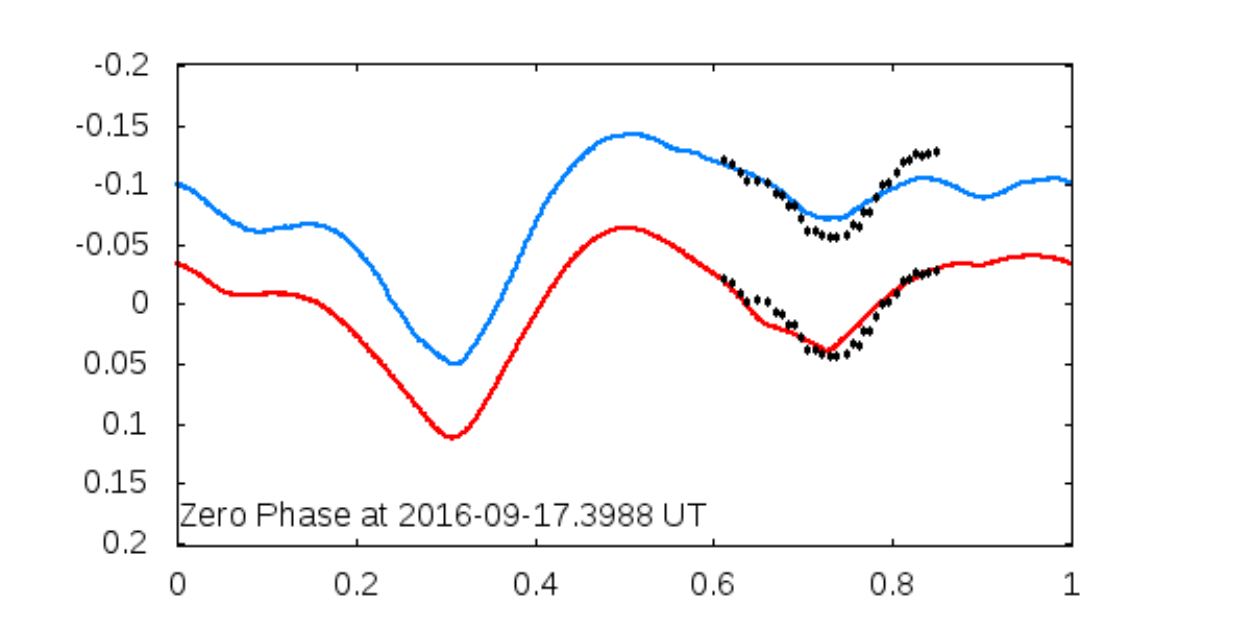}
\caption{Convex (upper curve) and non-convex (lower curve) model lightcurves of (227) Philosophia fitted to data 
from various apparitions (black points)}
\label{Philosophia_fit}
\end{figure*}

\begin{figure*}[h!]
\begin{tabular}{cc}
 \includegraphics[angle=90,width=0.4\textwidth]{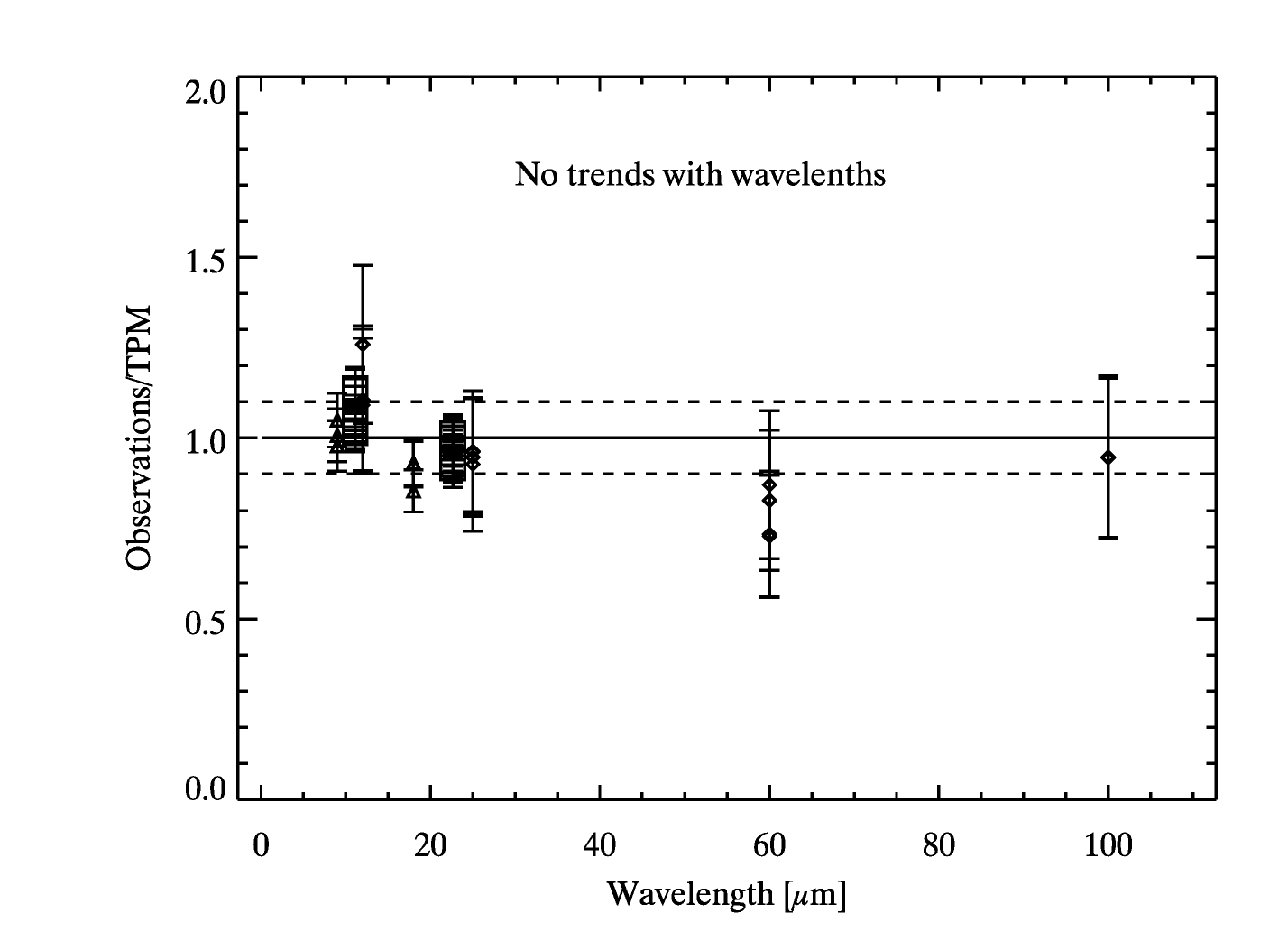}&
 \includegraphics[angle=90,width=0.4\textwidth]{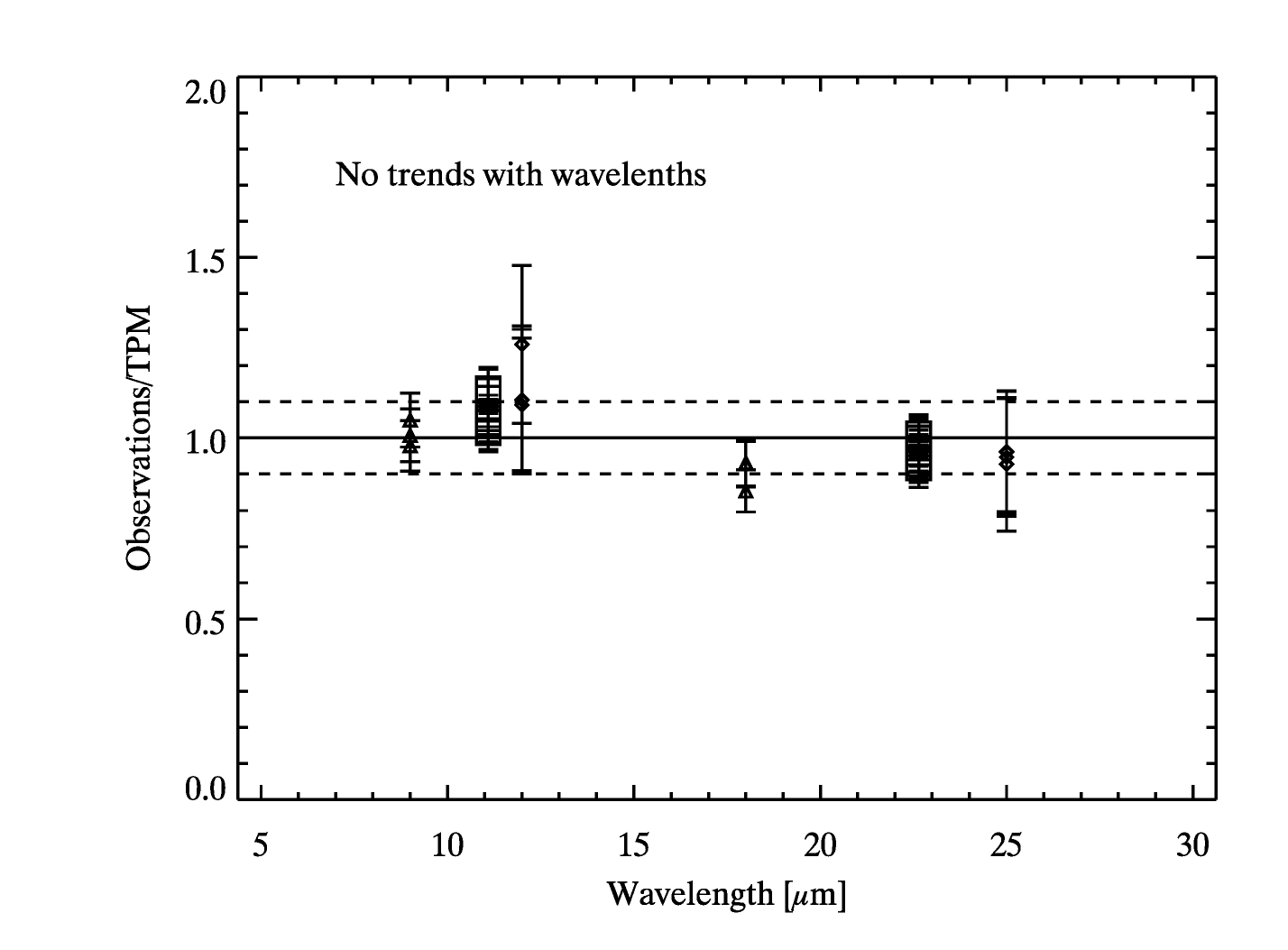}
\\
 \includegraphics[angle=90,width=0.4\textwidth]{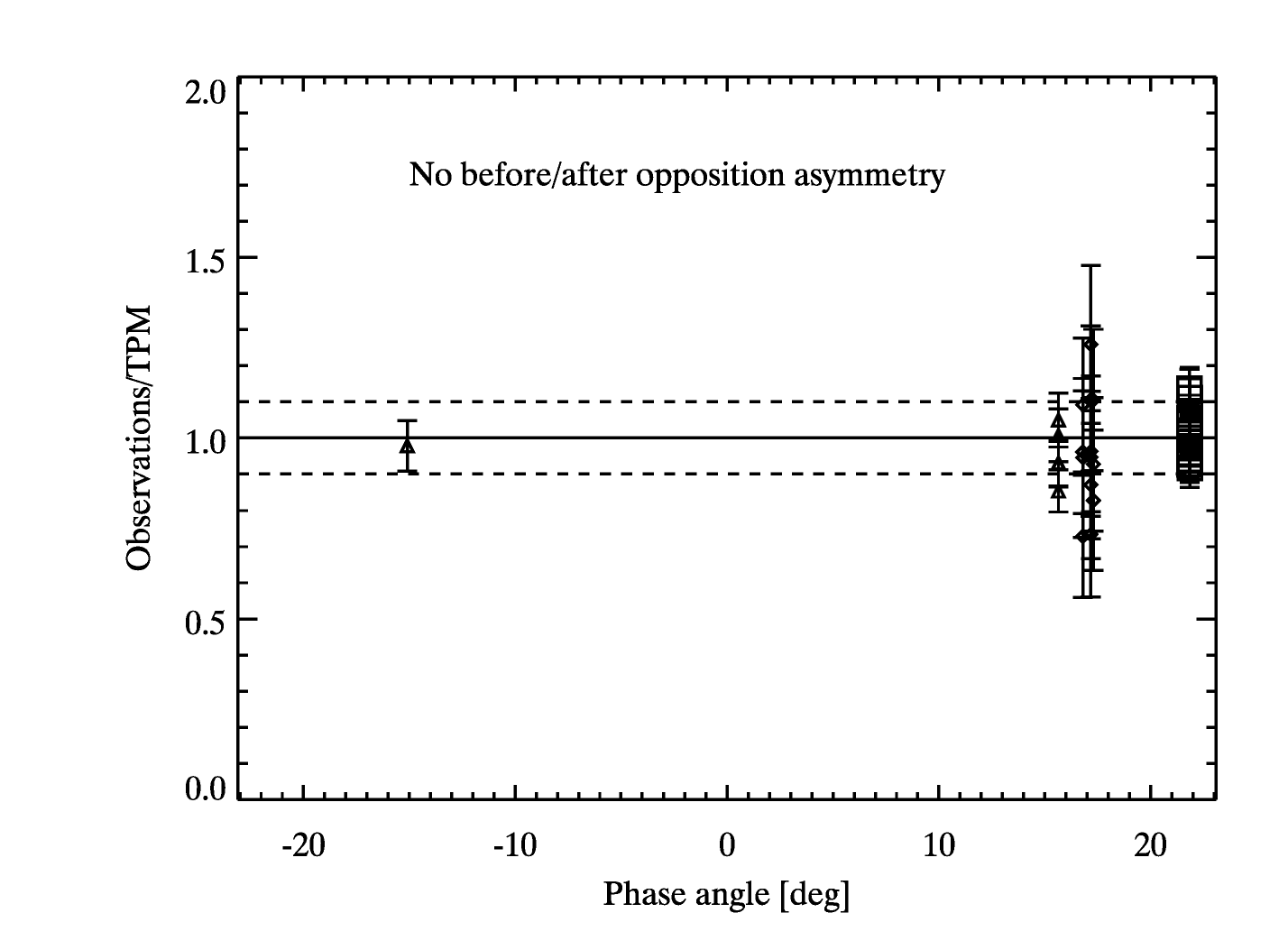}&
 \includegraphics[angle=90,width=0.4\textwidth]{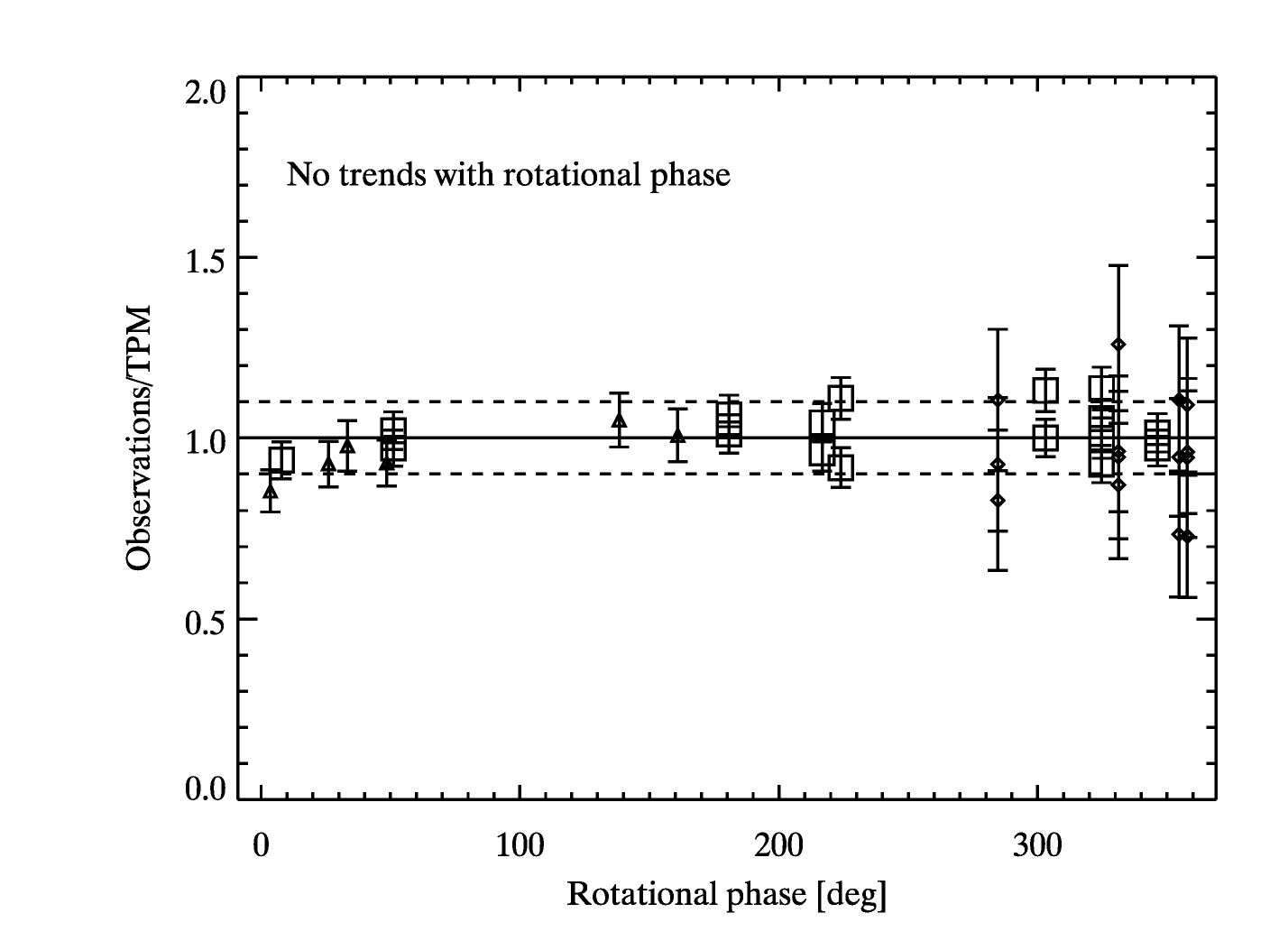}
\end{tabular}
\caption{O-C diagrams for the thermophysical model of (227) Philosphia, using convex model 1, 
illustrating that the spin/shape model works quite well against the thermal infrared data. 
There are no clear trends with wavelength, rotation, or pre- and post-opposition asymmetry.
For best fitting thermal parameters see Table \ref{TPMresults}.}
\label{Philosophia_TPM}
\end{figure*}

\clearpage

\begin{figure*}[h]
\includegraphics[width=0.33\textwidth]{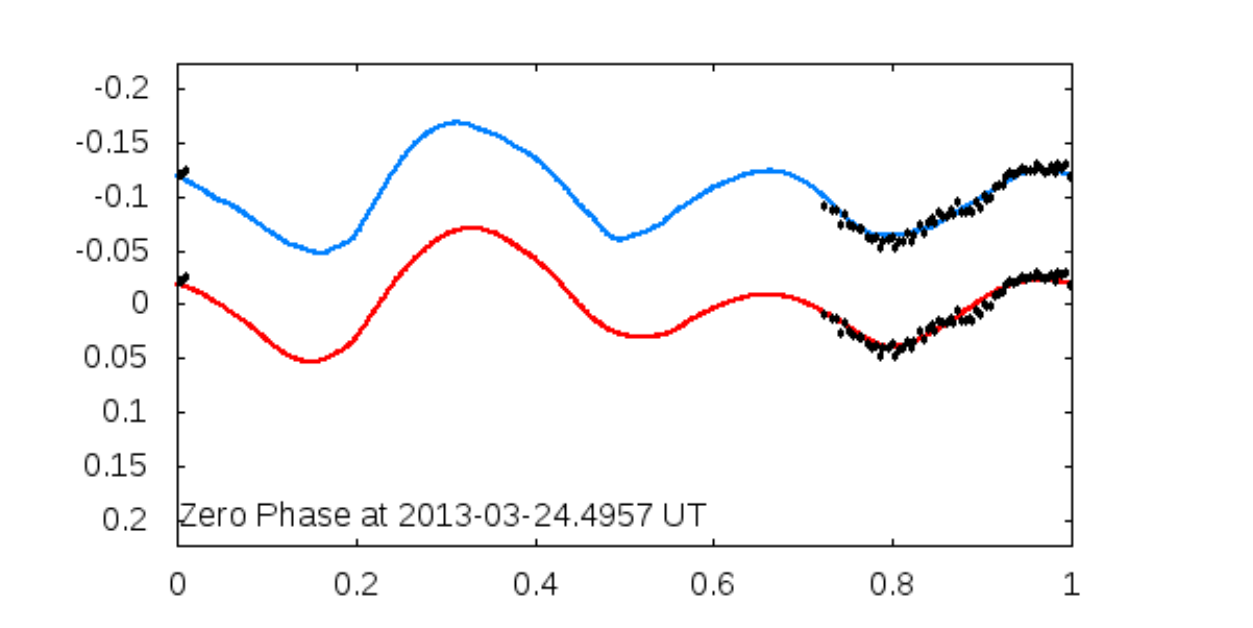}
\includegraphics[width=0.33\textwidth]{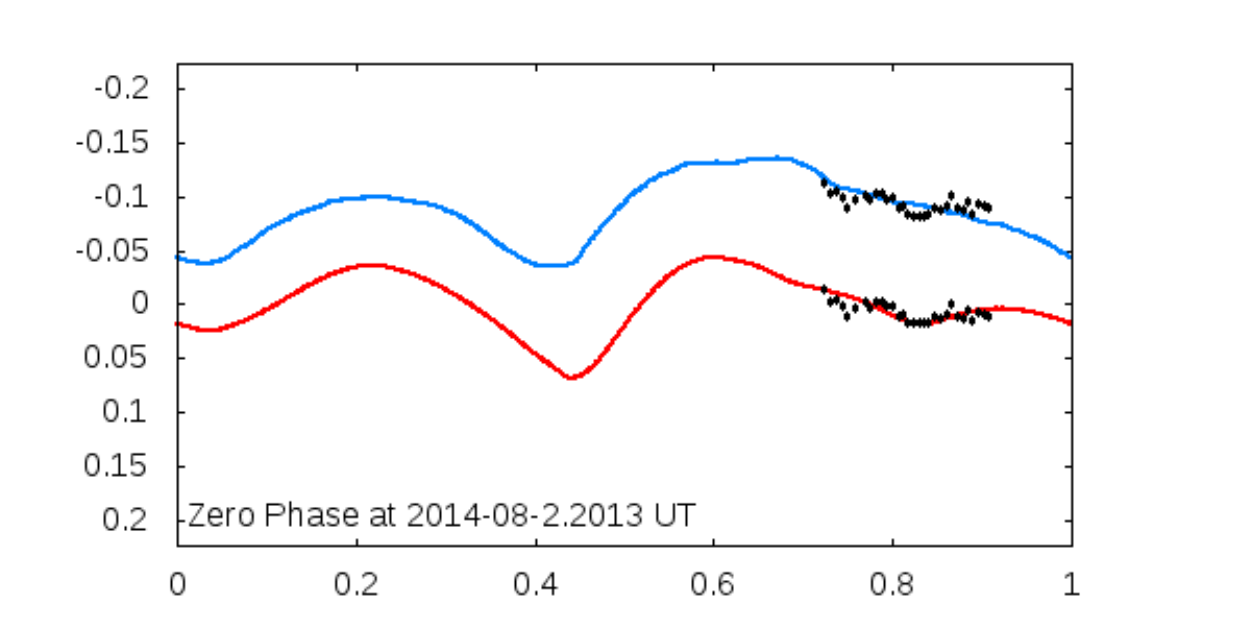}
\includegraphics[width=0.33\textwidth]{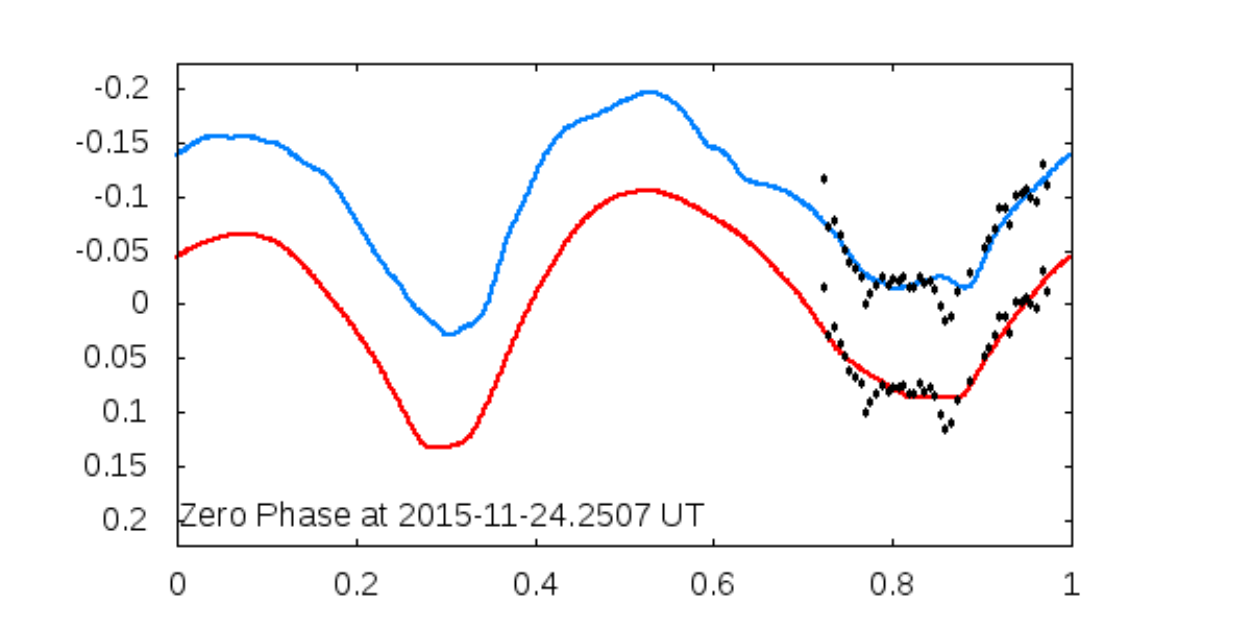}
\caption{Convex (upper curve) and non-convex (lower curve) model lightcurves of (329) Svea fitted to data 
from various apparitions (black points)}
\label{Svea_fit}
\end{figure*}

\begin{figure*}[h]
\begin{tabular}{cc}
 \includegraphics[width=0.4\textwidth]{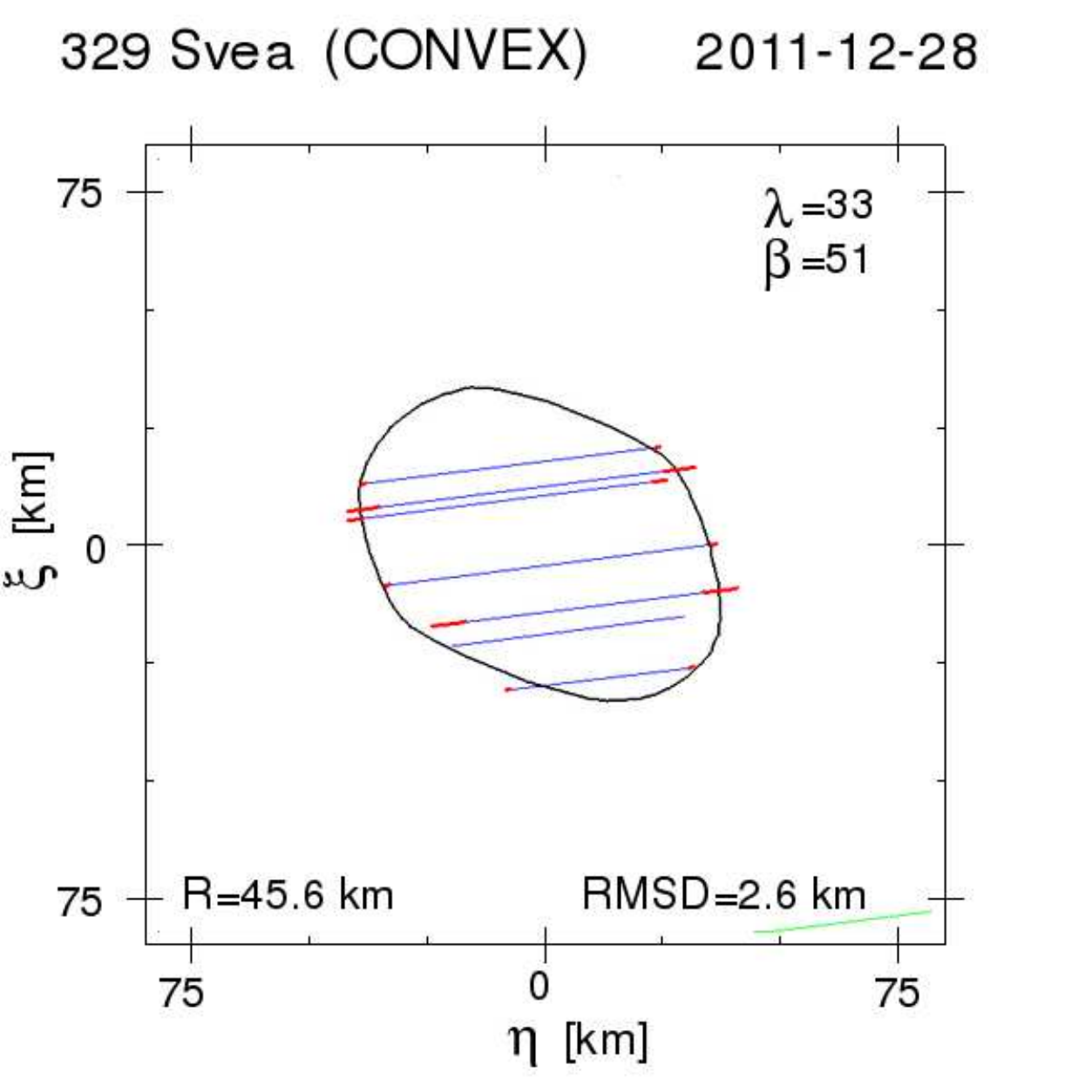}&
 \includegraphics[width=0.4\textwidth]{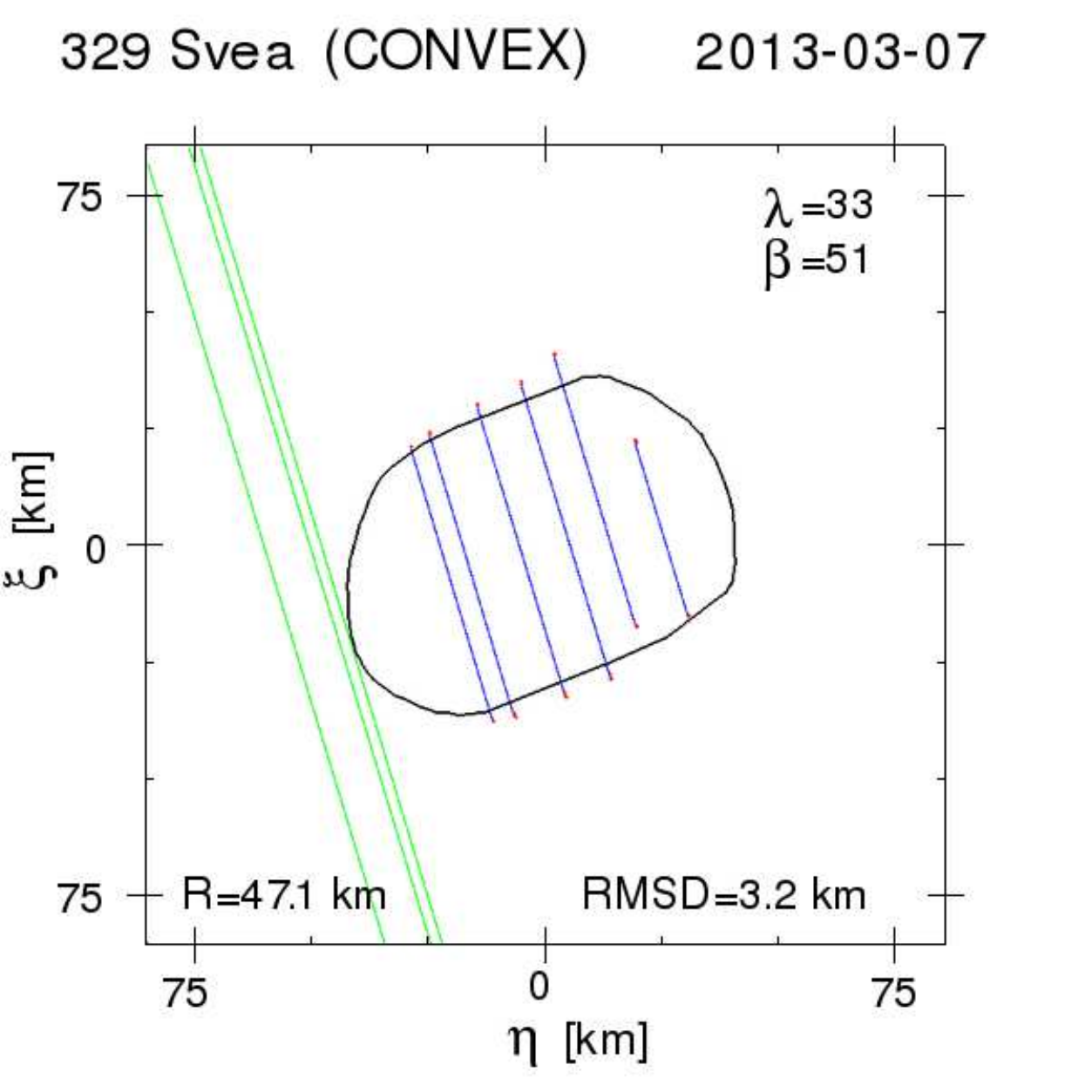}
\\
 \includegraphics[width=0.4\textwidth]{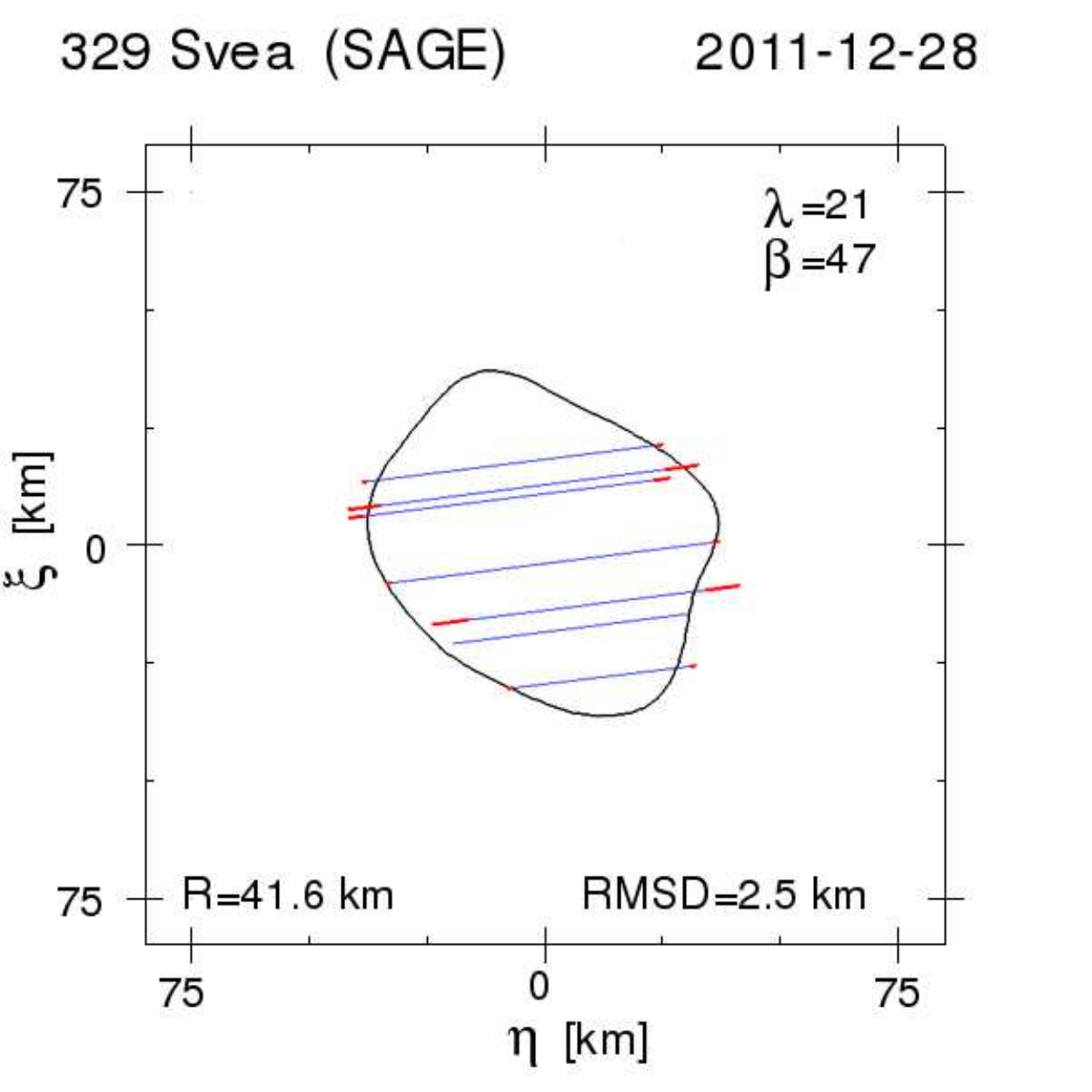}&
 \includegraphics[width=0.4\textwidth]{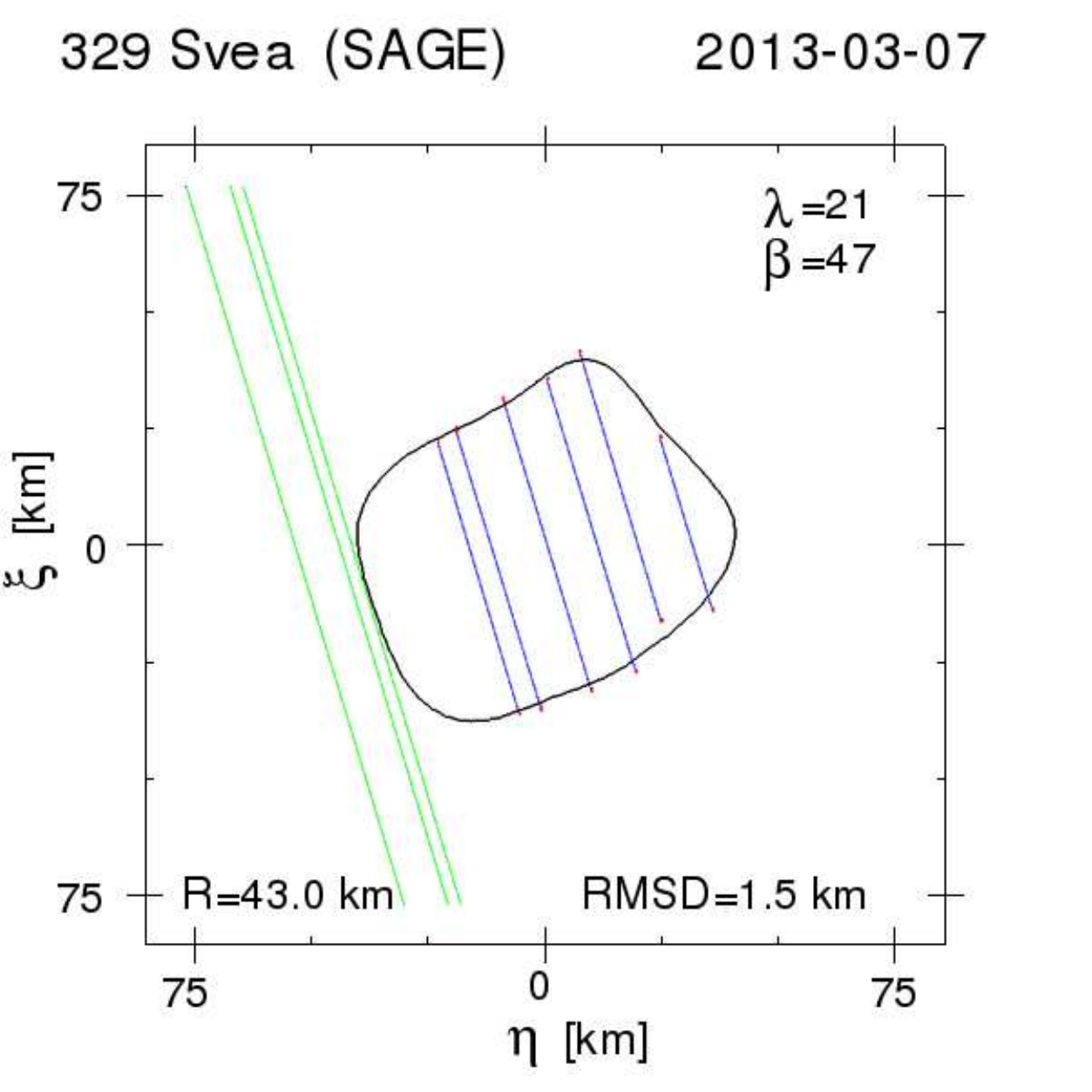}
\end{tabular}
\caption{Two stellar occultation fits of convex (top) and non-convex (bottom)
    models of (329) Svea, pole~1.\break At the end of each chord a timing uncertainty is marked.
    R is the radius of the largest model dimension.}
\label{Svea_occult}
\end{figure*}

\begin{figure*}[h]
\begin{tabular}{cc}
 \includegraphics[width=0.4\textwidth]{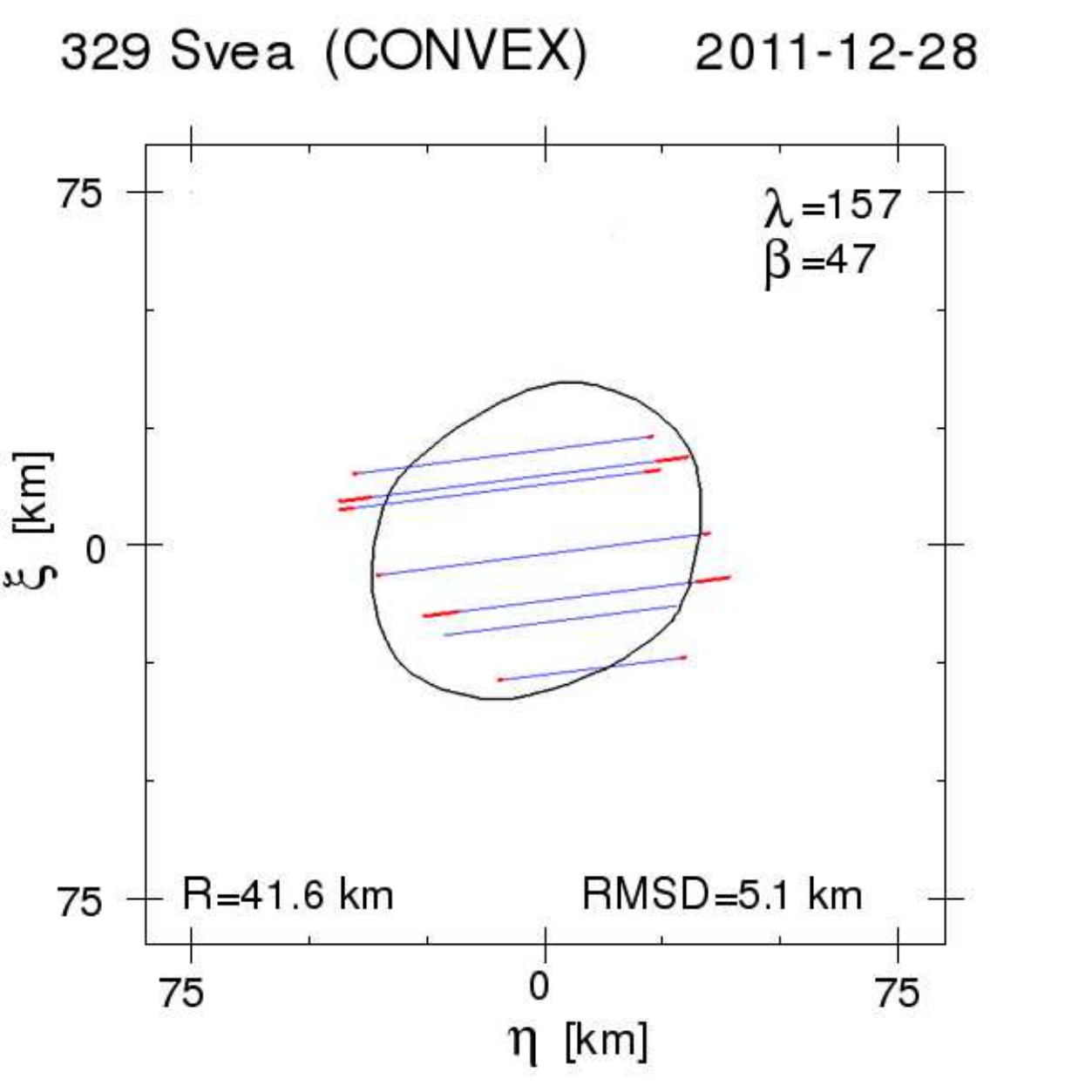}&
 \includegraphics[width=0.4\textwidth]{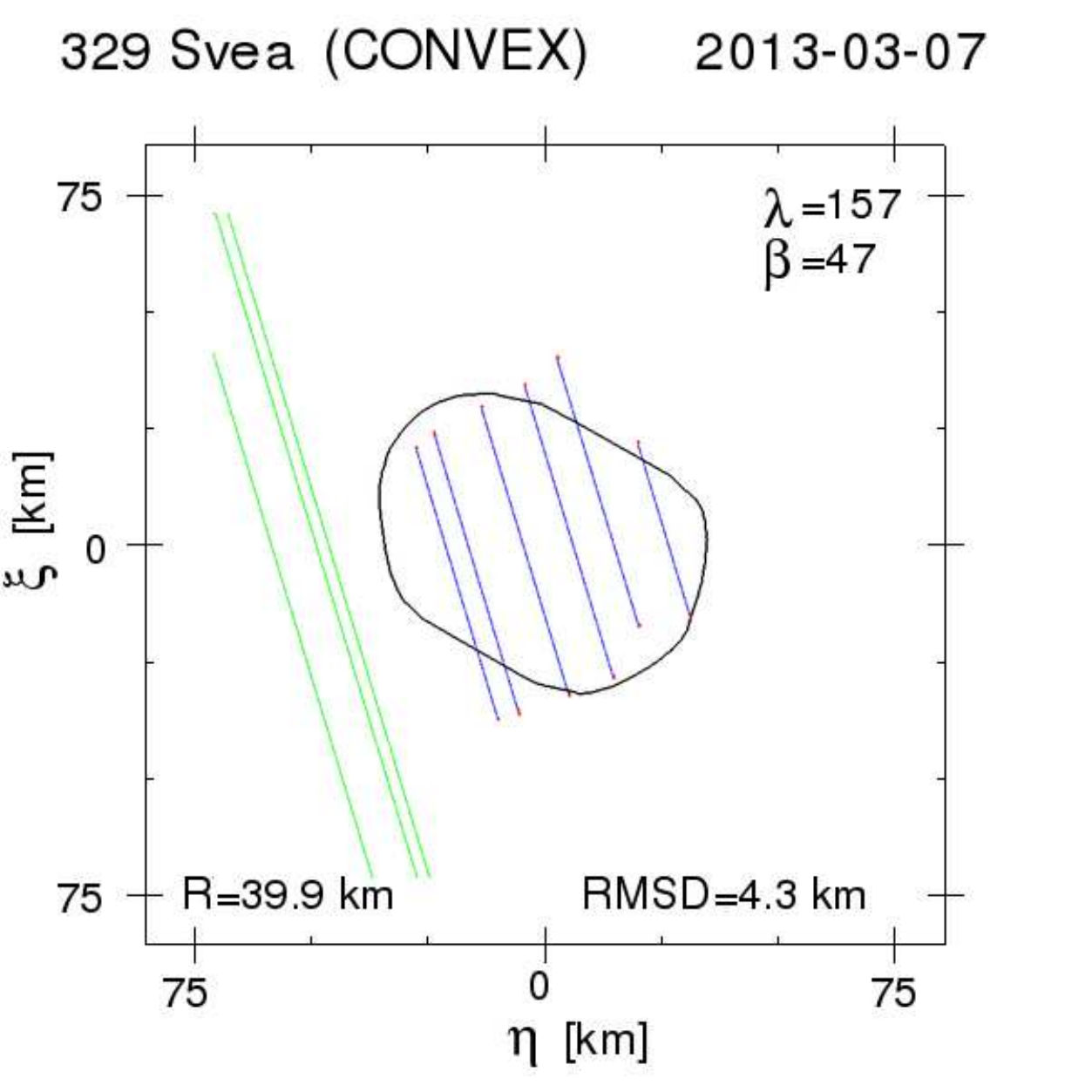}
\\
 \includegraphics[width=0.4\textwidth]{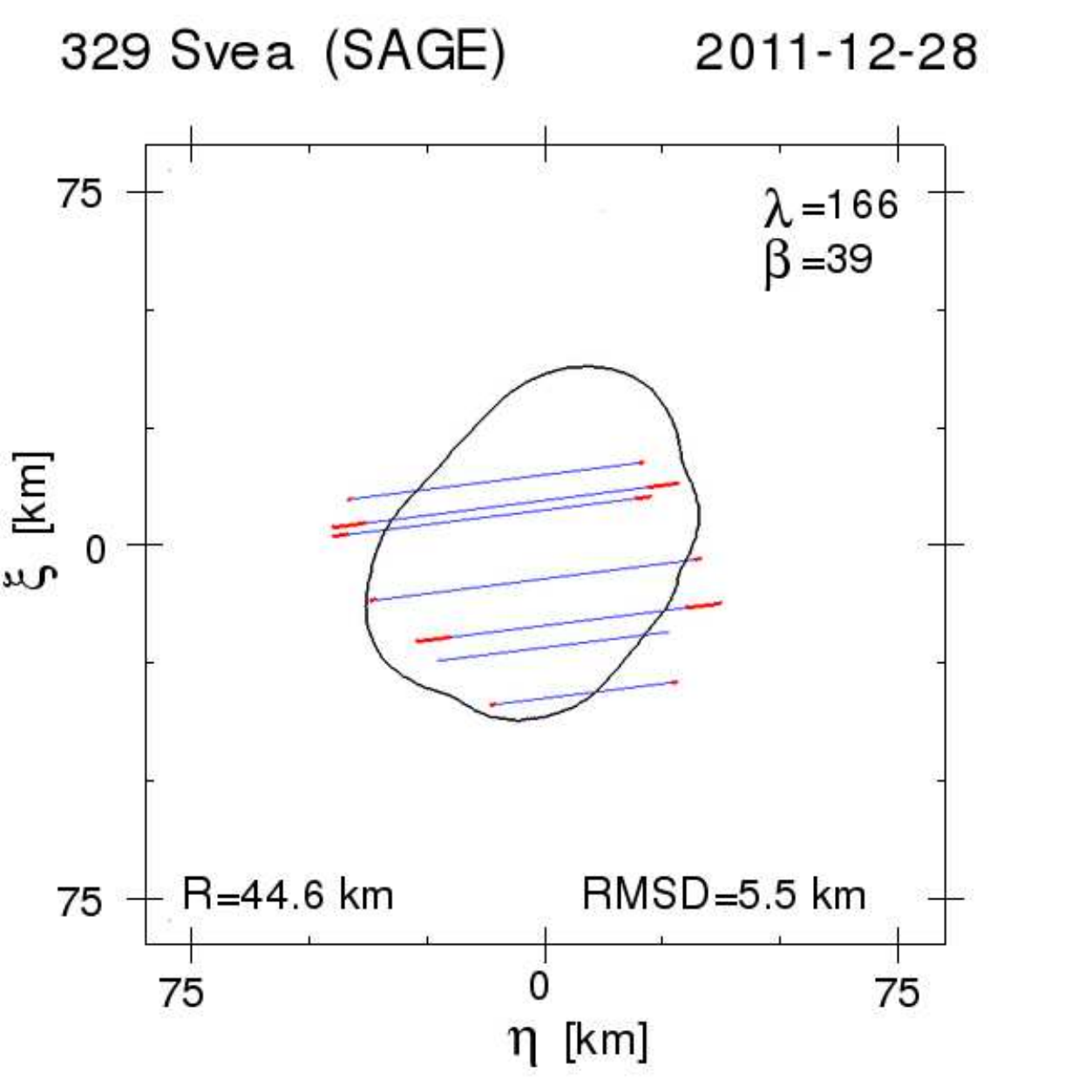}&
 \includegraphics[width=0.4\textwidth]{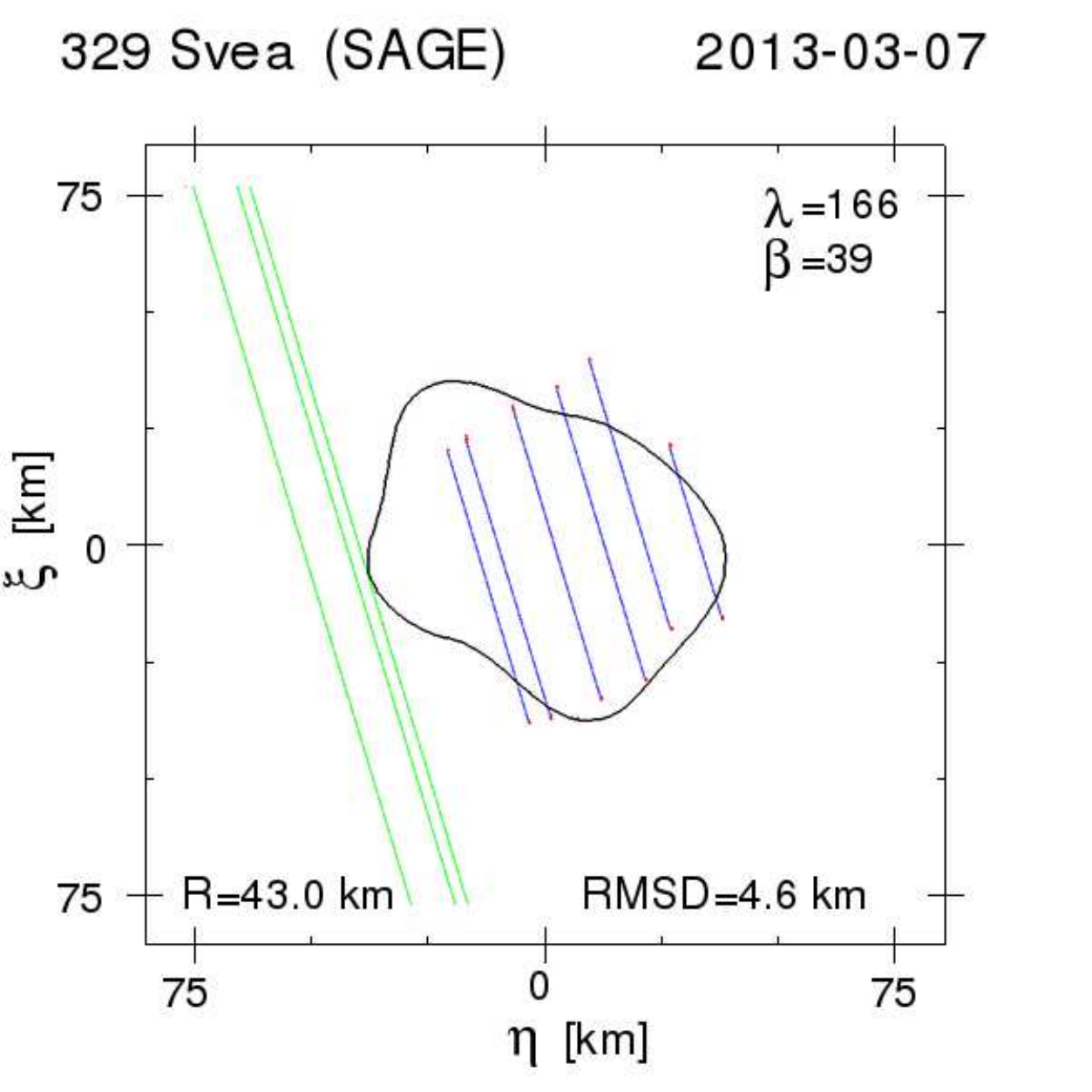}
\end{tabular}
\caption{Svea occultation fits for mirror pole solution (pole 2 from Table \ref{results}). 
The clear misfit of this pole solution allows it to be safely rejected  
in favour of the pole 1 solution (compare Fig. \ref{Svea_occult}).}
\label{Svea_occult_mirror}
\end{figure*}

\begin{figure*}[h!]
\begin{tabular}{cc}
 \includegraphics[angle=90,width=0.4\textwidth]{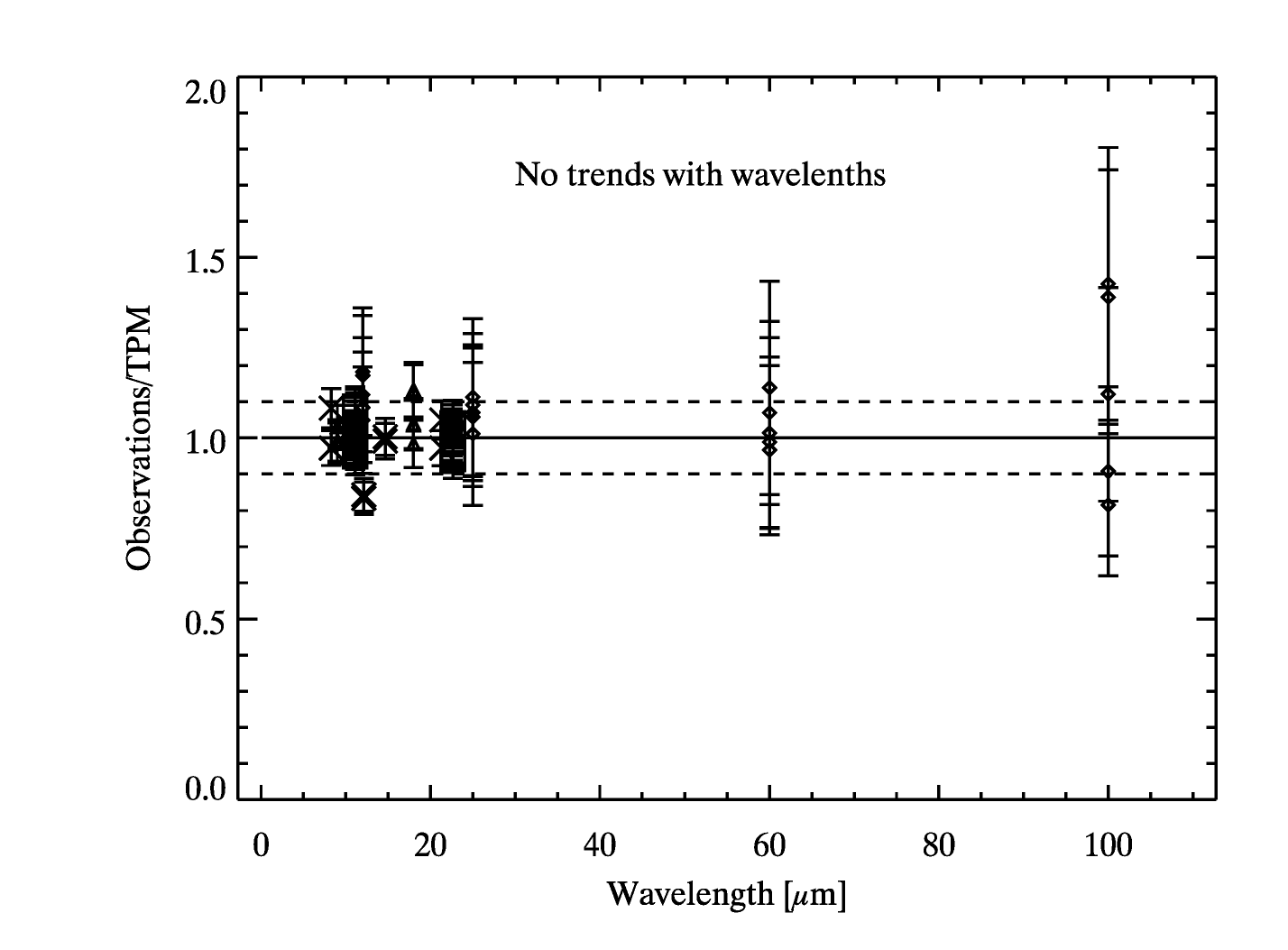}&
 \includegraphics[angle=90,width=0.4\textwidth]{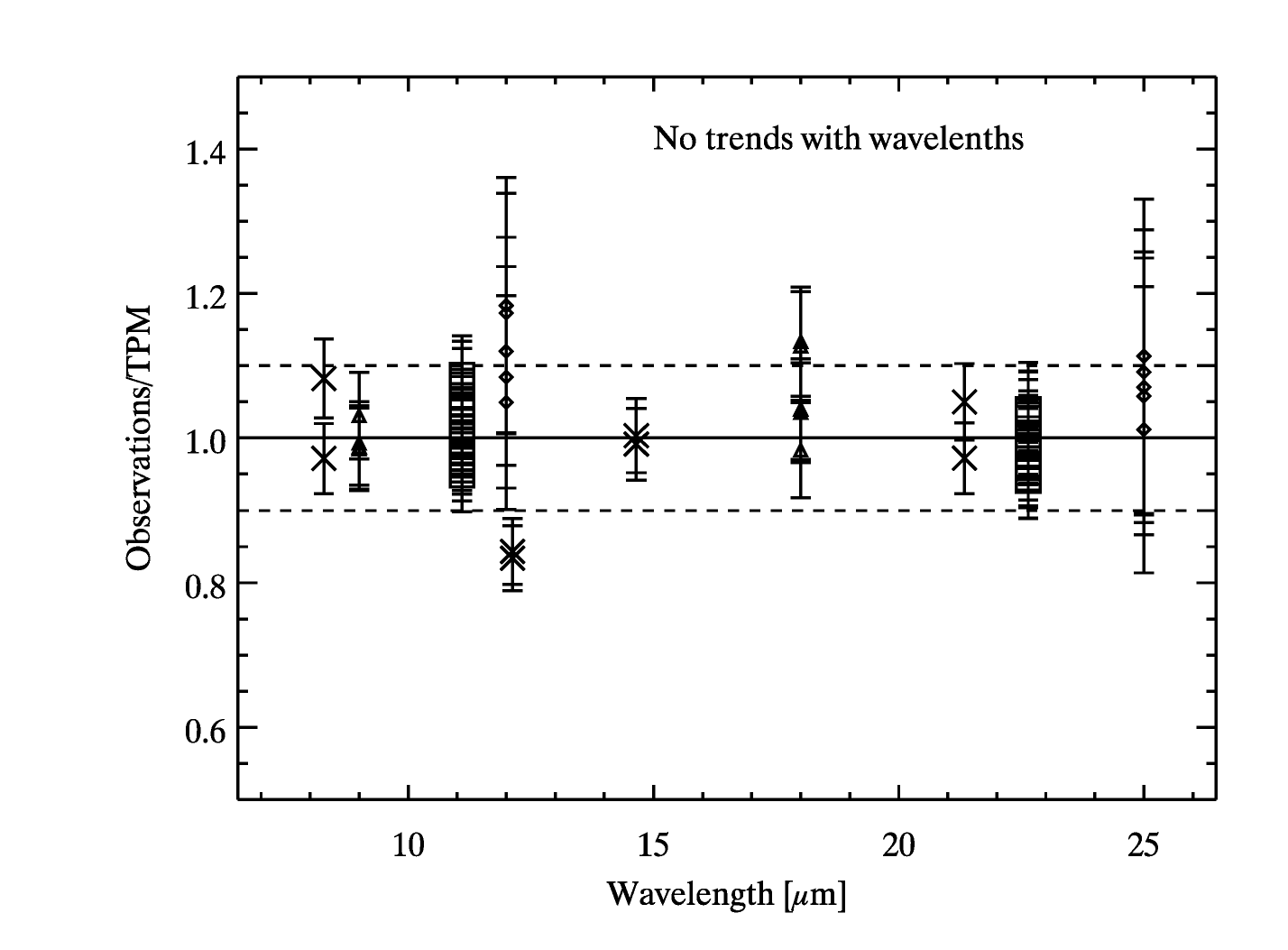}
\\
 \includegraphics[angle=90,width=0.4\textwidth]{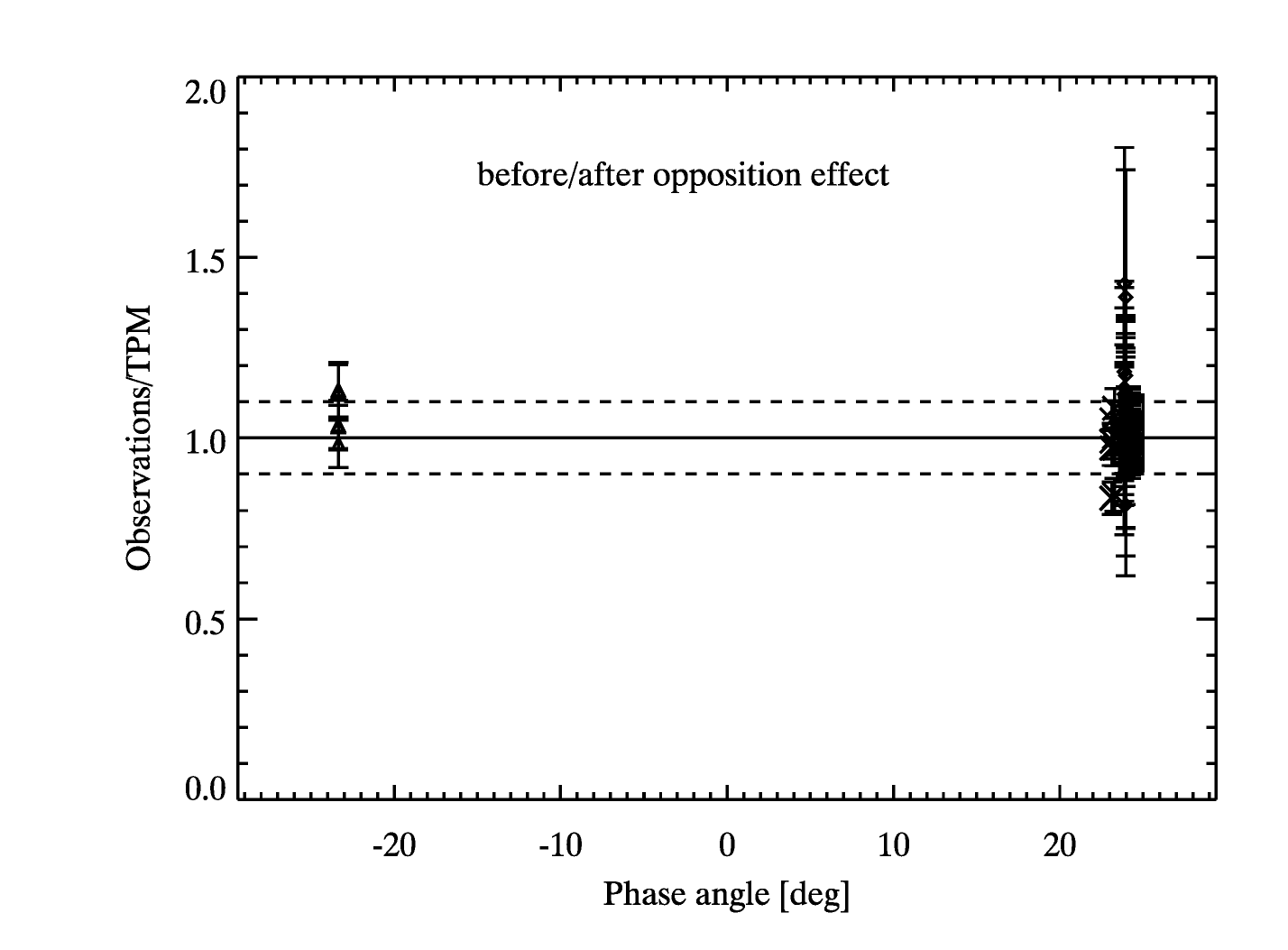}&
 \includegraphics[angle=90,width=0.4\textwidth]{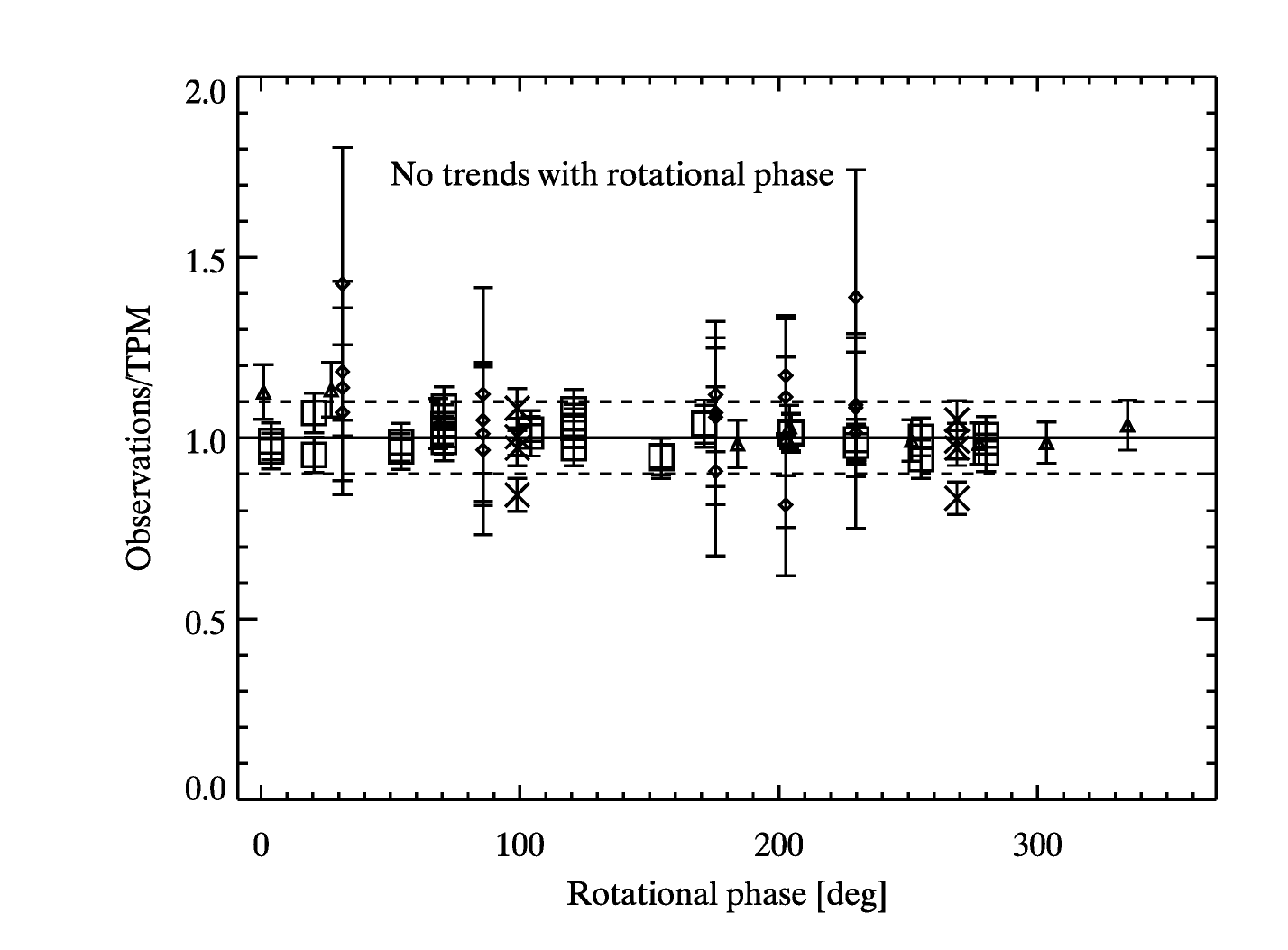}
\end{tabular}
\caption{O-C diagrams for the thermophysical model of (329) Svea using convex model 1. 
There are no trends with wavelength, rotation, or pre- and post-opposition asymmetry.
For the best fitting thermal parameters see Table \ref{TPMresults}. 
Triangles: data from AKARI, squares: WISE W3/W4, small diamonds: IRAS, X-symbols: MSX.}
\label{Svea_TPM}
\end{figure*}

\clearpage

\begin{figure*}[h]
\includegraphics[width=0.33\textwidth]{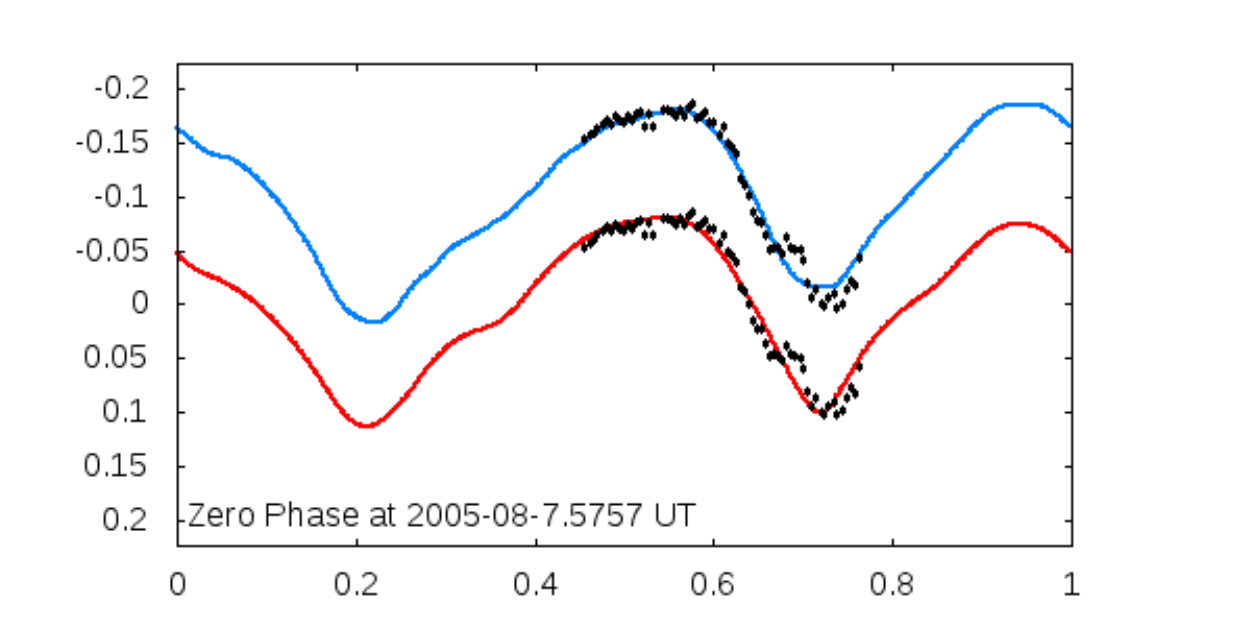}
\includegraphics[width=0.33\textwidth]{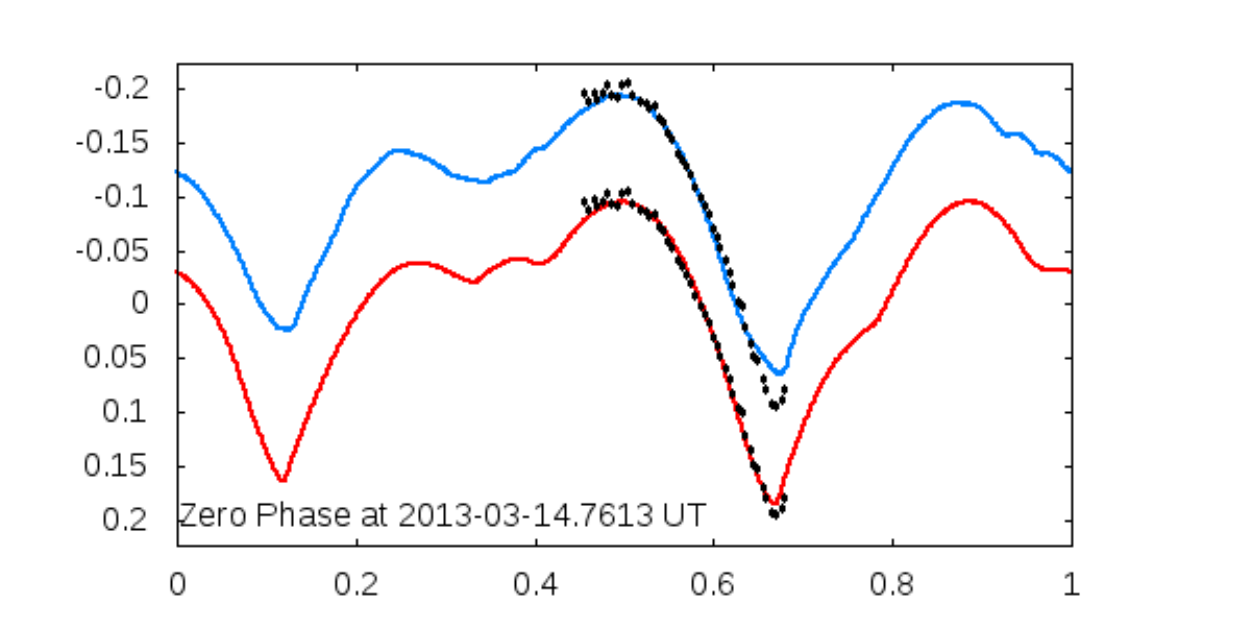}
\includegraphics[width=0.33\textwidth]{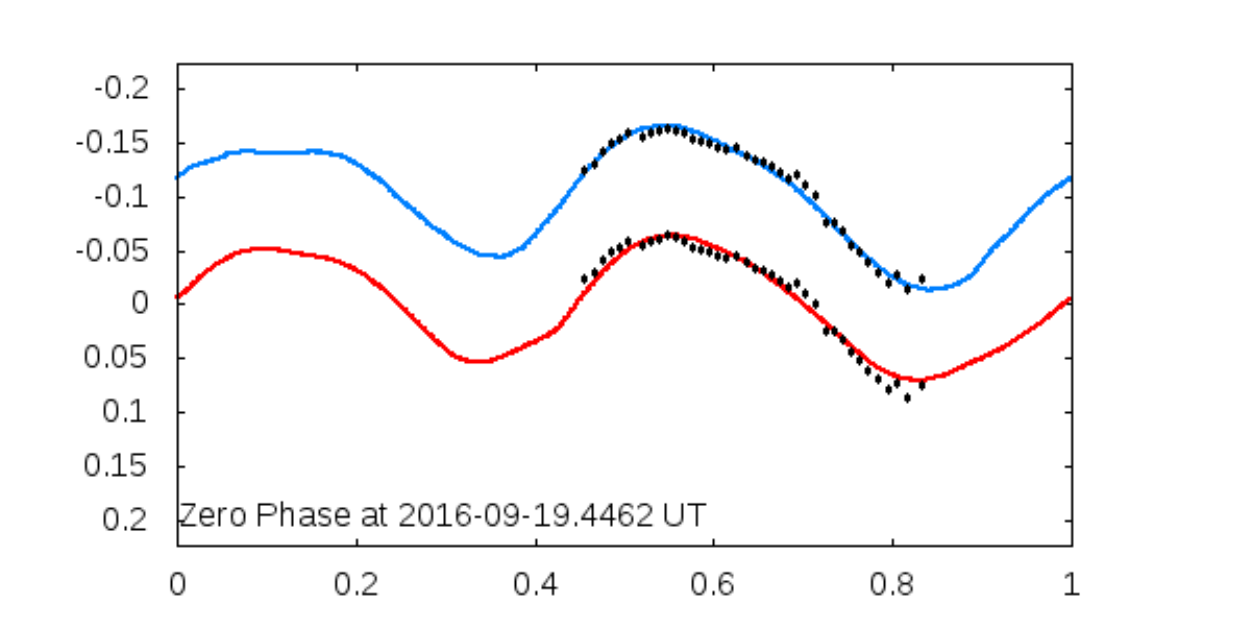}
\caption{Convex (upper curve) and non-convex (lower curve) model lightcurves of (478) Tergeste fitted to the data 
from various apparitions (black points)}
\label{Tergeste_fit}
\end{figure*}

\begin{figure*}[h]
\begin{tabular}{cc}
 \includegraphics[angle=90,width=0.4\textwidth]{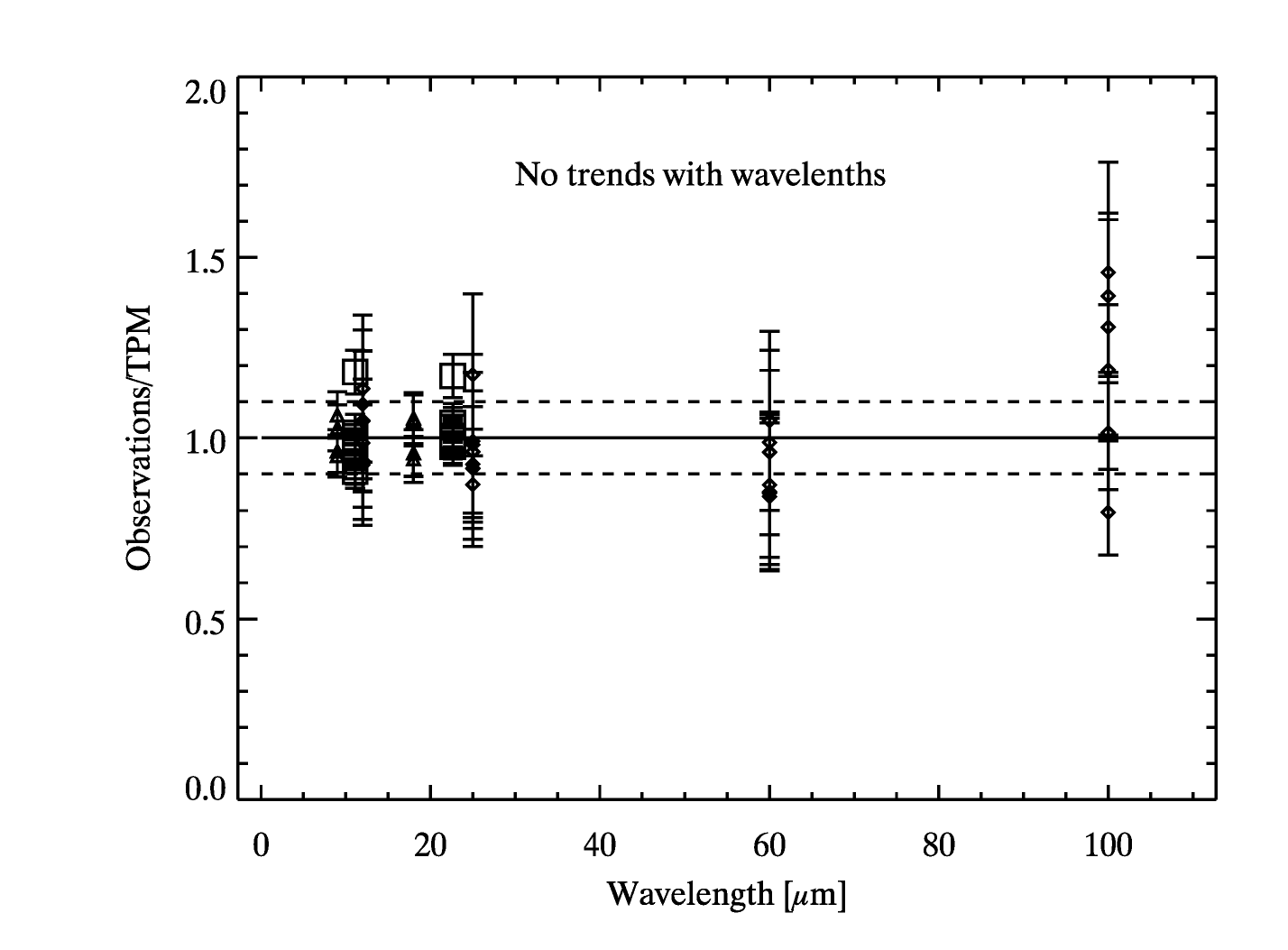}&
 \includegraphics[angle=90,width=0.4\textwidth]{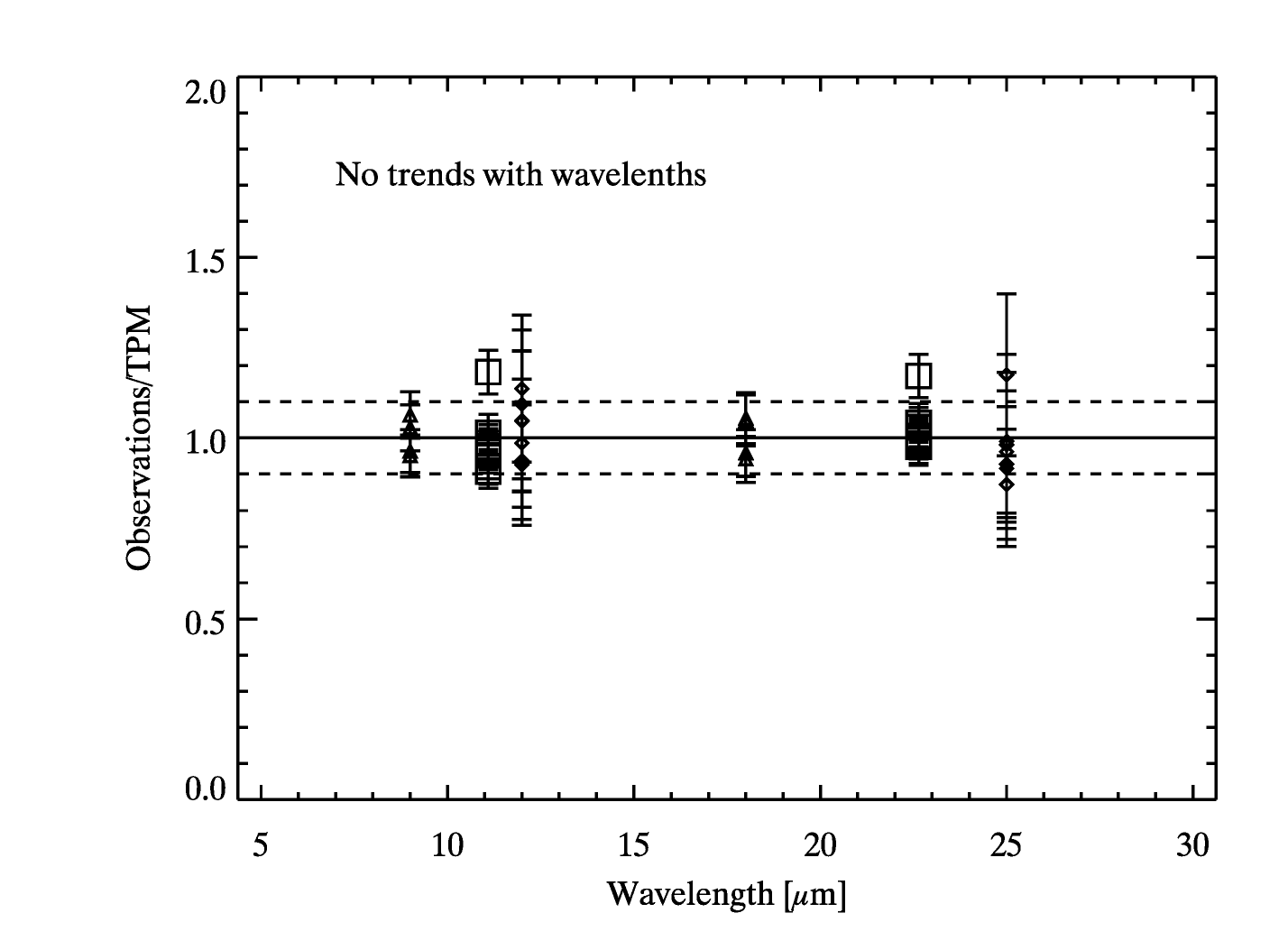}
\\
 \includegraphics[angle=90,width=0.4\textwidth]{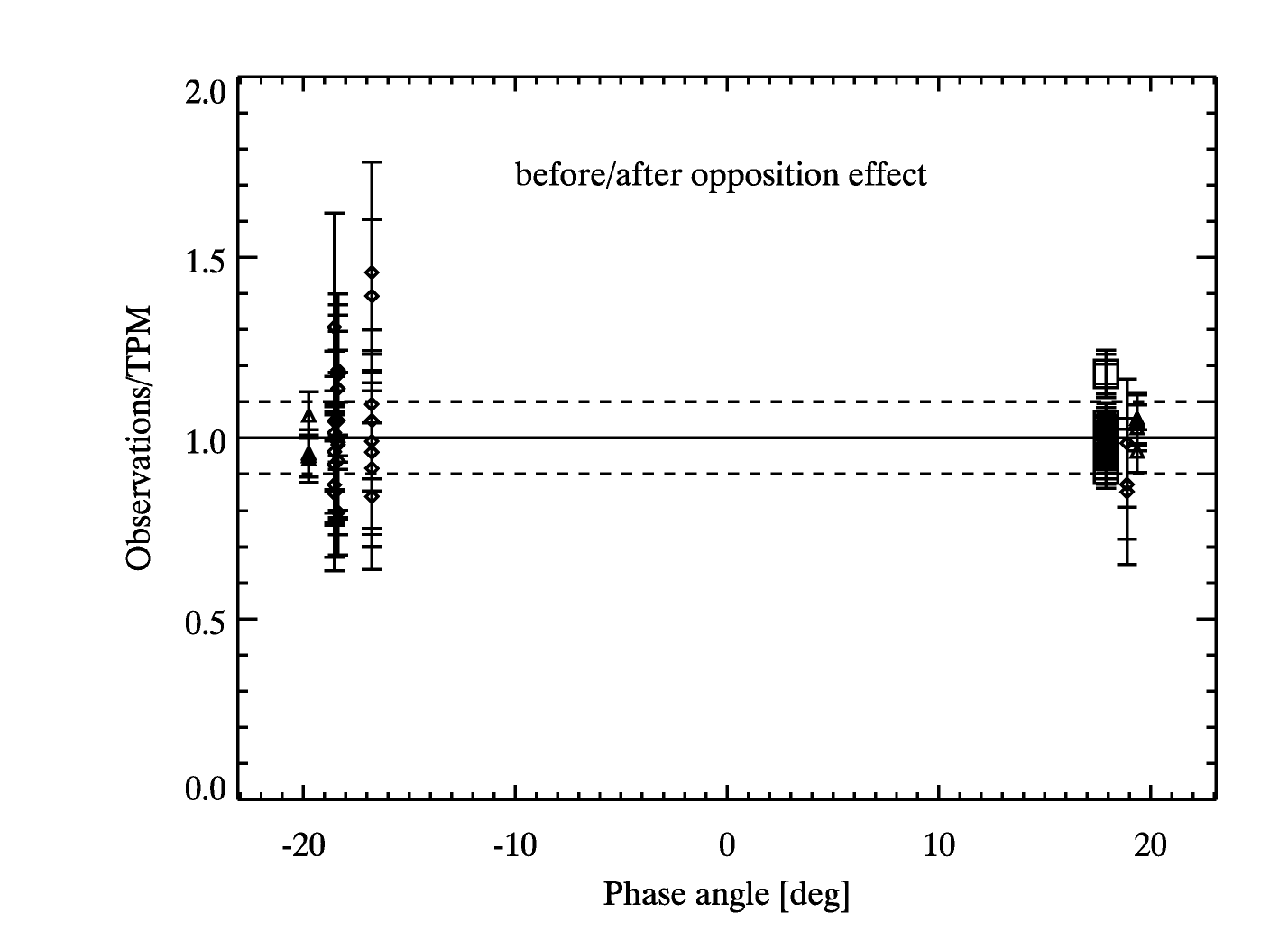}&
 \includegraphics[angle=90,width=0.4\textwidth]{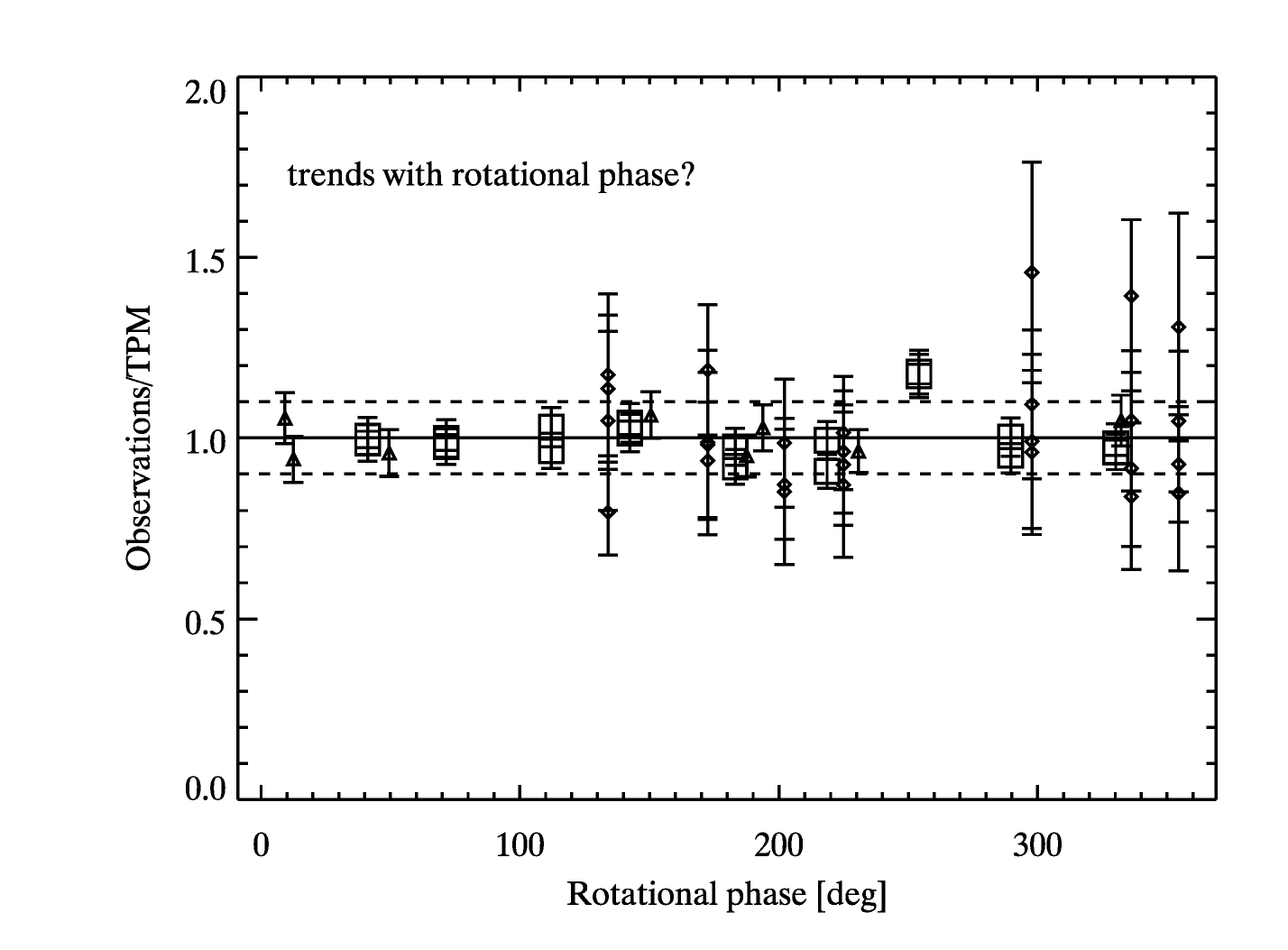}
\end{tabular}
\caption{O-C diagrams for thermophysical model of (478) Tergeste, using SAGE model 2. 
There are no trends with wavelength or pre- and post-opposition
asymmetry. The two outliers at rotational phase 250 deg might
be an indication for a small-scale shape problem, but could
also be connected to a wrong flux (single WISE W3/W4 epoch where
a bright background source might have influenced the photometry).
For best fitting thermal parameters see Table \ref{TPMresults}.}
\label{Tergeste_TPM}
\end{figure*}

\clearpage

\begin{figure*}[h]
\includegraphics[width=0.33\textwidth]{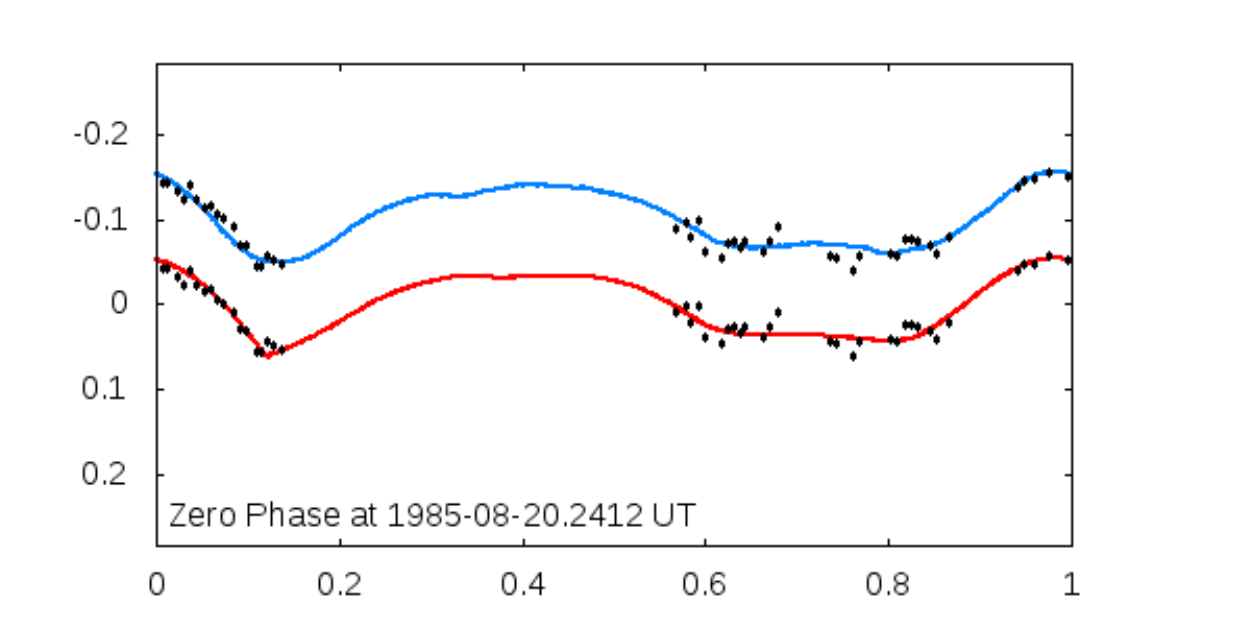}
\includegraphics[width=0.33\textwidth]{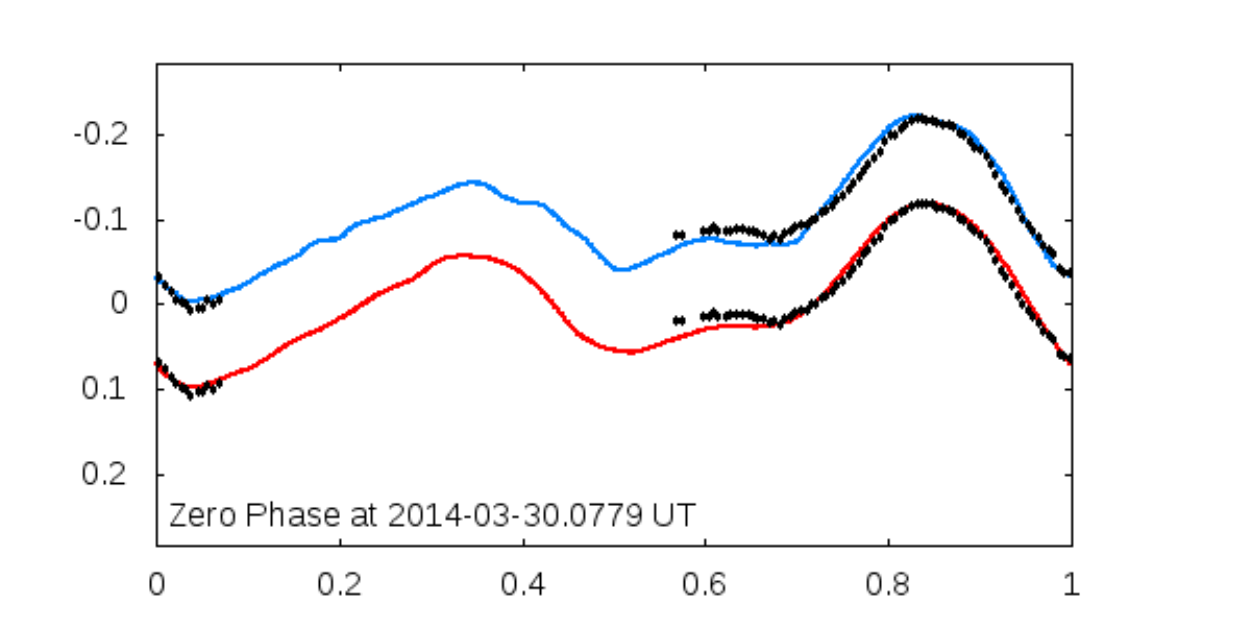}
\includegraphics[width=0.33\textwidth]{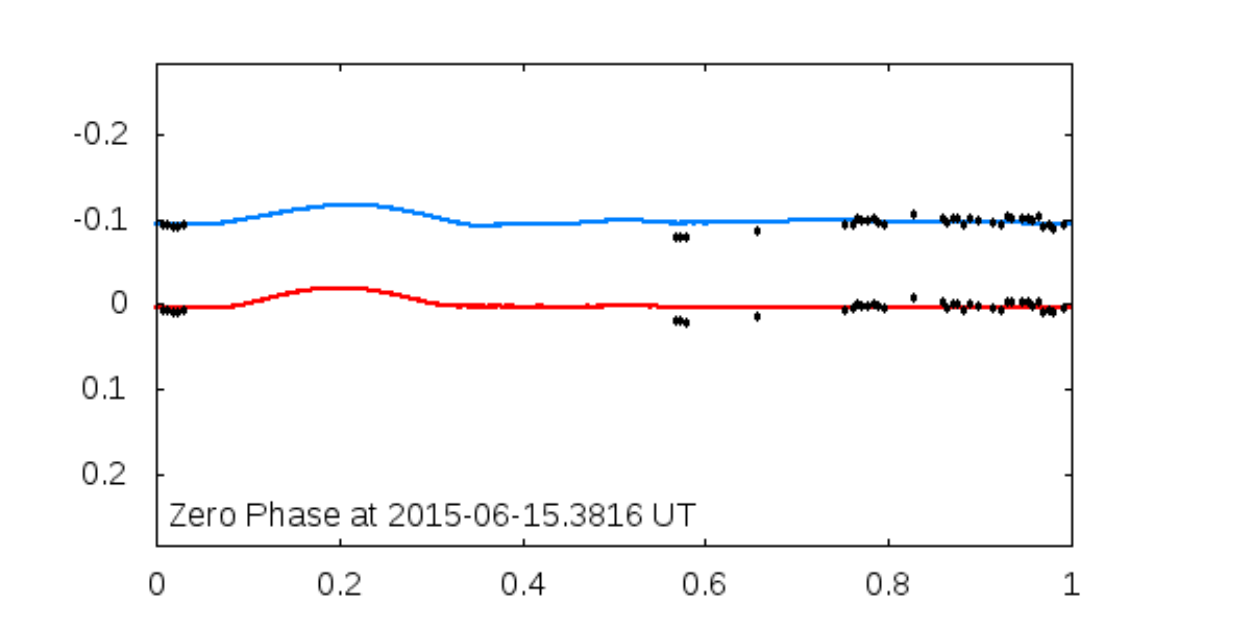}
\caption{Convex (upper curve) and non-convex (lower curve) model lightcurves of (487) Venetia fitted to the data 
from various apparitions (black points)}
\label{Venetia_fit}
\end{figure*}

\begin{figure*}[h]
\begin{tabular}{cc}
 \includegraphics[angle=90,width=0.4\textwidth]{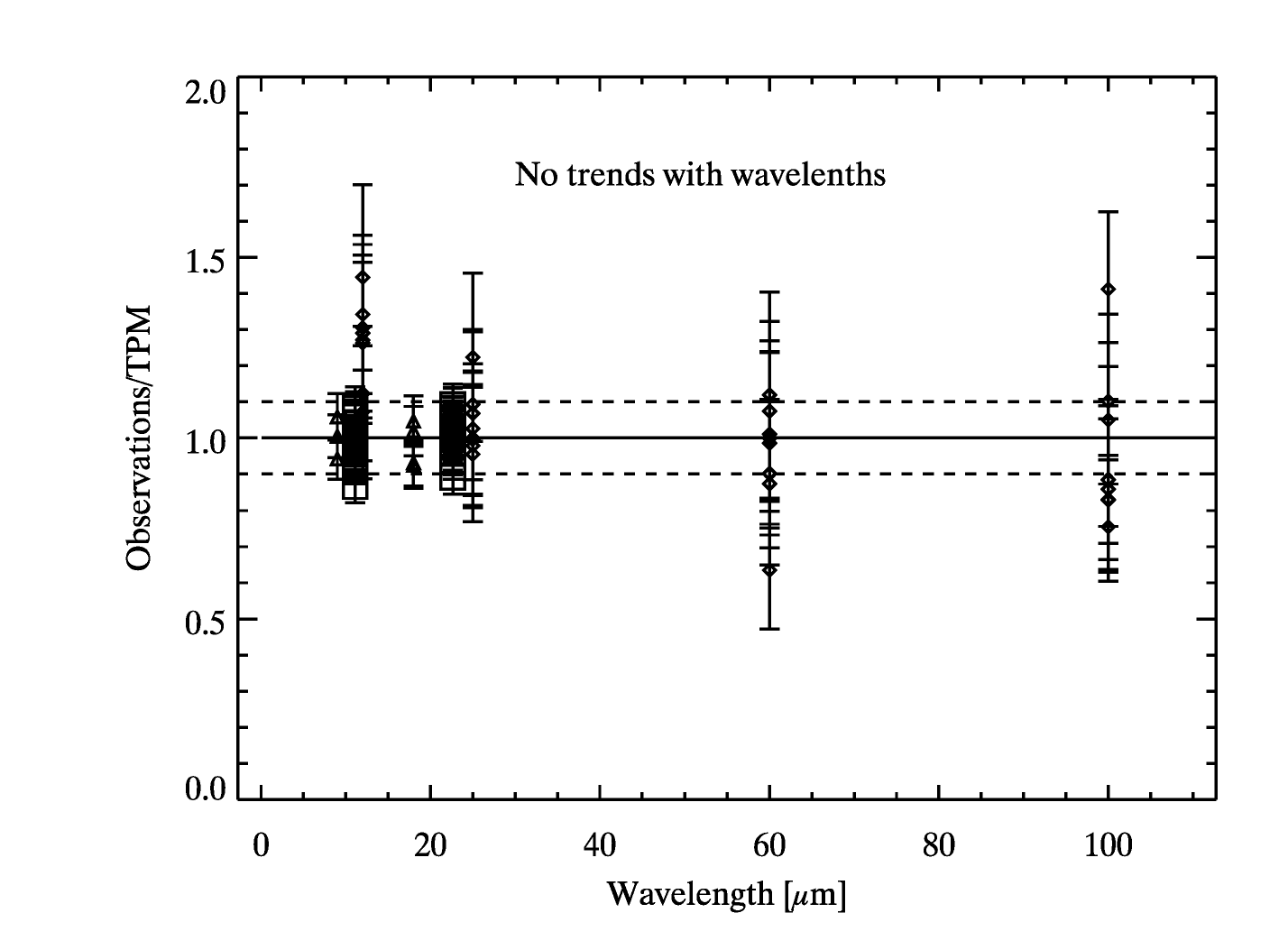}&
 \includegraphics[angle=90,width=0.4\textwidth]{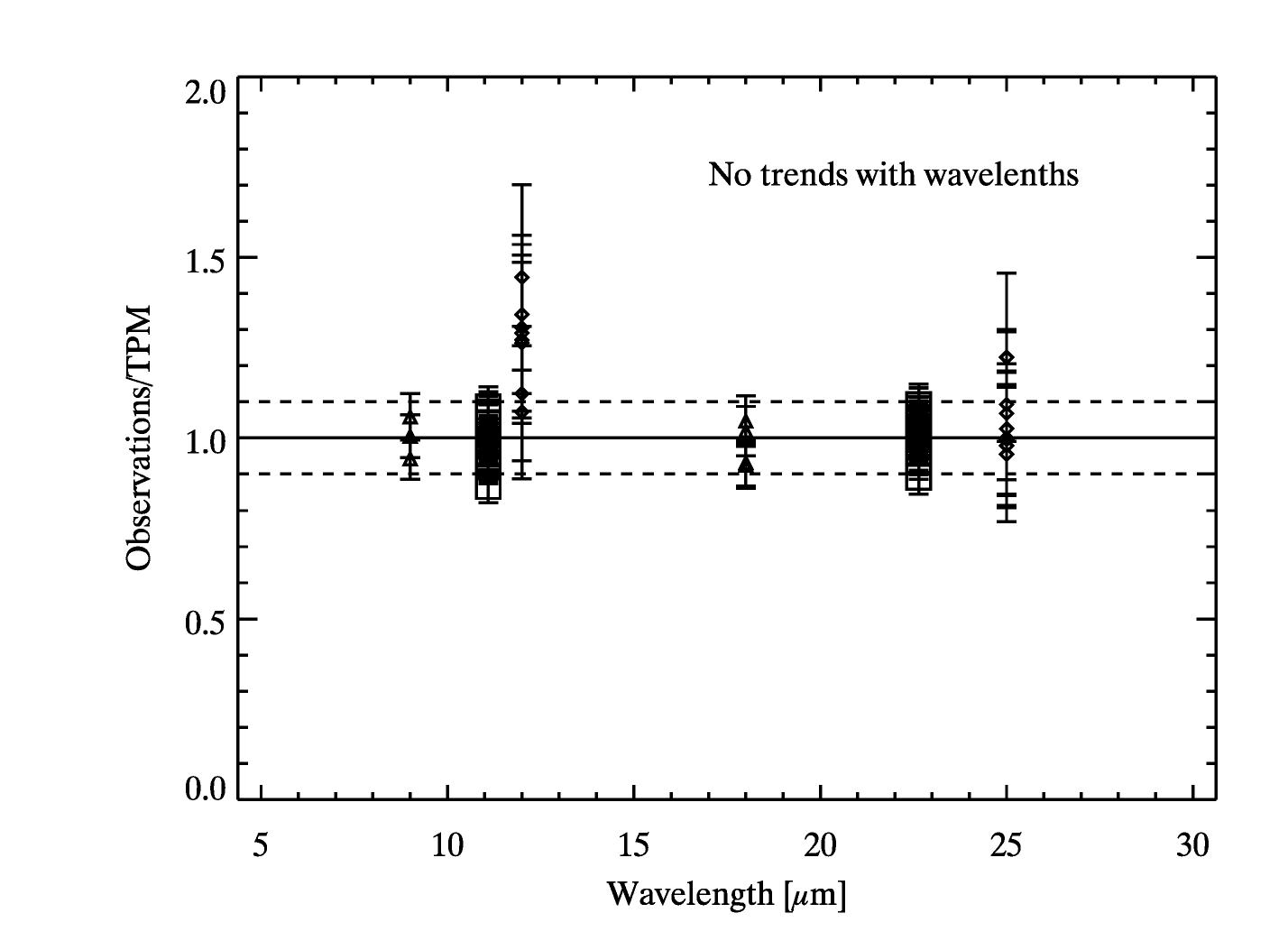}
\\
 \includegraphics[angle=90,width=0.4\textwidth]{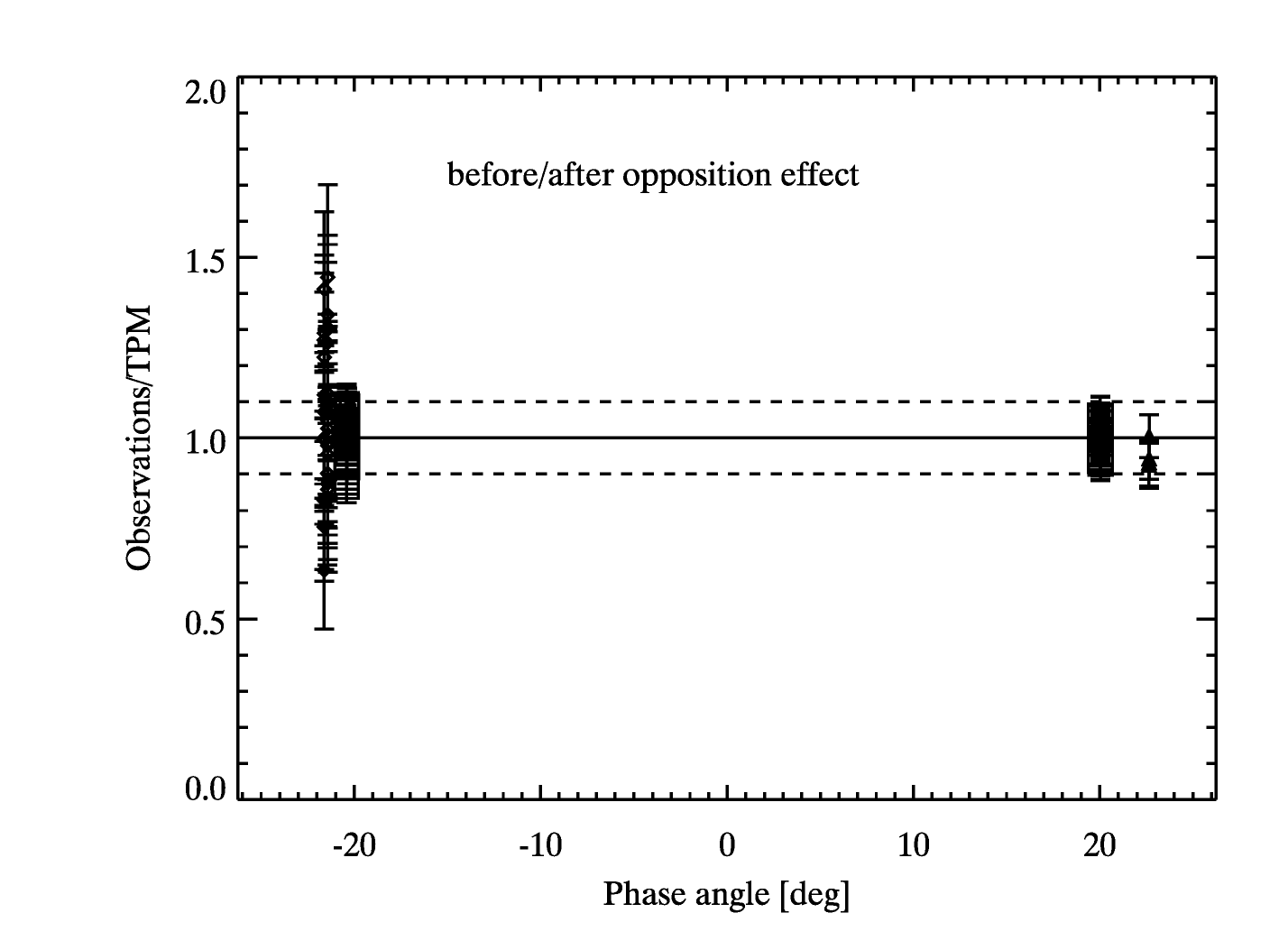}&
 \includegraphics[angle=90,width=0.4\textwidth]{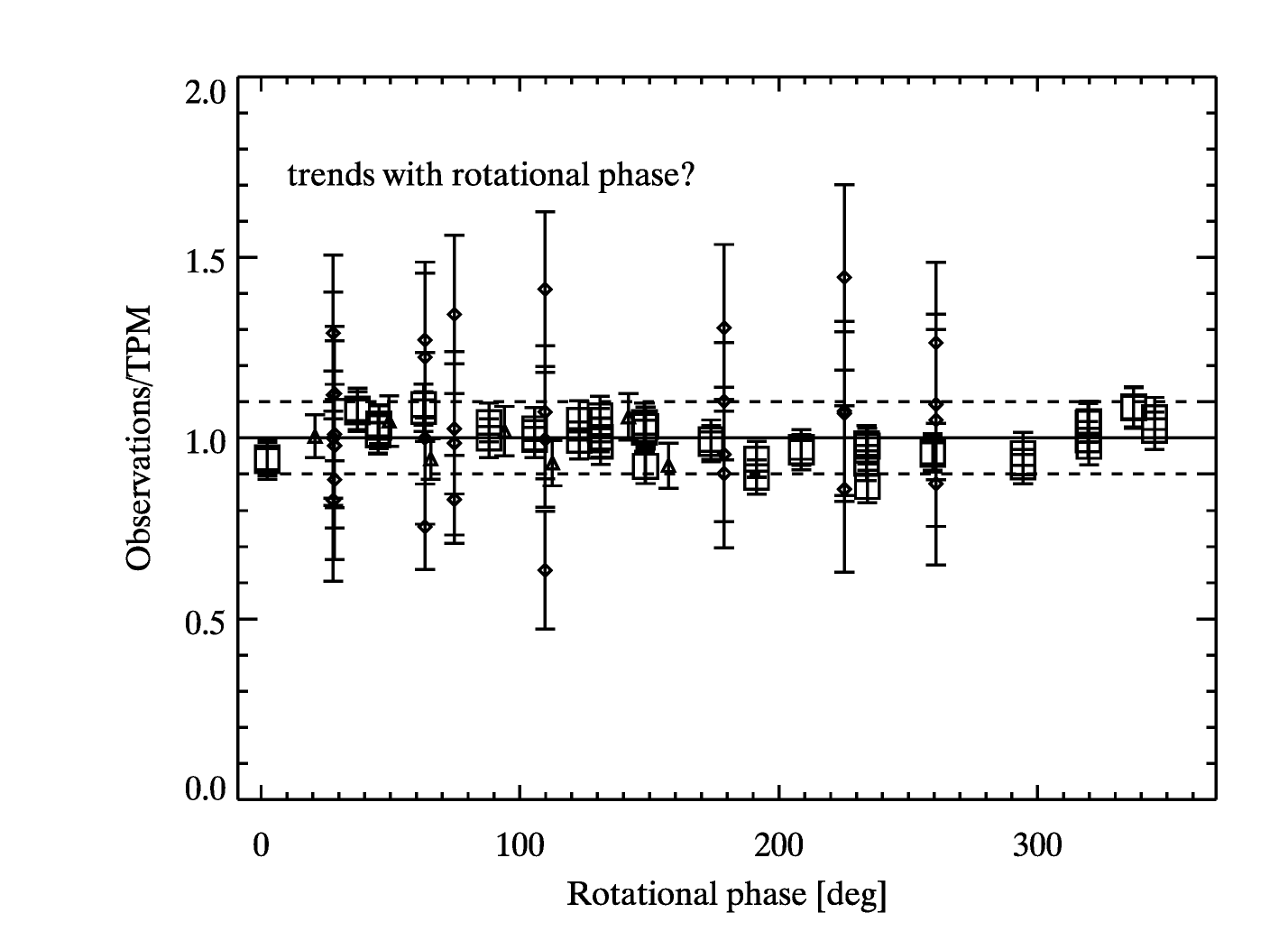}
\end{tabular}
\caption{O-C diagrams for the thermophysical model of (487) Venetia using SAGE model 2. 
There are no trends with wavelength or pre- and post-opposition asymmetry, but 
some trends with rotation can be noticed in the WISE data (box symbol).
These data cover the object's full rotation during two separate
epochs in January and July 2010, and residual trends can only be
explained by shape effects. 
For the best fitting thermal parameters see Table \ref{TPMresults}.}
\label{Venetia_TPM}
\end{figure*}

\clearpage

\section*{Appendix A}
\begin{figure}[h]
\includegraphics[width=0.4\textwidth]{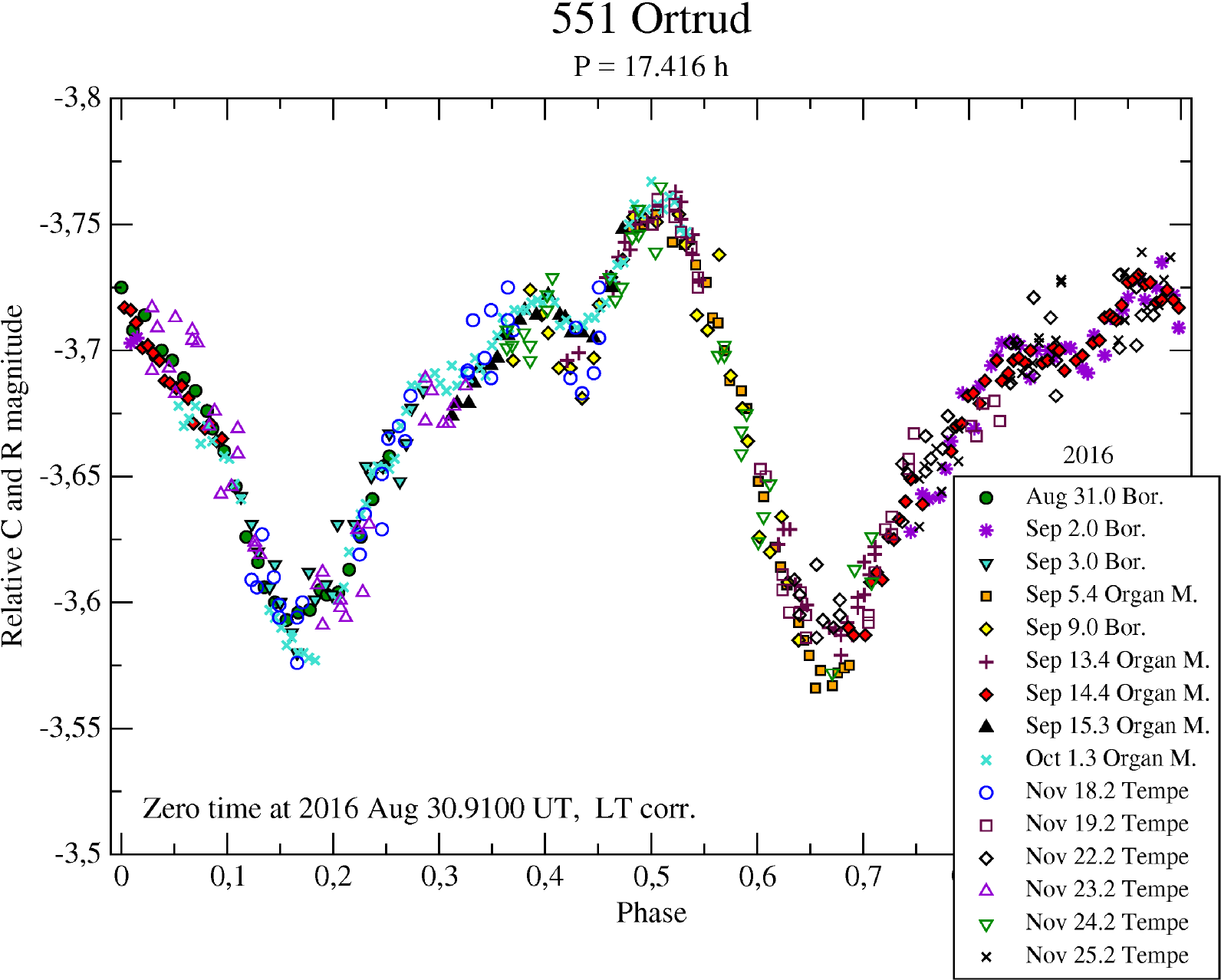}
\caption{Composite lightcurve of (551) Ortrud in the year 2016}
\label{Ortrud2016}
\end{figure}

\begin{figure}[h]
\includegraphics[width=0.4\textwidth]{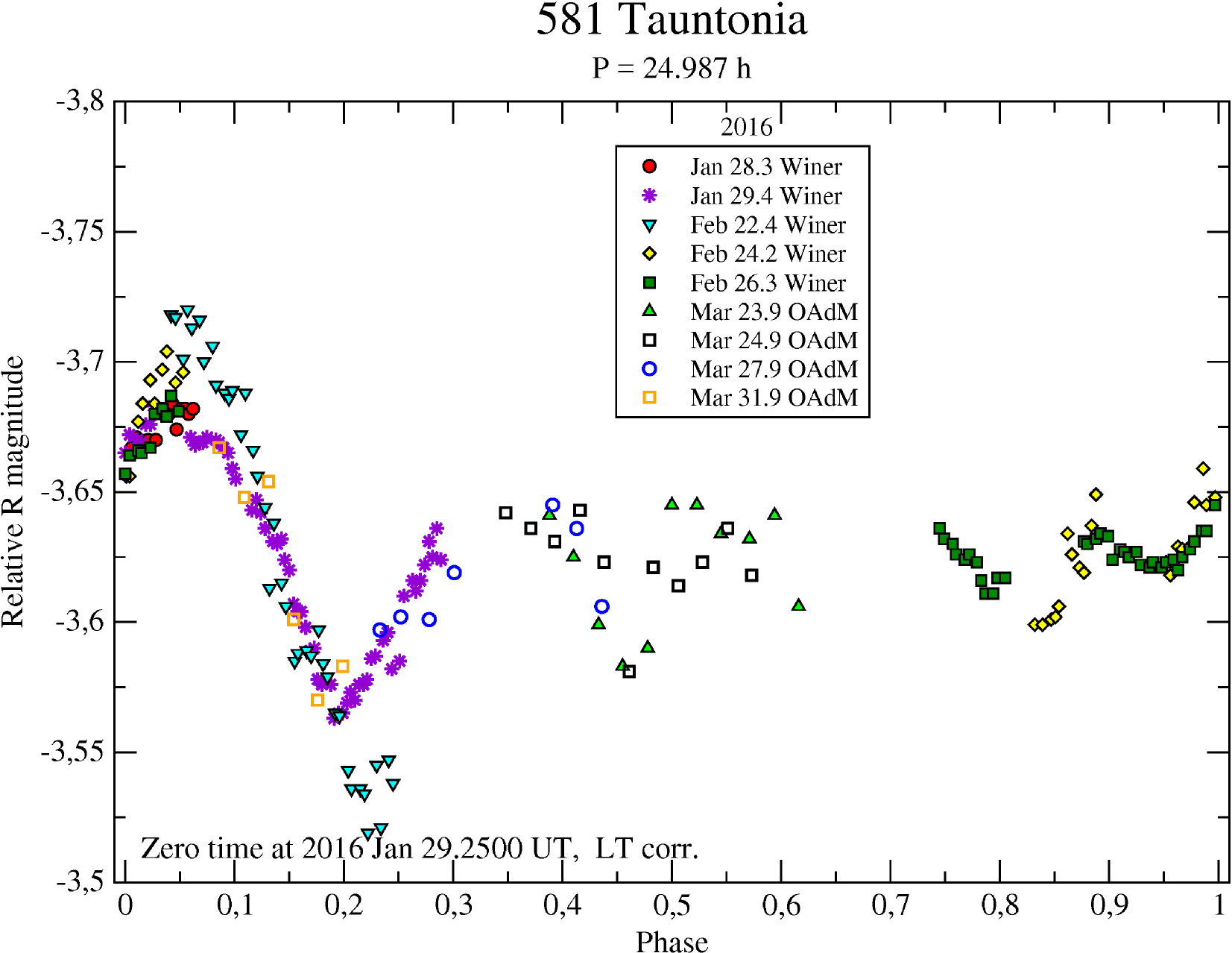}
\caption{Composite lightcurve of (581) Tauntonia in 2016}
\label{Tauntonia2016}
\end{figure}

\begin{figure}[h]
\includegraphics[width=0.45\textwidth]{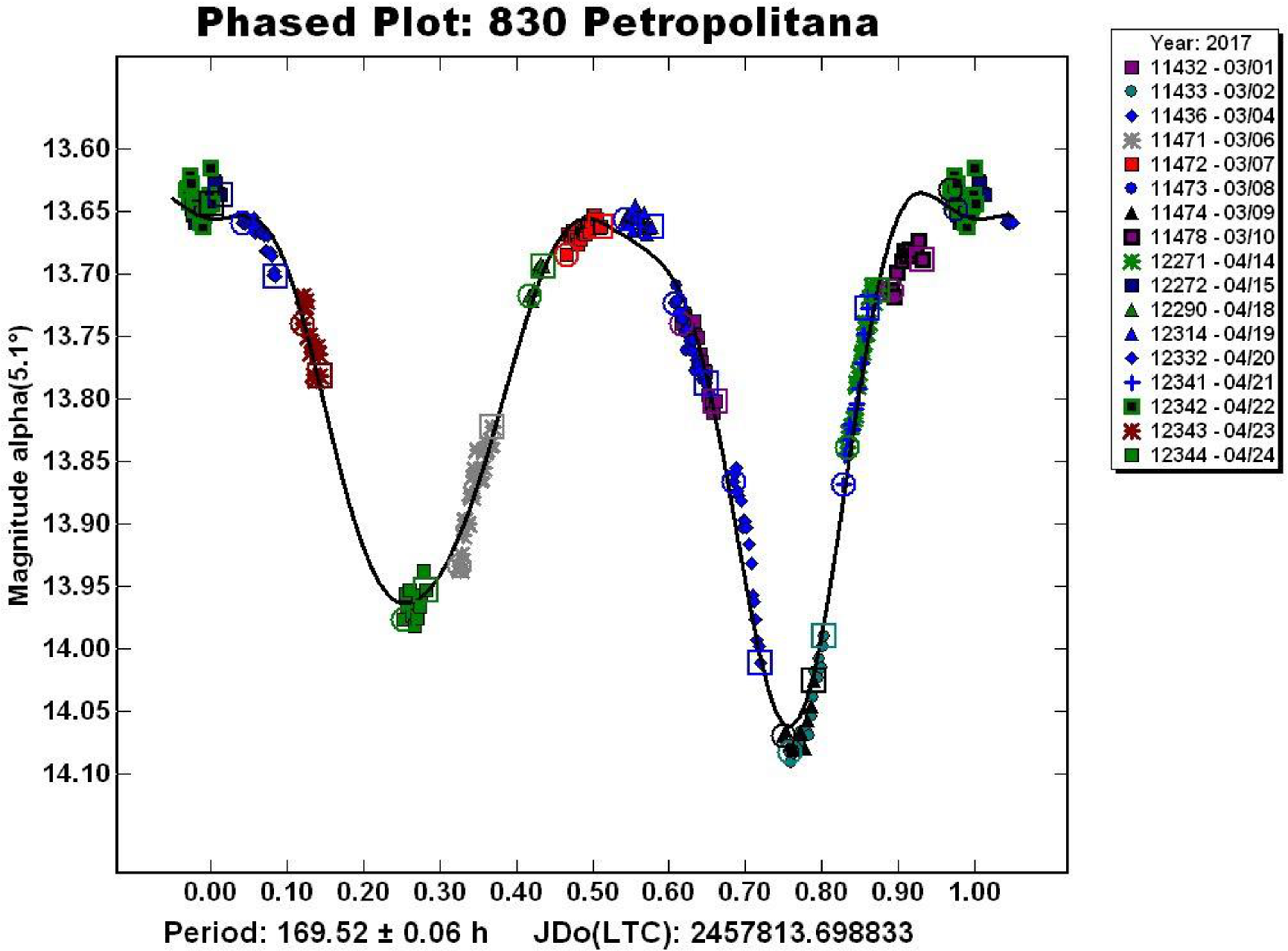}
\caption{Calibrated composite lightcurve of (830) Petropolitana in 2017}
\label{Petropolitana2017}
\end{figure}

\begin{figure}[h]
\includegraphics[width=0.45\textwidth]{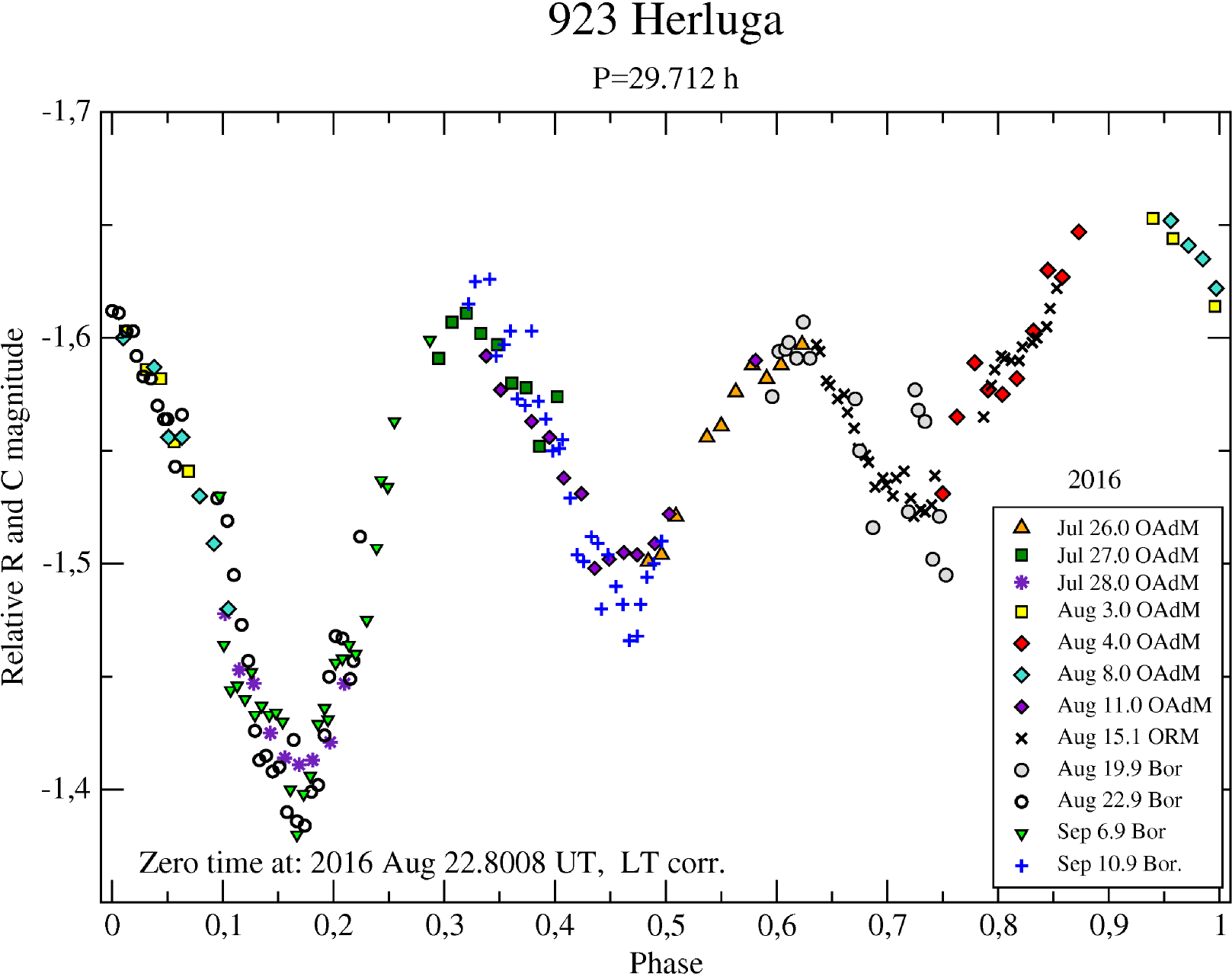}
\caption{Composite lightcurve of (923) Herluga in 2016}
\label{Herluga2016}
\end{figure}

\begin{figure}[h]
\includegraphics[width=0.45\textwidth]{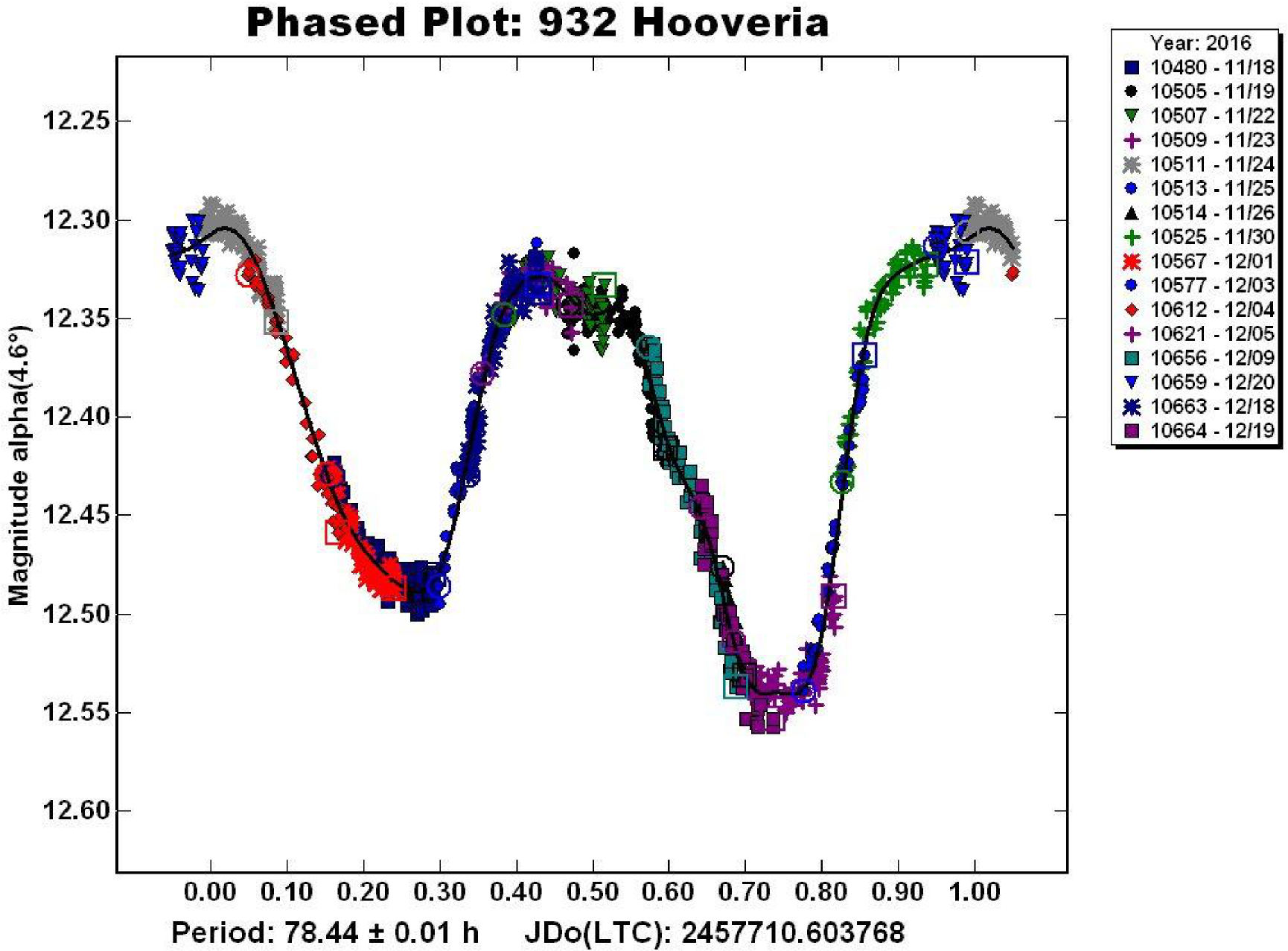}
\caption{Calibrated composite lightcurve of (932) Hooveria in 2016}
\label{Hooveria2016}
\end{figure}

\begin{figure}[h]
\includegraphics[width=0.45\textwidth]{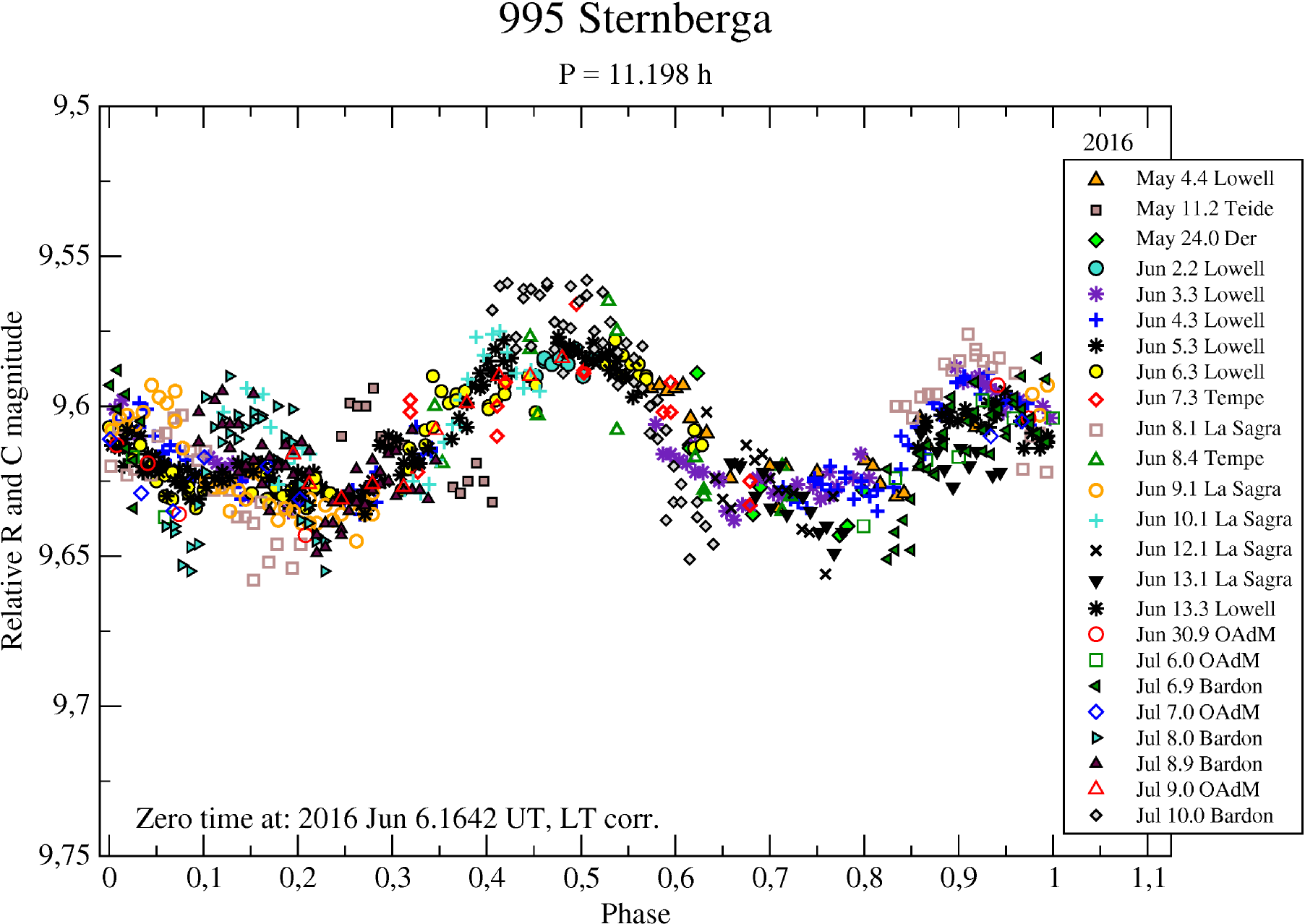}
\caption{Composite lightcurve of (995) Sternberga in 2016}
\label{Sternberga2016}
\end{figure}

\begin{figure*}[h]
\includegraphics[width=0.75\textwidth]{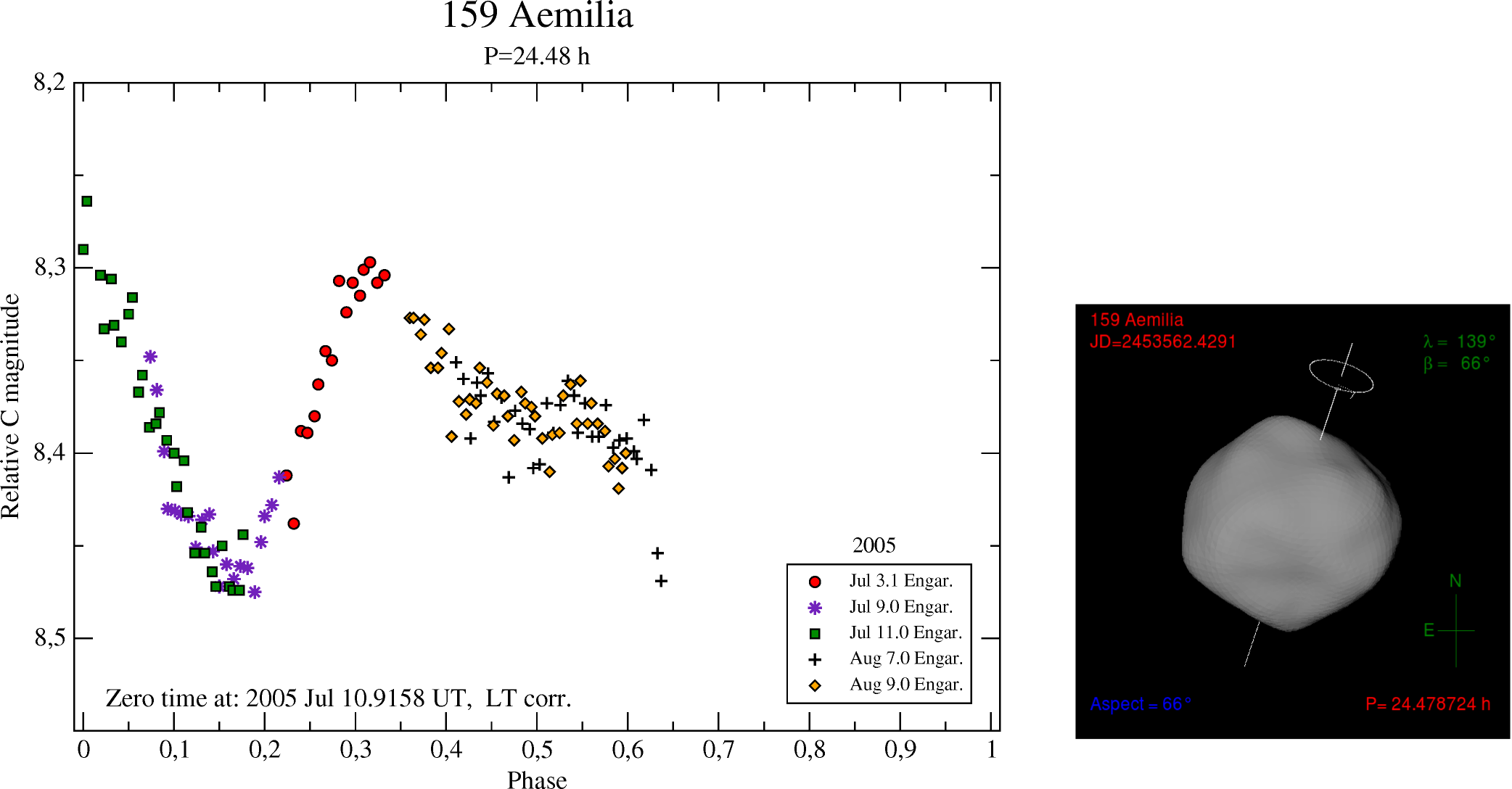} 
\caption{Composite lightcurve of (159) Aemilia in the year 2005 with the
orientation of SAGE model 1 for the zero phase}
\label{Aemilia2005}
\end{figure*}

\begin{figure*}[h]
\includegraphics[width=0.75\textwidth]{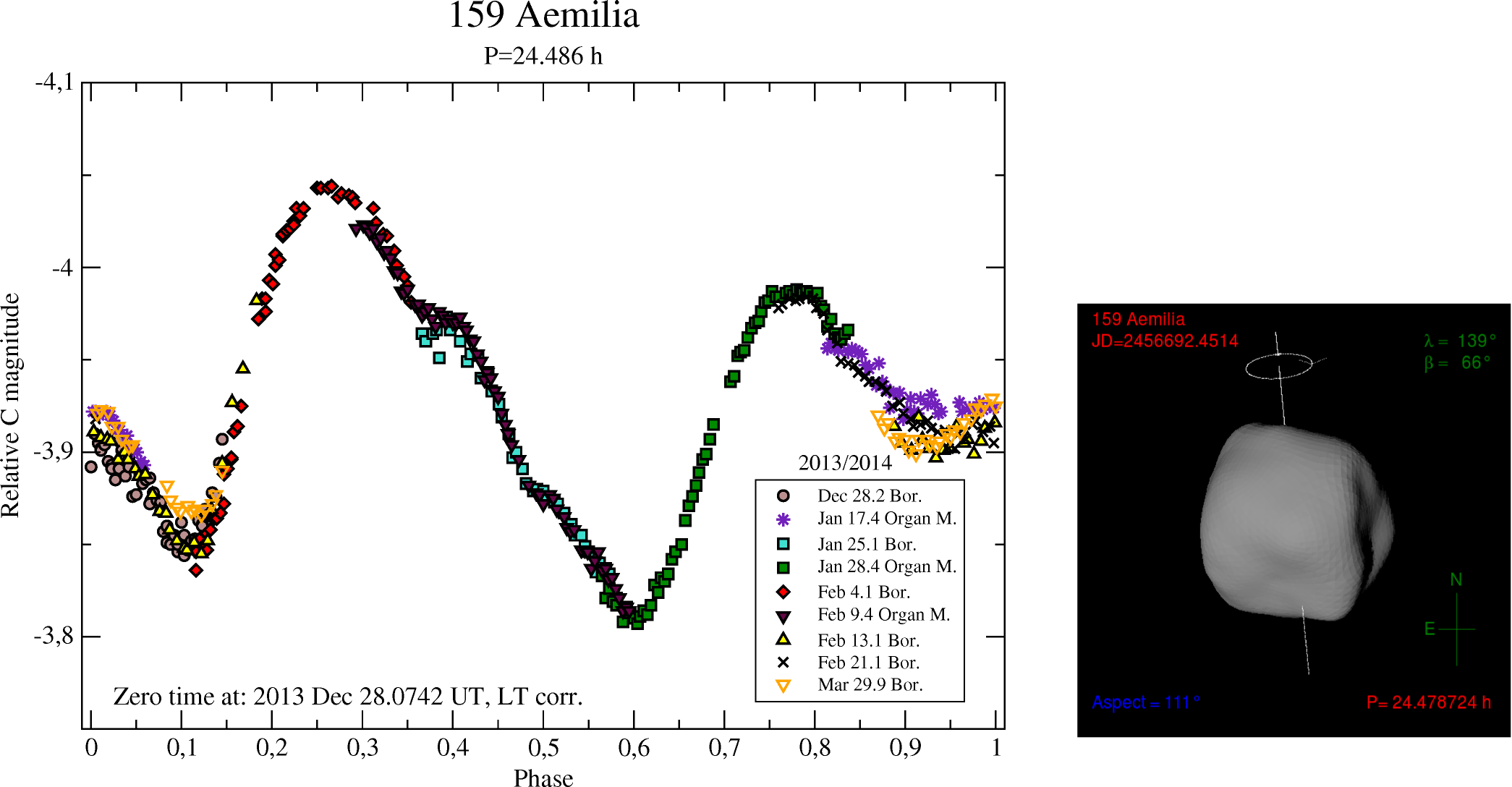} 
\caption{Composite lightcurve of (159) Aemilia in the years 2013-2014 with the
orientation of SAGE model 1 for the zero phase}
\label{Aemilia2014}
\end{figure*}

\begin{figure*}[h]
\includegraphics[width=0.75\textwidth]{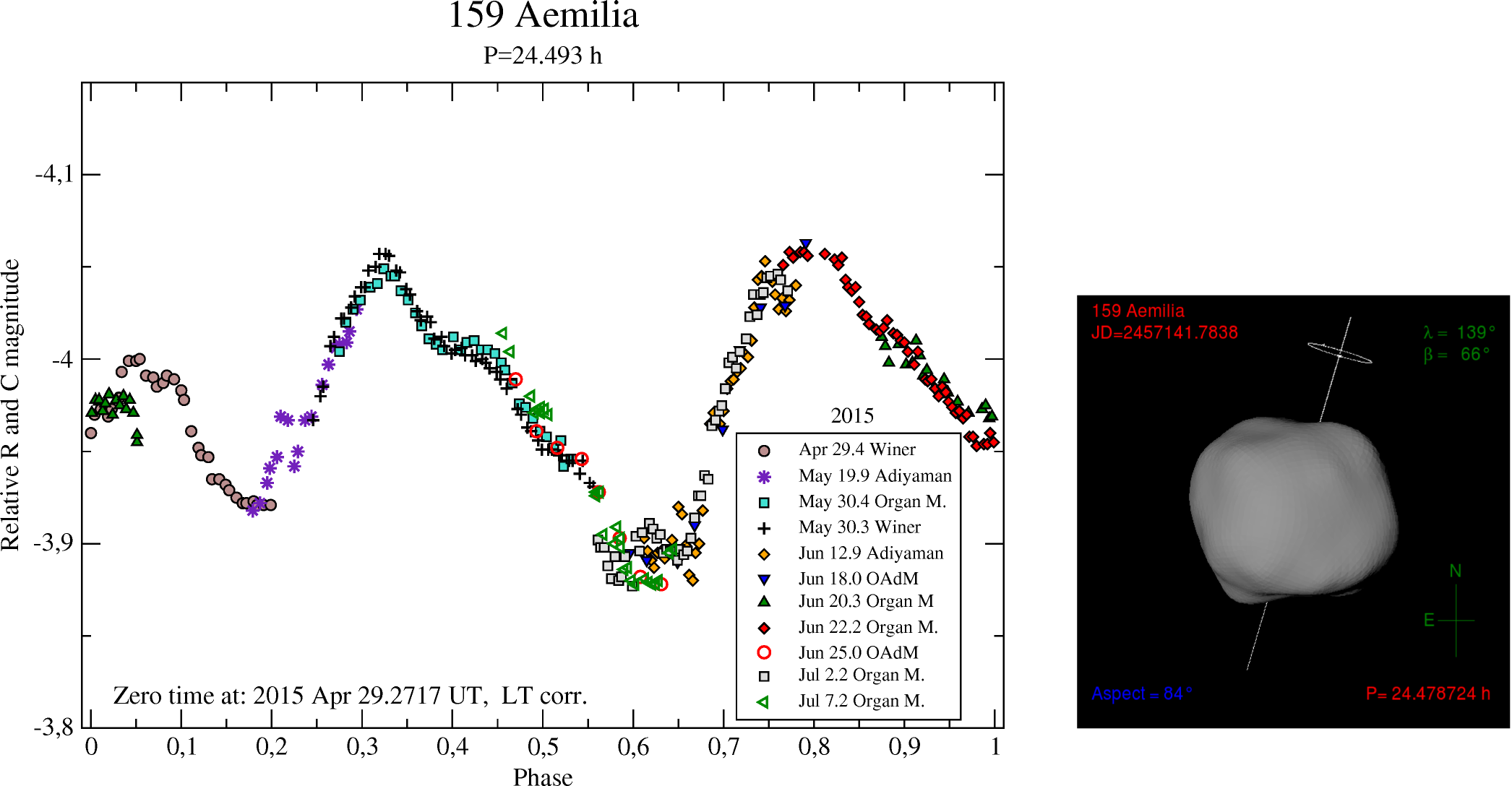}
\caption{Composite lightcurve of (159) Aemilia in the year 2015 with the
orientation of SAGE model 1 for the zero phase}
\label{Aemilia2015}
\end{figure*}

\begin{figure*}[h]
\includegraphics[width=0.75\textwidth]{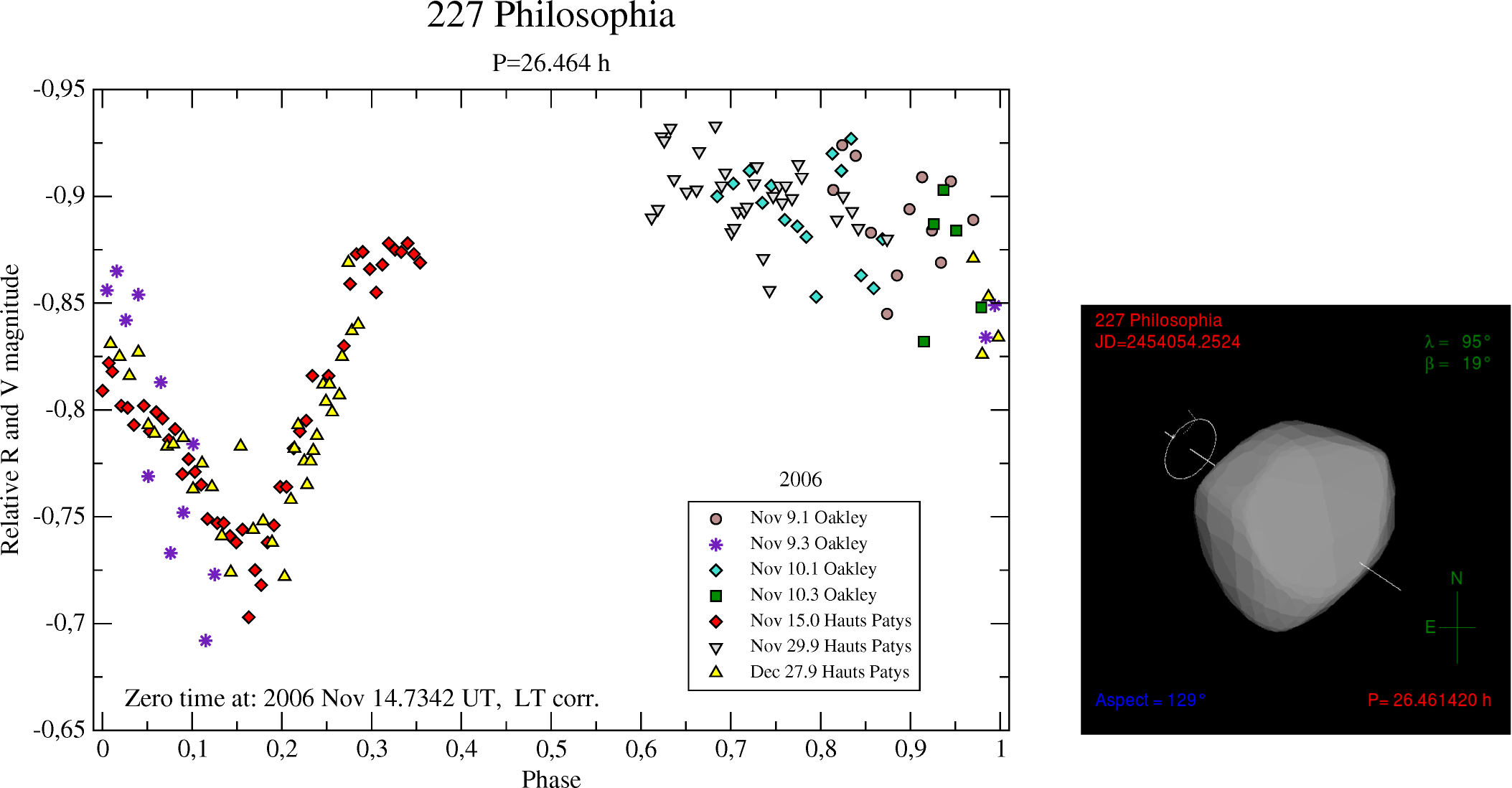} 
\caption{Composite lightcurve of (227) Philosophia in the year 2006 with the
orientation of convex model 1 for the zero phase}
\label{Philosophia2006}
\end{figure*}

\begin{figure*}[h]
\includegraphics[width=0.75\textwidth]{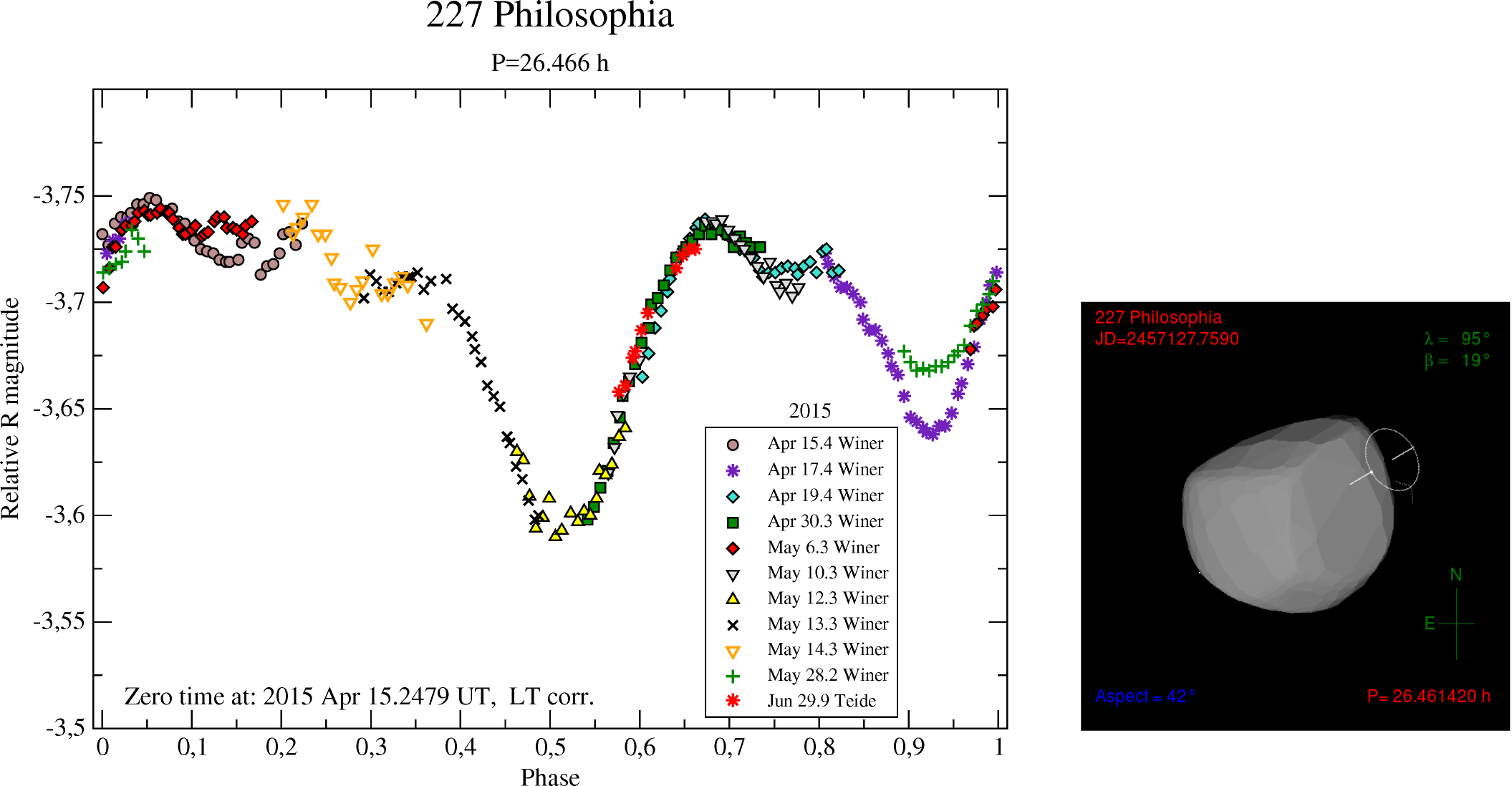}
\caption{Composite lightcurve of (227) Philosophia in the year 2015 with the
orientation of convex model 1 for the zero phase}
\label{Philosophia2015}
\end{figure*}

\begin{figure*}[h]
\includegraphics[width=0.75\textwidth]{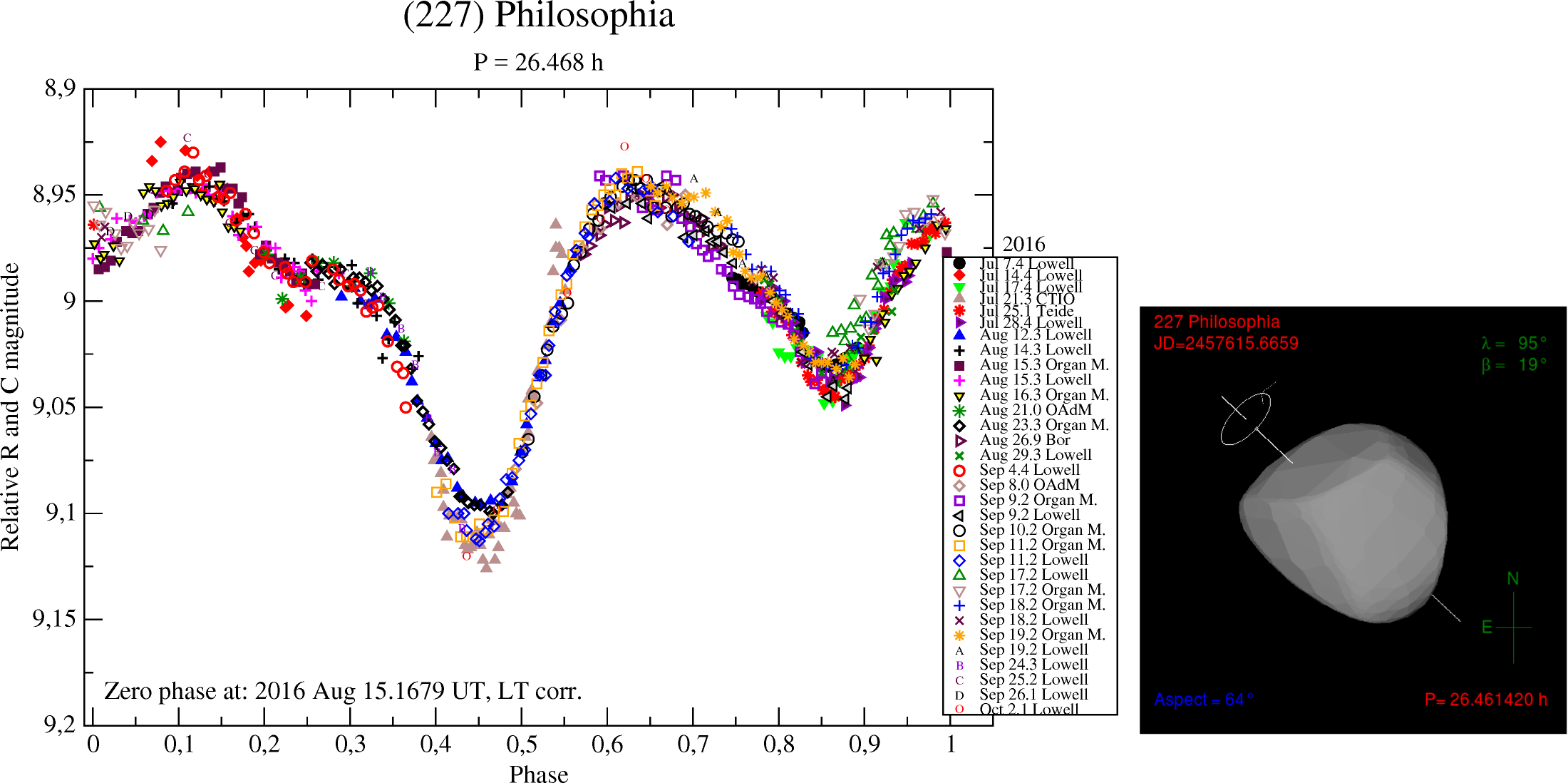} 
\caption{Composite lightcurve of (227) Philosophia in the year 2016 with the
orientation of convex model 1 for the zero phase}
\label{Philosophia2016}
\end{figure*}

\begin{figure*}[h]
\includegraphics[width=0.75\textwidth]{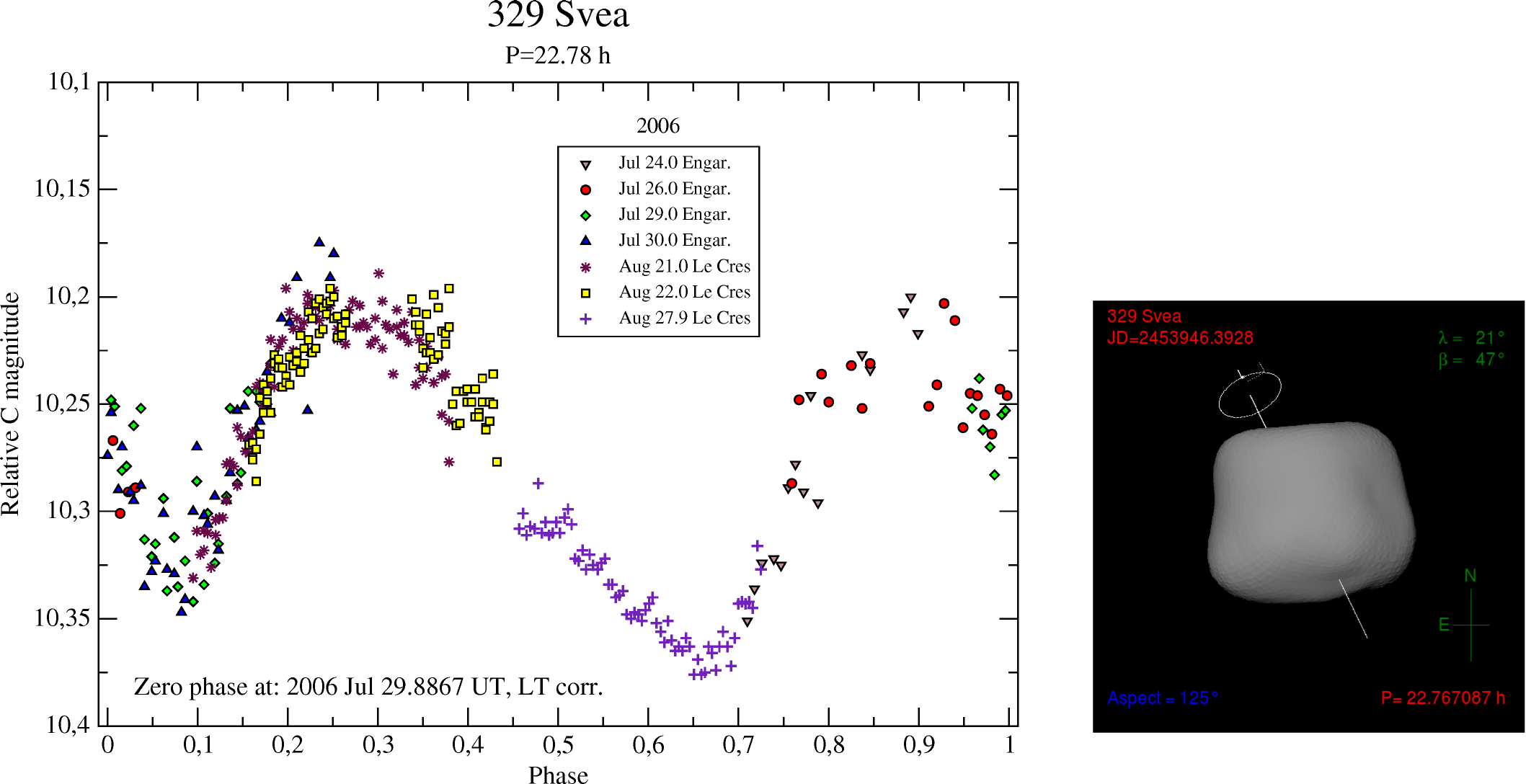} 
\caption{Composite lightcurve of (329) Svea in the year 2006 with the
orientation of SAGE model 1 for the zero phase}
\label{Svea2006}
\end{figure*}

\begin{figure*}[h]
\includegraphics[width=0.75\textwidth]{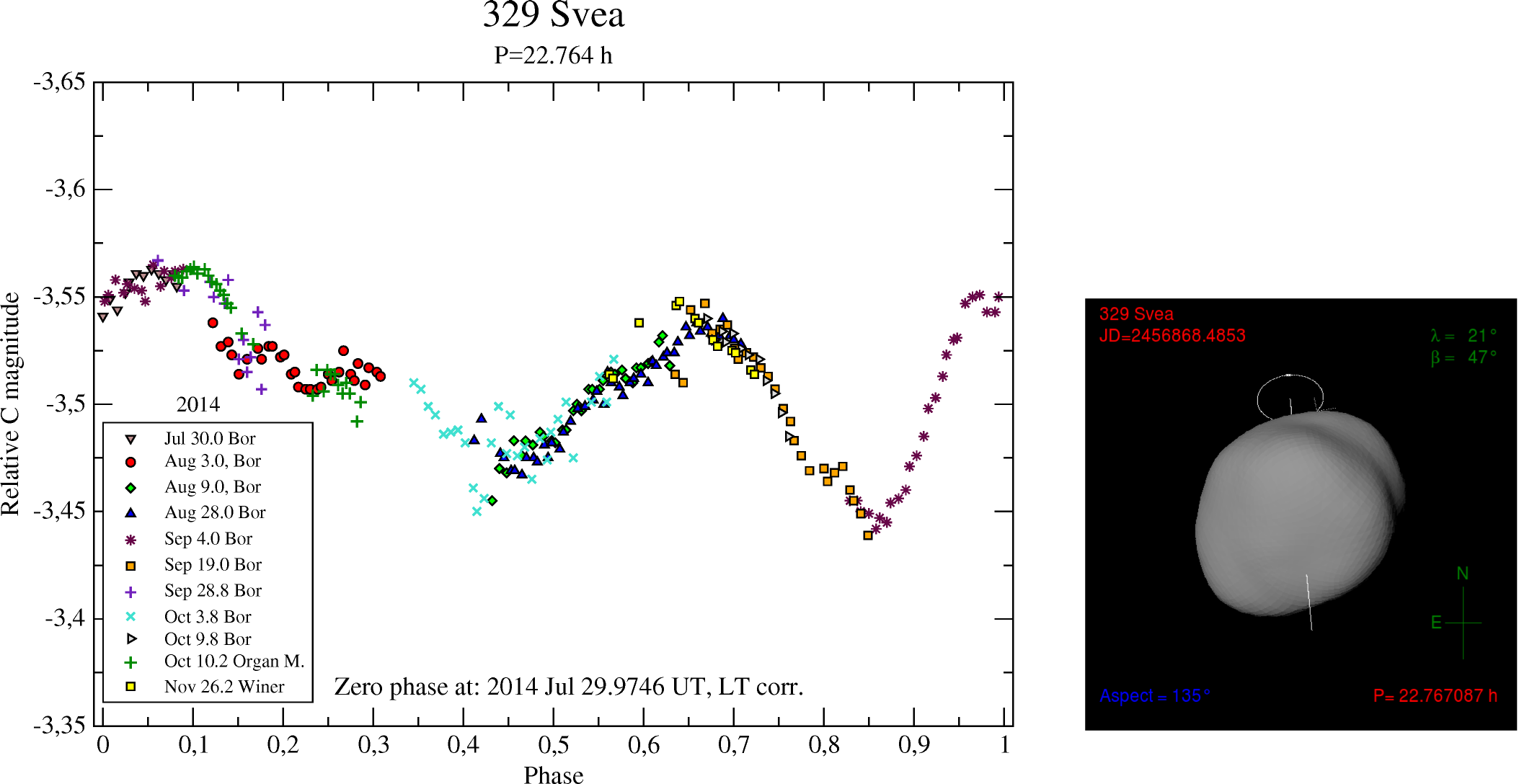} 
\caption{Composite lightcurve of (329) Svea in the year 2014 with the
orientation of SAGE model 1 for the zero phase}
\label{Svea2014}
\end{figure*}

\begin{figure*}[h]
\includegraphics[width=0.75\textwidth]{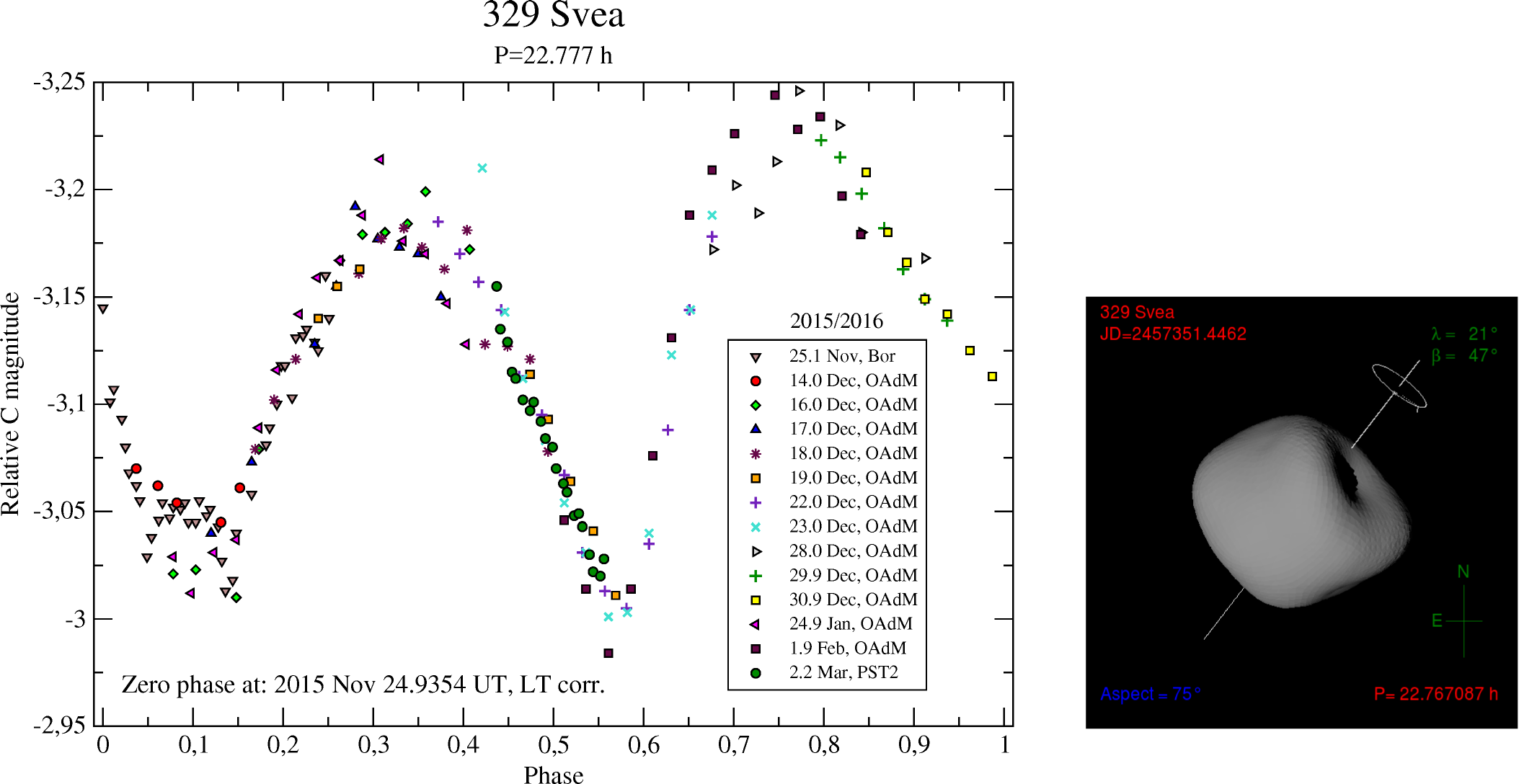} 
\caption{Composite lightcurve of (329) Svea in the years 2015-2016 with the
orientation of SAGE model 1 for the zero phase}
\label{Svea2015}
\end{figure*}

\begin{figure*}[h]
\includegraphics[width=0.75\textwidth]{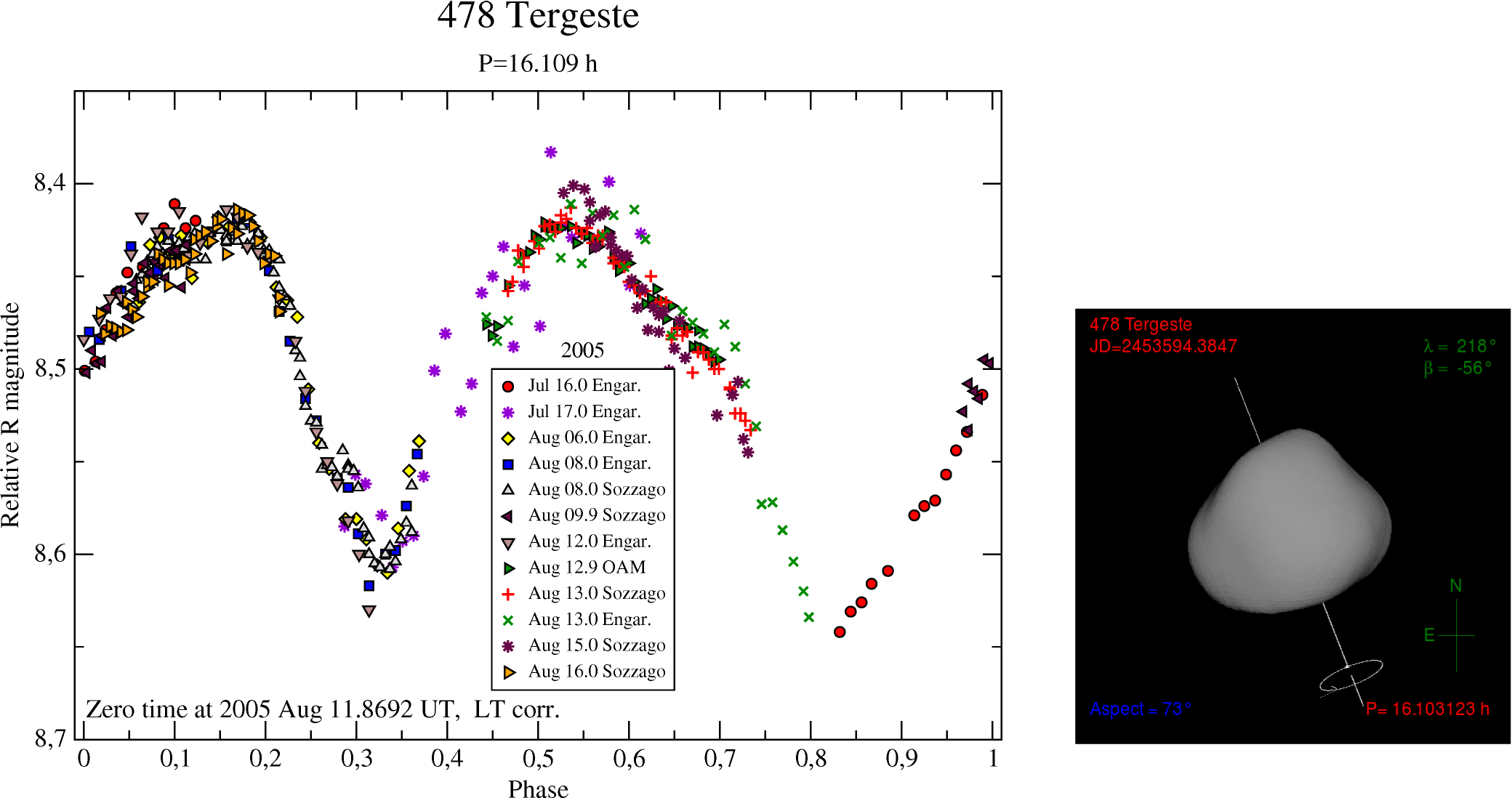} 
\caption{Composite lightcurve of (478) Tergeste in the year 2005 with the
orientation of SAGE model 2 for the zero phase}
\label{Tergeste2005}
\end{figure*}

\begin{figure*}[h]
\includegraphics[width=0.75\textwidth]{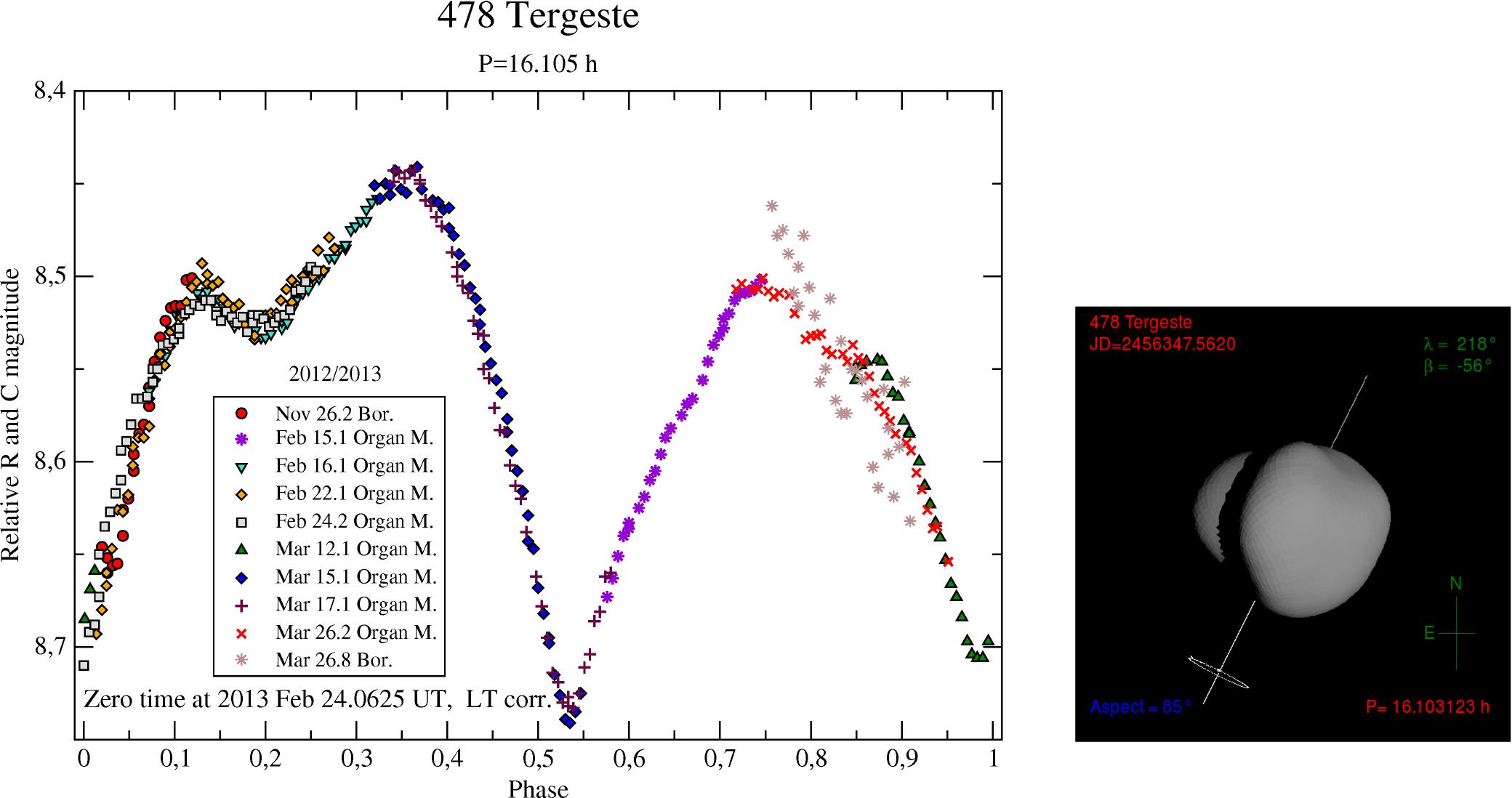} 
\caption{Composite lightcurve of (478) Tergeste in the years 2012-2013 with the
orientation of SAGE model 2 for the zero phase}
\label{Tergeste2013}
\end{figure*}

\begin{figure*}[h]
\includegraphics[width=0.75\textwidth]{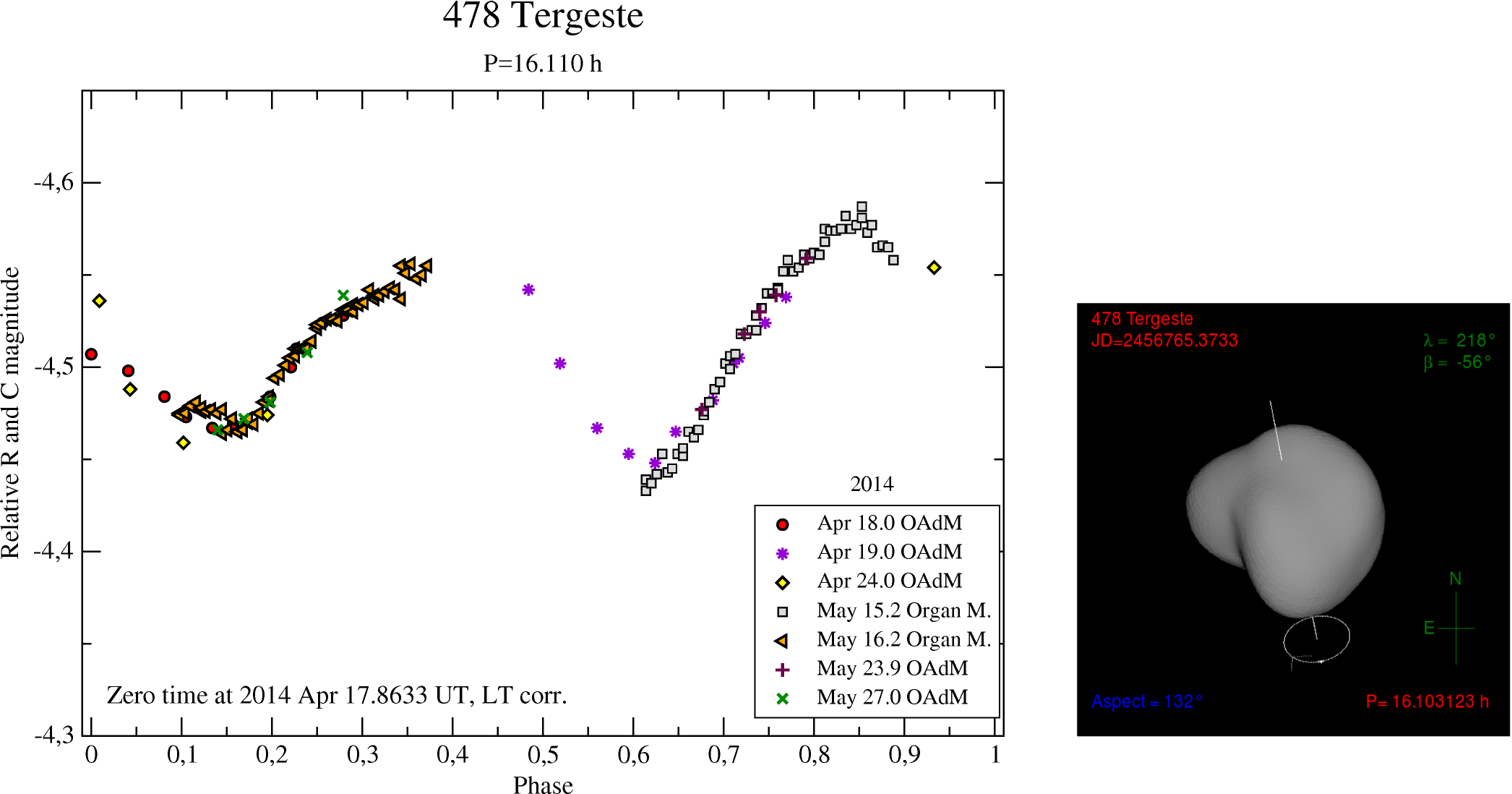} 
\caption{Composite lightcurve of (478) Tergeste in the year 2014 with the
orientation of SAGE model 2 for the zero phase}
\label{Tergeste2014}
\end{figure*}

\begin{figure*}[h]
\includegraphics[width=0.75\textwidth]{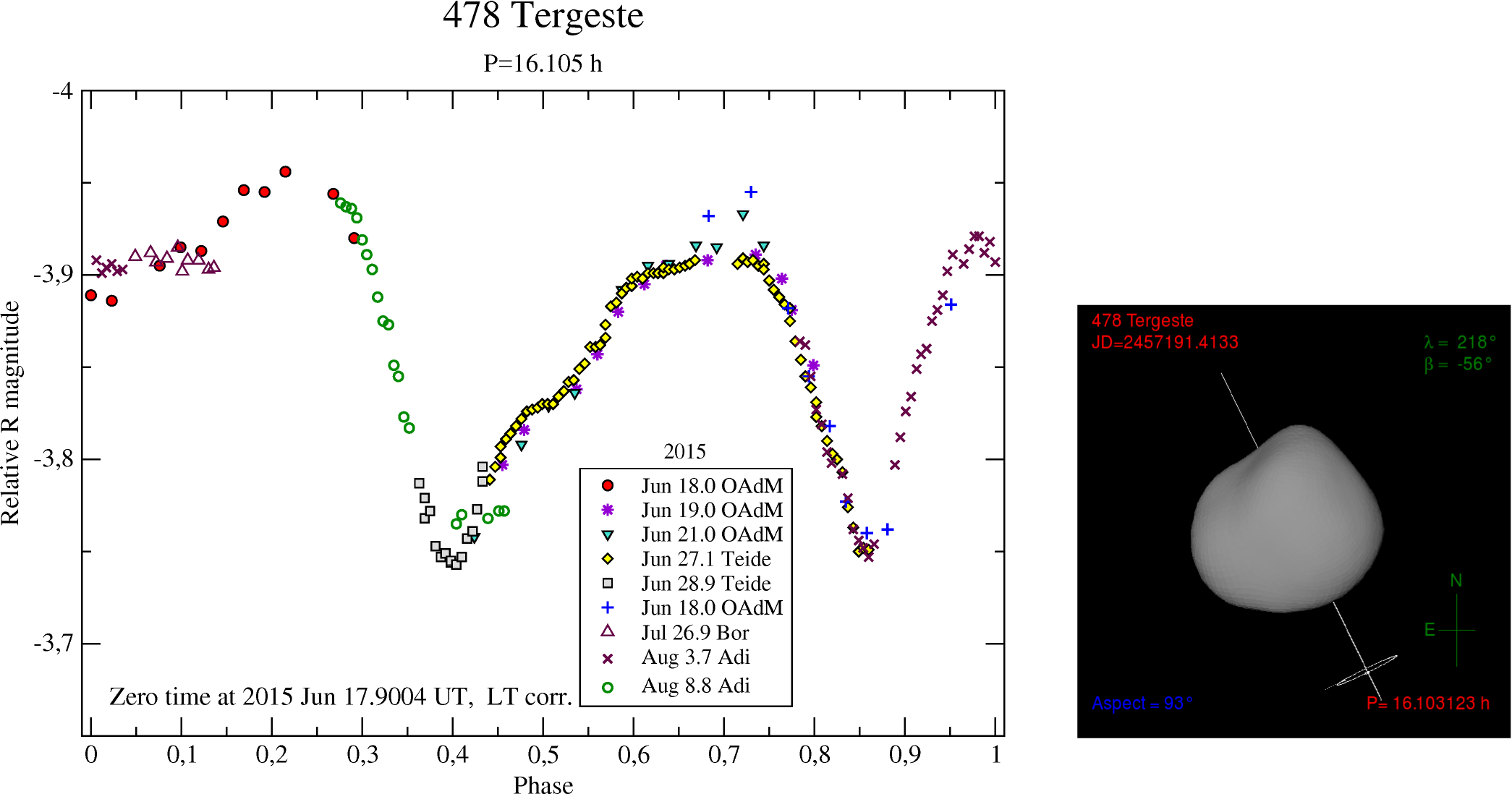} 
\caption{Composite lightcurve of (478) Tergeste in the year 2015 with the
orientation of SAGE model 2 for the zero phase}
\label{Tergeste2015}
\end{figure*}

\begin{figure*}[h]
\includegraphics[width=0.75\textwidth]{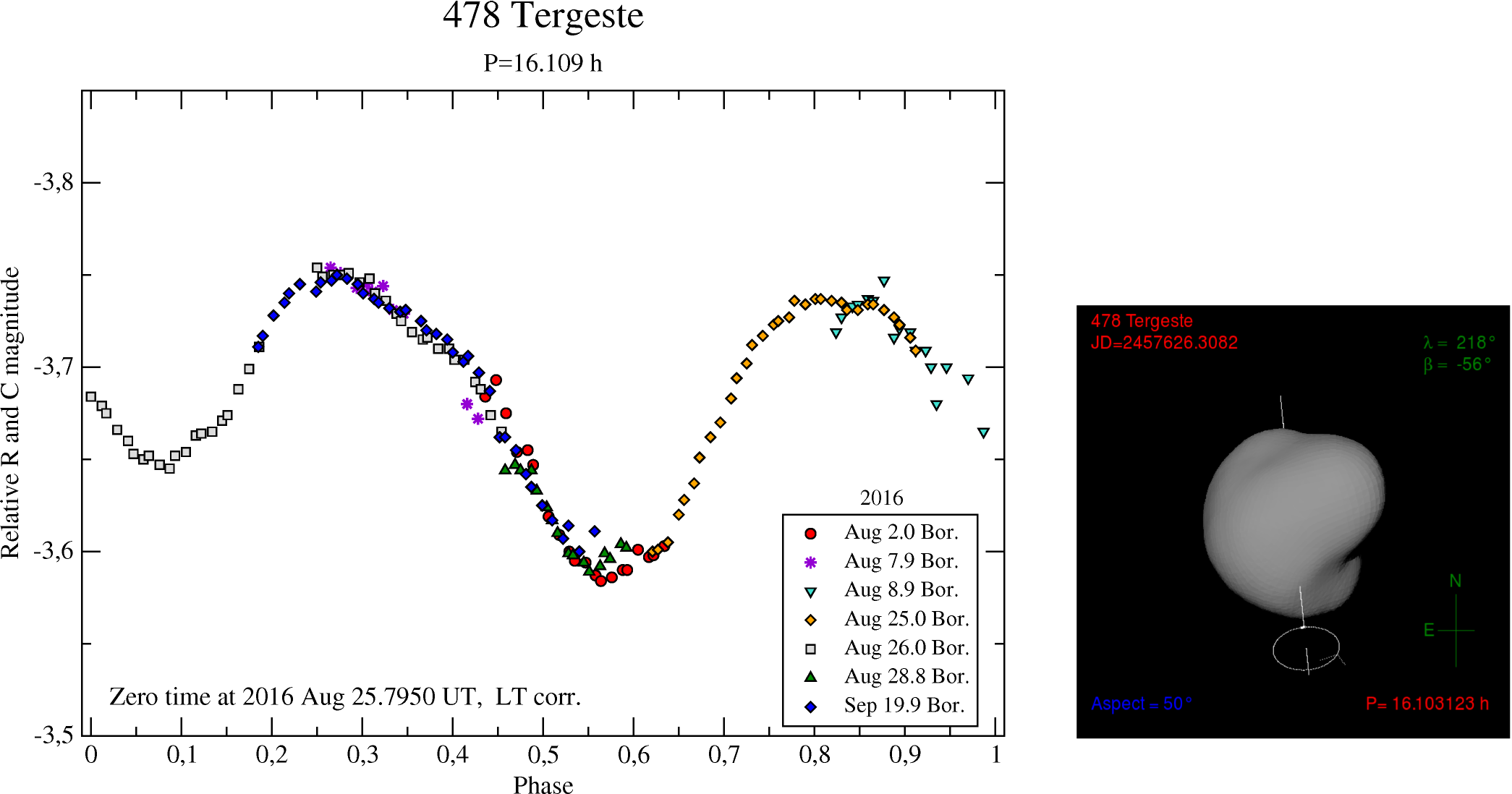} 
\caption{Composite lightcurve of (478) Tergeste in the year 2016 with the
orientation of SAGE model 2 for the zero phase}
\label{Tergeste2016}
\end{figure*}

\begin{figure*}[h]
\includegraphics[width=0.75\textwidth]{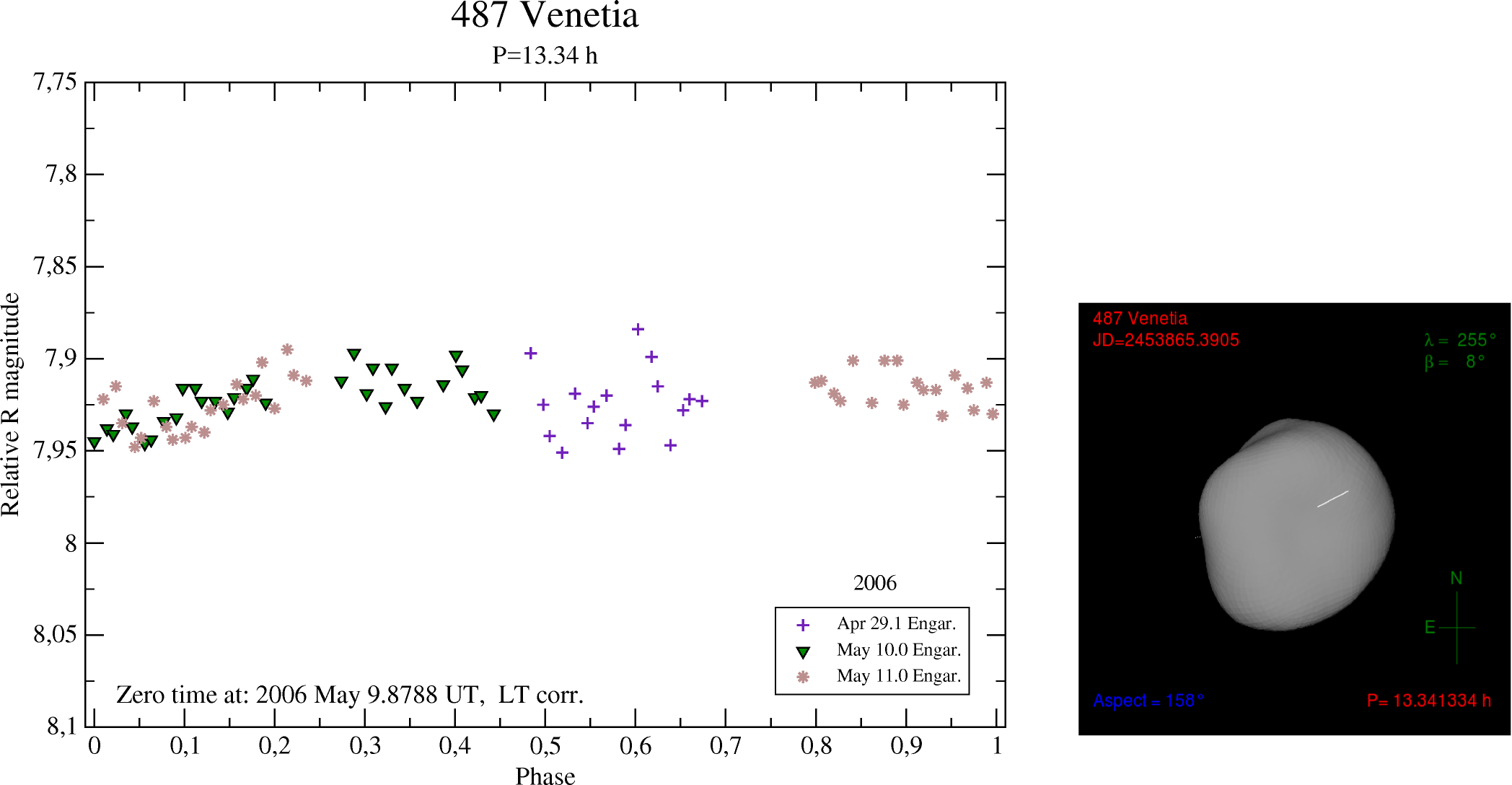} 
\caption{Composite lightcurve of (487) Venetia in the year 2006 with the
orientation of SAGE model 2 for the zero phase}
\label{Venetia2006}
\end{figure*}

\begin{figure*}[h]
\includegraphics[width=0.75\textwidth]{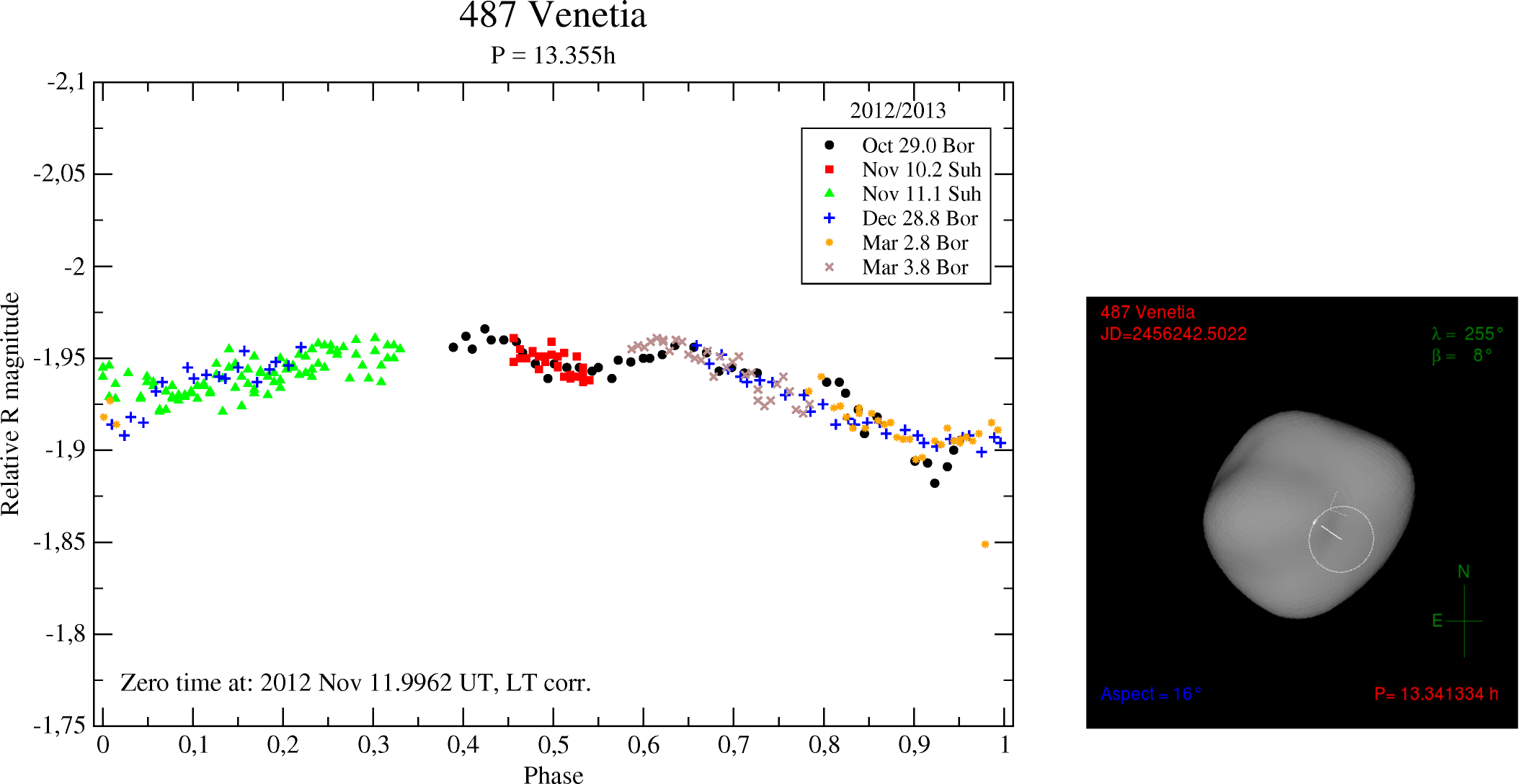} 
\caption{Composite lightcurve of (487) Venetia in the years 2012-2013 with the
orientation of SAGE model 2 for the zero phase}
\label{Venetia2012}
\end{figure*}

\begin{figure*}[h]
\includegraphics[width=0.75\textwidth]{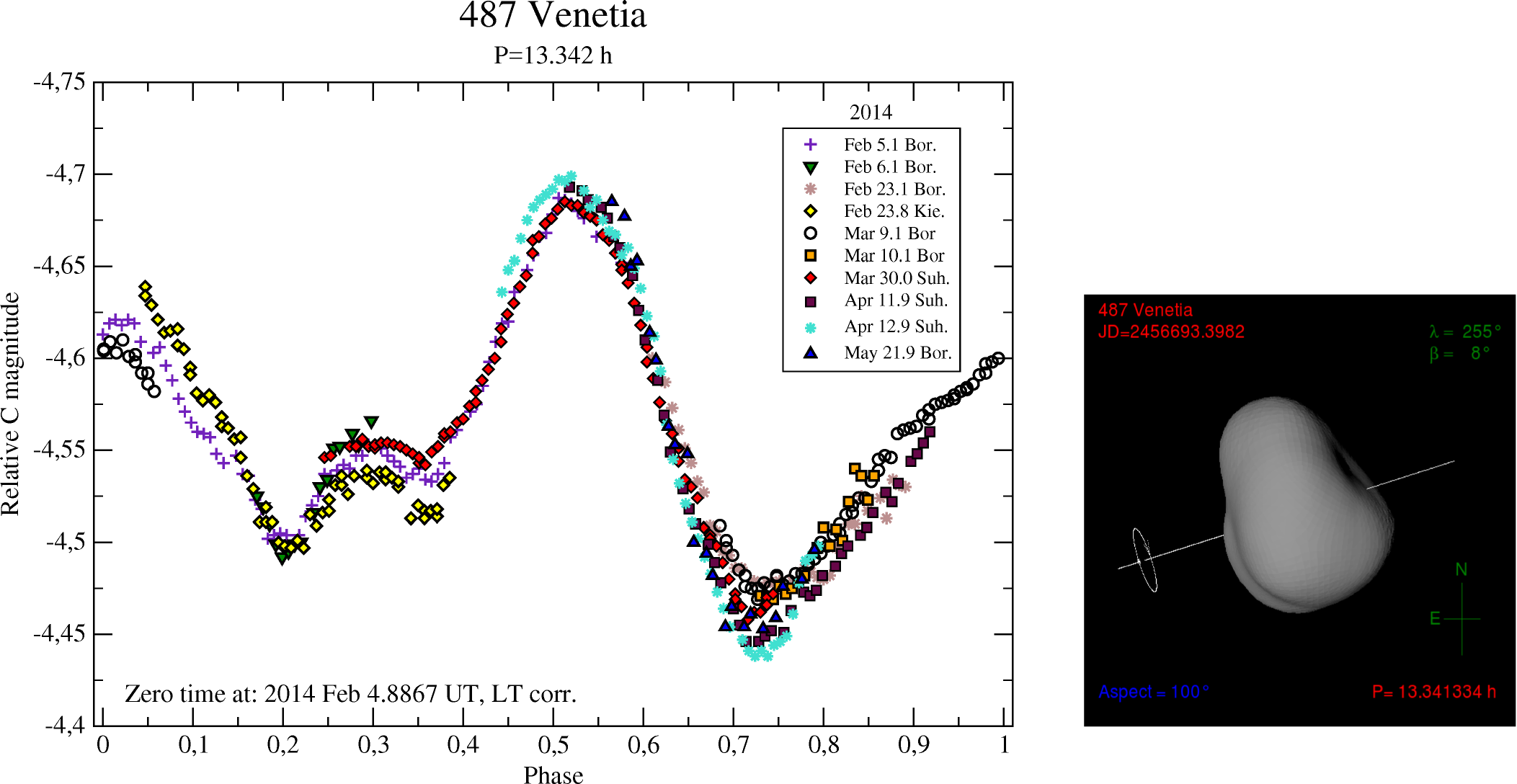} 
\caption{Composite lightcurve of (487) Venetia in the year 2014 with the
orientation of SAGE model 2 for the zero phase}
\label{Venetia2014}
\end{figure*}

\begin{figure*}[h]
\includegraphics[width=0.75\textwidth]{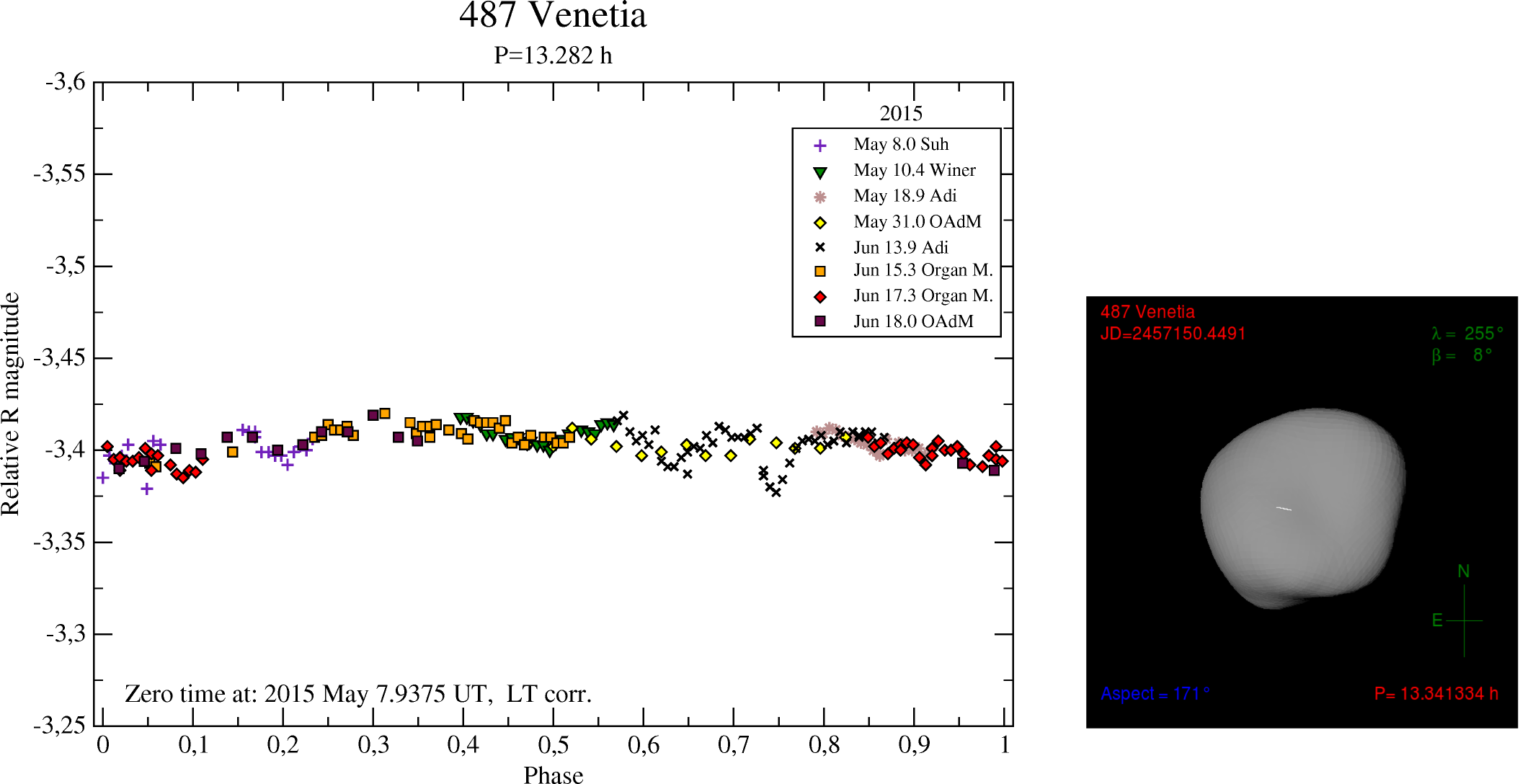} 
\caption{Composite lightcurve of (487) Venetia in the year 2015 with the
orientation of SAGE model 2 for the zero phase}
\label{Venetia2015}
\end{figure*}

Observing runs details (Table \ref{obs}) and
composite lightcurves of asteroids with new period determinations (Figures \ref{Ortrud2016} 
- \ref{Hooveria2016}) and asteroids with  spin and shape models presented here 
(Figures \ref{Aemilia2005} - \ref{Venetia2015}).
\hspace{2cm}
\begin{table}[h]
\begin{scriptsize}
\noindent 
\begin{tabularx}{\textwidth}{|l|l|p{10mm}|l|p{10mm}|p{40mm}|l|}
\hline
 Date & $\lambda$ & Phase angle & Duration  & $\sigma$ & Observer & Site \\
      &   [deg]   &   [deg]     & [hours]   &  [mag]   &          &      \\
\hline
\multicolumn{7}{l}{(551) Ortrud}\\
\hline
 2016 Aug 31.0 &  32.9 & 17.5 & 4.7 & 0.005 & K. {\.Z}ukowski & Borowiec\\
 2016 Sep 02.0 &  33.0 & 17.1 & 4.8 & 0.008 & A. Marciniak & Borowiec\\
 2016 Sep 03.0 &  33.0 & 16.8 & 3.7 & 0.008 & R. Hirsch & Borowiec\\
 2016 Sep 05.4 &  32.9 & 16.3 & 3.6 & 0.011 & F. Pilcher & Organ Mesa Obs.\\
 2016 Sep 09.0 &  32.8 & 15.4 & 4.8 & 0.014 & M.~Butkiewicz B\k{a}k & Borowiec\\
 2016 Sep 13.4 &  32.6 & 14.1 & 5.3 & 0.006 & F. Pilcher & Organ Mesa Obs.\\
 2016 Sep 14.4 &  32.5 & 13.8 & 7.2 & 0.007 & F. Pilcher & Organ Mesa Obs.\\
 2016 Sep 15.3 &  32.4 & 13.6 & 2.8 & 0.013 & F. Pilcher & Organ Mesa Obs.\\
 2016 Oct 01.3 &  30.4 &  7.8 & 8.4 & 0.005 & F. Pilcher & Organ Mesa Obs.\\
 2016 Nov 18.2 &  21.4 & 12.2 & 5.7 & 0.016 & T. Polakis & Tempe \\
 2016 Nov 19.2 &  21.3 & 12.5 & 5.7 & 0.016 & T. Polakis & Tempe \\
 2016 Nov 22.2 &  21.0 & 13.5 & 5.9 & 0.017 & T. Polakis & Tempe \\
 2016 Nov 23.2 &  20.9 & 13.8 & 5.2 & 0.016 & T. Polakis & Tempe \\
 2016 Nov 24.2 &  20.9 & 14.1 & 6.0 & 0.018 & T. Polakis & Tempe \\
 2016 Nov 25.2 &  20.8 & 14.4 & 4.1 & 0.016 & T. Polakis & Tempe \\
               &       &      &77.9 total &&&\\
 \hline
 \multicolumn{7}{l}{(581) Tauntonia}\\
 \hline
 2016 Jan 28.3 & 120.2 &  4.2 & 1.7 & 0.002 & K. Kamiński & Winer Obs. \\ 
 2016 Jan 29.4 & 119.9 &  4.5 & 7.2 & 0.006 & K. Kamiński & Winer Obs. \\ 
 2016 Feb 22.4 & 115.6 & 11.6 & 6.0 & 0.011 & K. Kamiński & Winer Obs. \\ 
 2016 Feb 24.2 & 115.4 & 12.1 & 5.7 & 0.012 & K. Kamiński & Winer Obs. \\ 
 2016 Feb 26.3 & 115.2 & 12.6 & 7.6 & 0.006 & K. Kamiński & Winer Obs. \\ 
 2016 Mar 23.9 & 114.5 & 17.5 & 5.8 & 0.007 & - & Montsec Obs.\\
 2016 Mar 24.9 & 114.5 & 17.6 & 5.8 & 0.007 & - & Montsec Obs.\\
 2016 Mar 27.9 & 114.7 & 17.9 & 5.8 & 0.005 & - & Montsec Obs.\\
 2016 Mar 31.9 & 115.1 & 18.2 & 3.4 & 0.006 & - & Montsec Obs.\\
               &       &      &49.0 total &&&\\
 \hline
 \multicolumn{7}{l}{(830) Petropolitana}\\
 \hline
 2017 Mar 01.4 & 178.0 &  5.1 & 7.6 & 0.012 & T. Polakis & Tempe \\
 2017 Mar 02.4 & 177.8 &  4.8 & 7.5 & 0.013 & T. Polakis & Tempe \\
 2017 Mar 04.4 & 177.5 &  4.1 & 7.6 & 0.011 & T. Polakis & Tempe \\
 2017 Mar 06.4 & 177.1 &  3.3 & 7.5 & 0.010 & T. Polakis & Tempe \\
 2017 Mar 07.4 & 176.9 &  3.0 & 7.5 & 0.009 & T. Polakis & Tempe \\
 2017 Mar 08.4 & 176.7 &  2.6 & 7.2 & 0.010 & T. Polakis & Tempe \\
 2017 Mar 09.4 & 176.5 &  2.3 & 6.9 & 0.014 & T. Polakis & Tempe \\
 2017 Mar 10.4 & 176.3 &  1.9 & 6.9 & 0.011 & T. Polakis & Tempe \\
 2017 Apr 14.2 & 170.2 &  9.9 & 6.8 & 0.013 & T. Polakis & Tempe \\
 2017 Apr 15.2 & 170.1 & 10.2 & 6.6 & 0.011 & T. Polakis & Tempe \\
 2017 Apr 18.3 & 169.8 & 11.0 & 3.0 & 0.017 & T. Polakis & Tempe \\
 2017 Apr 19.2 & 169.7 & 11.2 & 6.1 & 0.017 & T. Polakis & Tempe \\
 2017 Apr 20.2 & 169.6 & 11.5 & 6.1 & 0.019 & T. Polakis & Tempe \\
 2017 Apr 21.2 & 169.5 & 11.7 & 5.7 & 0.019 & T. Polakis & Tempe \\
 2017 Apr 22.2 & 169.4 & 12.0 & 5.9 & 0.016 & T. Polakis & Tempe \\
 2017 Apr 23.3 & 169.3 & 12.2 & 4.1 & 0.017 & T. Polakis & Tempe \\
 2017 Apr 24.2 & 169.2 & 12.5 & 5.2 & 0.023 & T. Polakis & Tempe \\
               &        &     &108.2  total  &&&\\
\hline
\end{tabularx}
\end{scriptsize}
\end{table}

\clearpage
\newpage

\begin{table}
\begin{scriptsize}
\begin{tabularx}{\textwidth}{|l|l|p{10mm}|l|p{10mm}|p{40mm}|l|}
\hline
 Date & $\lambda$ & Phase angle & Duration  & $\sigma$ & Observer & Site \\
      &   [deg]   &   [deg]     & [hours]   &  [mag]   &          &      \\
 \hline
 \multicolumn{7}{l}{(923) Herluga}\\
 \hline
 2016 Jul 26.0 & 330.2 & 13.1 & 4.2 & 0.007 & - & Montsec Obs.\\
 2016 Jul 27.0 & 330.0 & 12.8 & 3.2 & 0.006 & - & Montsec Obs.\\
 2016 Jul 28.0 & 329.8 & 12.6 & 3.2 & 0.005 & - & Montsec Obs.\\
 2016 Aug 03.0 & 328.7 & 10.8 & 3.8 & 0.003 & - & Montsec Obs.\\
 2016 Aug 04.0 & 328.5 & 10.6 & 3.6 & 0.019 & - & Montsec Obs.\\
 2016 Aug 08.0 & 327.6 &  9.6 & 4.4 & 0.003 & - & Montsec Obs.\\
 2016 Aug 11.0 & 326.9 &  9.0 & 7.3 & 0.006 & - & Montsec Obs.\\
 2016 Aug 15.1 & 326.0 &  8.4 & 7.6 & 0.002 & S. Geier & JKT, ORM\\
 2016 Aug 19.9 & 324.8 &  8.3 & 6.7 & 0.022 & R. Hirsch & Borowiec\\
 2016 Aug 22.9 & 324.0 &  8.5 & 7.0 & 0.014 & K. {\.Z}ukowski & Borowiec\\
 2016 Sep 06.9 & 320.7 & 12.2 & 5.7 & 0.008 & A. Marciniak & Borowiec\\
 2016 Sep 10.9 & 320.0 & 13.6 & 5.3 & 0.013 & A. Marciniak & Borowiec\\
               &        &      &62.0 total  &&&\\
 \hline
 \multicolumn{7}{l}{(932) Hooveria}\\
 \hline
 2016 Nov 18.3 &  58.8 &  4.6 & 10.2 & 0.007 & T. Polakis & Tempe \\
 2016 Nov 19.3 &  58.6 &  4.5 & 10.2 & 0.006 & T. Polakis & Tempe \\
 2016 Nov 22.3 &  57.8 &  4.7 & 10.2 & 0.005 & T. Polakis & Tempe \\
 2016 Nov 23.3 &  57.5 &  4.9 & 11.9 & 0.007 & T. Polakis & Tempe \\
 2016 Nov 24.2 &  57.3 &  5.1 &  7.5 & 0.007 & T. Polakis & Tempe \\
 2016 Nov 25.3 &  57.0 &  5.4 & 10.1 & 0.006 & T. Polakis & Tempe \\
 2016 Nov 26.4 &  56.8 &  5.7 &  2.1 & 0.010 & T. Polakis & Tempe \\
 2016 Nov 30.2 &  55.8 &  7.2 &  8.7 & 0.007 & T. Polakis & Tempe \\
 2016 Dec 01.3 &  55.6 &  7.7 &  6.9 & 0.007 & T. Polakis & Tempe \\
 2016 Dec 03.3 &  55.1 &  8.5 &  6.3 & 0.008 & T. Polakis & Tempe \\
 2016 Dec 04.3 &  54.9 &  9.0 &  9.2 & 0.007 & T. Polakis & Tempe \\ 
 2016 Dec 05.3 &  54.7 &  9.4 &  9.2 & 0.007 & T. Polakis & Tempe \\ 
 2016 Dec 09.2 &  53.9 & 11.2 &  9.0 & 0.008 & T. Polakis & Tempe \\ 
 2016 Dec 18.2 &  52.5 & 15.0 &  7.4 & 0.009 & T. Polakis & Tempe \\ 
 2016 Dec 19.2 &  52.4 & 15.4 &  7.4 & 0.009 & T. Polakis & Tempe \\ 
 2016 Dec 20.1 &  52.3 & 15.8 &  3.3 & 0.015 & T. Polakis & Tempe \\  
               &        &      &129.6 total  &&&\\
\hline
 \multicolumn{7}{l}{(995) Sternberga}\\
 \hline
 2016 May 04.4 & 265.0 & 15.1 & 3.9 & 0.006 & B. Skiff & Lowell Obs.\\ 
 2016 May 11.2 & 264.3 & 13.0 & 2.0 & 0.003 & S. Geier & Teide\\
 2016 May 24.0 & 262.4 &  8.7 & 2.1 & 0.028 & V. Kudak & Derenivka\\
 2016 Jun 02.2 & 260.4 &  6.0 & 0.8 & 0.006 & B. Skiff & Lowell Obs.\\ 
 2016 Jun 03.3 & 260.1 &  5.8 & 7.0 & 0.004 & B. Skiff & Lowell Obs.\\ 
 2016 Jun 04.3 & 259.9 &  5.6 & 7.1 & 0.004 & B. Skiff & Lowell Obs.\\ 
 2016 Jun 05.3 & 259.7 &  5.5 & 7.0 & 0.005 & B. Skiff & Lowell Obs.\\ 
 2016 Jun 06.3 & 259.4 &  5.4 & 7.0 & 0.006 & B. Skiff & Lowell Obs.\\ 
 2016 Jun 07.3 & 259.2 &  5.3 & 4.0 & 0.007 & T. Polakis & Tempe\\
 2016 Jun 08.1 & 259.0 &  5.3 & 4.1 & 0.020 & R. Duffard & La Sagra\\
 2016 Jun 08.4 & 259.0 &  5.3 & 4.2 & 0.008 & T. Polakis & Tempe\\
 2016 Jun 09.1 & 258.8 &  5.3 & 3.4 & 0.016 & R. Duffard & La Sagra\\
 2016 Jun 10.1 & 258.5 &  5.4 & 3.7 & 0.014 & R. Duffard & La Sagra\\
 2016 Jun 12.1 & 258.0 &  5.6 & 2.8 & 0.018 & R. Duffard & La Sagra\\
 2016 Jun 13.1 & 258.0 &  5.8 & 3.2 & 0.021 & R. Duffard & La Sagra\\
 2016 Jun 13.3 & 257.7 &  5.8 & 6.7 & 0.005 & B. Skiff & Lowell Obs.\\ 
 2016 Jun 30.9 & 253.8 & 11.6 & 3.3 & 0.004 & - & Montsec Obs.\\
 2016 Jul 06.0 & 253.0 & 13.5 & 3.0 & 0.008 & - & Montsec Obs.\\
 2016 Jul 06.9 & 252.8 & 13.8 & 2.4 & 0.004 & S. Fauvaud & Bardon Obs.\\
 2016 Jul 07.0 & 252.8 & 13.8 & 3.0 & 0.008 & - & Montsec Obs.\\
 2016 Jul 08.0 & 252.6 & 14.2 & 1.9 & 0.005 & S. Fauvaud & Bardon Obs.\\
 2016 Jul 08.9 & 252.5 & 14.5 & 2.8 & 0.005 & S. Fauvaud & Bardon Obs.\\
 2016 Jul 09.0 & 252.5 & 14.5 & 3.2 & 0.006 & - & Montsec Obs.\\
 2016 Jul 10.0 & 252.4 & 14.9 & 3.9 & 0.005 & S. Fauvaud & Bardon Obs.\\
               &&&92.5 total  &&&\\
\hline
\end{tabularx}
\end{scriptsize}
\end{table}

\clearpage
\newpage

\begin{table}
\begin{scriptsize}
\begin{tabularx}{\textwidth}{|l|l|p{10mm}|l|p{10mm}|p{40mm}|l|}
\hline
 Date & $\lambda$ & Phase angle & Duration  & $\sigma$ & Observer & Site \\
      &   [deg]   &   [deg]     & [hours]   &  [mag]   &          &      \\
 \hline
 \multicolumn{7}{l}{(159) Aemilia}\\
 \hline
 2005 Jul 03.1 & 317.6 & 10.2 & 2.6  & 0.013 & L. Bernasconi & Obs. des Engarouines \\
 2005 Jul 09.0 & 316.9 &  8.5 & 3.5  & 0.011 & L. Bernasconi & Obs. des Engarouines \\
 2005 Jul 11.0 & 316.6 &  8.0 & 4.3  & 0.018 & L. Bernasconi & Obs. des Engarouines \\
 2005 Aug 07.0 & 311.8 &  0.8 & 5.6  & 0.014 & L. Bernasconi & Obs. des Engarouines \\
 2005 Aug 09.0 & 311.4 &  1.5 & 5.9  & 0.015 & L. Bernasconi & Obs. des Engarouines \\
 &&&&&&\\                     
 2013 Dec 28.2 & 180.3 & 19.9 & 3.5  & 0.004 & R. Hirsch & Borowiec \\
 2014 Jan 17.4 & 182.6 & 18.0 & 6.0  & 0.006 & F. Pilcher & Organ Mesa Obs.\\
 2014 Jan 25.1 & 182.9 & 16.7 & 5.5  & 0.006 & K. Sobkowiak & Borowiec \\
 2014 Jan 28.4 & 182.9 & 16.1 & 6.9  & 0.003 & F. Pilcher & Organ Mesa Obs.\\
 2014 Feb 04.1 & 182.7 & 14.5 & 6.5  & 0.005 & A. Marciniak & Borowiec \\
 2014 Feb 09.4 & 182.4 & 13.1 & 7.5  & 0.006 & F. Pilcher & Organ Mesa Obs.\\
 2014 Feb 13.1 & 182.0 & 12.0 & 7.5  & 0.005 & A. Marciniak & Borowiec \\
 2014 Feb 21.1 & 181.0 &  9.5 & 6.2  & 0.005 & I. Konstanciak & Borowiec \\
 2014 Mar 29.9 & 173.9 &  5.5 & 7.0  & 0.004 & A. Marciniak & Borowiec \\
 &&&&&&\\                     
 2015 Apr 29.4 & 259.0 & 11.4 & 5.0  & 0.007 & K. Kamiński & Winer Obs. \\
 2015 May 19.9 & 256.1 &  5.6 & 2.9  & 0.005 & M. Żejmo & Adiyaman Obs. \\
 2015 May 30.3 & 254.1 &  2.8 & 7.6  & 0.010 & K. Kamiński & Winer Obs. \\
 2015 May 30.4 & 254.1 &  2.8 & 6.4  & 0.008 & F. Pilcher & Organ Mesa Obs.\\
 2015 Jun 12.9 & 251.5 &  3.8 & 4.2  & 0.007 & M. Żejmo & Adiyaman Obs. \\
 2015 Jun 18.0 & 250.6 &  5.2 & 4.9  & 0.005 &  -  & Montsec Obs. \\
 2015 Jun 20.3 & 250.2 &  5.9 & 4.3  & 0.004 & F. Pilcher & Organ Mesa Obs.\\
 2015 Jun 22.2 & 249.8 &  6.4 & 5.7  & 0.005 & F. Pilcher & Organ Mesa Obs.\\
 2015 Jun 25.0 & 249.4 &  7.2 & 4.4  & 0.003 &  -  & Montsec Obs. \\
 2015 Jul 02.2 & 248.3 &  9.3 & 5.1  & 0.010 & F. Pilcher & Organ Mesa Obs.\\
 2015 Jul 07.2 & 247.8 & 10.6 & 4.7  & 0.006 & F. Pilcher & Organ Mesa Obs.\\
               &&&133.7 total &&&\\
\hline
\end{tabularx}
\end{scriptsize}
\end{table}

\clearpage
\newpage

\begin{table}
\begin{scriptsize}
\begin{tabularx}{\textwidth}{|l|l|p{10mm}|l|p{10mm}|p{40mm}|l|}
\hline
 Date & $\lambda$ & Phase angle & Duration  & $\sigma$ & Observer & Site \\
      &   [deg]   &   [deg]     & [hours]   &  [mag]   &          &      \\
\hline
\multicolumn{7}{l}{(227) Philosophia}\\
\hline
 2006 Nov 09.1 &  43.6 &  3.2 & 4.2  & 0.030  & R. Ditteon & Oakley Obs. \\
 2006 Nov 09.3 &  43.6 &  3.2 & 4.3  & 0.035  & R. Ditteon & Oakley Obs. \\
 2006 Nov 10.1 &  43.4 &  3.3 & 5.1  & 0.028  & R. Ditteon & Oakley Obs. \\
 2006 Nov 10.3 &  43.4 &  3.3 & 3.3  & 0.038  & R. Ditteon & Oakley Obs. \\
 2006 Nov 15.0 &  42.5 &  4.0 & 9.6  & 0.011  & P. Antonini & Obs. Hauts Patys \\
 2006 Nov 29.9 &  39.9 &  7.5 & 7.2  & 0.020  & P. Antonini & Obs. Hauts Patys \\
 2006 Dec 27.9 &  37.3 & 12.9 & 8.4  & 0.026  & P. Antonini & Obs. Hauts Patys \\
 &&&&&&\\                            
 2015 Apr 15.4 & 223.0 &  8.5 & 6.0  & 0.004  & K. Kamiński & Winer Obs. \\
 2015 Apr 17.4 & 222.7 &  7.9 & 6.0  & 0.003  & K. Kamiński & Winer Obs. \\
 2015 Apr 19.4 & 222.3 &  7.4 & 6.0  & 0.003  & K. Kamiński & Winer Obs. \\
 2015 Apr 30.3 & 220.2 &  5.6 & 5.0  & 0.004  & K. Kamiński & Winer Obs. \\
 2015 May 06.3 & 219.0 &  6.1 & 5.3  & 0.009  & K. Kamiński & Winer Obs. \\
 2015 May 10.3 & 218.2 &  7.0 & 5.8  & 0.003  & K. Kamiński & Winer Obs. \\
 2015 May 12.3 & 217.8 &  7.6 & 3.3  & 0.008  & K. Kamiński & Winer Obs. \\
 2015 May 13.3 & 217.6 &  7.9 & 5.3  & 0.004  & K. Kamiński & Winer Obs. \\
 2015 May 14.3 & 217.5 &  8.2 & 5.0  & 0.007  & K. Kamiński & Winer Obs. \\
 2015 May 28.2 & 215.4 & 12.8 & 4.1  & 0.003  & K. Kamiński & Winer Obs. \\
 2015 Jun 29.9 & 215.1 & 20.8 & 2.5  & 0.005  & A. Marciniak & Teide Obs. \\
 &&&&&&\\                            
 2016 Jul 07.4 & 337.6 & 15.3 & 3.1  & 0.003  & D. Oszkiewicz, B. Skiff & Lowell Obs.\\
 2016 Jul 14.4 & 337.2 & 13.6 & 4.7  & 0.004  & D. Oszkiewicz, B. Skiff & Lowell Obs.\\
 2016 Jul 17.4 & 337.0 & 12.8 & 5.1  & 0.007  & D. Oszkiewicz, B. Skiff & Lowell Obs.\\
 2016 Jul 21.3 & 336.6 & 11.7 & 4.7  & 0.015  & D. Oszkiewicz & Cerro Tololo \\
 2016 Jul 25.1 & 336.1 & 10.5 & 5.7  & 0.004  & A. Marciniak & Teide Obs. \\
 2016 Jul 28.4 & 335.6 &  9.5 & 4.7  & 0.005  & D. Oszkiewicz, B. Skiff & Lowell Obs.\\
 2016 Aug 12.3 & 333.0 &  4.2 & 7.2  & 0.004  & B. Skiff & Lowell Obs.\\
 2016 Aug 14.3 & 332.6 &  3.5 & 7.7  & 0.006  & B. Skiff & Lowell Obs.\\
 2016 Aug 15.3 & 332.4 &  3.1 & 7.7  & 0.008  & F. Pilcher & Organ Mesa Obs.\\
 2016 Aug 15.3 & 332.4 &  3.1 & 7.0  & 0.005  & B. Skiff & Lowell Obs.\\
 2016 Aug 16.3 & 332.2 &  2.7 & 7.3  & 0.008  & F. Pilcher & Organ Mesa Obs.\\
 2016 Aug 21.0 & 331.3 &  1.0 & 4.3  & 0.009  &  -  & Montsec Obs. \\
 2016 Aug 23.3 & 330.8 &  0.3 & 6.4  & 0.011  & F. Pilcher & Organ Mesa Obs.\\
 2016 Aug 26.9 & 330.1 &  1.3 & 4.2  & 0.007  & R. Hirsch & Borowiec \\
 2016 Aug 29.3 & 329.6 &  2.2 & 5.8  & 0.004  & B. Skiff & Lowell Obs.\\
 2016 Sep 04.4 & 328.4 &  4.4 & 7.4  & 0.004  & B. Skiff & Lowell Obs.\\
 2016 Sep 08.0 & 327.8 &  5.7 & 5.2  & 0.006  &  -  & Montsec Obs. \\
 2016 Sep 09.2 & 327.6 &  6.1 & 7.0  & 0.005  & F. Pilcher & Organ Mesa Obs.\\
 2016 Sep 09.2 & 327.6 &  6.1 & 7.3  & 0.004  & B. Skiff & Lowell Obs.\\
 2016 Sep 10.2 & 327.4 &  6.4 & 6.9  & 0.006  & F. Pilcher & Organ Mesa Obs.\\
 2016 Sep 11.2 & 327.2 &  6.8 & 6.9  & 0.006  & F. Pilcher & Organ Mesa Obs.\\
 2016 Sep 11.2 & 327.2 &  6.8 & 7.4  & 0.005  & B. Skiff & Lowell Obs.\\
 2016 Sep 17.2 & 326.3 &  8.7 & 6.6  & 0.008  & B. Skiff & Lowell Obs.\\
 2016 Sep 17.2 & 326.3 &  8.7 & 5.0  & 0.013  & F. Pilcher & Organ Mesa Obs.\\
 2016 Sep 18.2 & 326.2 &  9.0 & 6.5  & 0.009  & F. Pilcher & Organ Mesa Obs.\\
 2016 Sep 18.2 & 326.2 &  9.0 & 6.5  & 0.008  & B. Skiff & Lowell Obs.\\
 2016 Sep 19.2 & 326.1 &  9.3 & 6.5  & 0.009  & F. Pilcher & Organ Mesa Obs.\\
 2016 Sep 19.2 & 326.1 &  9.3 & 6.5  & 0.008  & B. Skiff & Lowell Obs.\\
 2016 Sep 24.3 & 325.5 & 10.7 & 2.8  & 0.007  & B. Skiff & Lowell Obs.\\
 2016 Sep 25.2 & 325.4 & 11.0 & 5.6  & 0.006  & B. Skiff & Lowell Obs.\\
 2016 Sep 26.1 & 325.3 & 11.2 & 2.0  & 0.005  & B. Skiff & Lowell Obs.\\
 2016 Oct 02.1 & 324.8 & 12.7 & 6.2  & 0.006  & B. Skiff & Lowell Obs.\\
  &&&  284.3 total &&&\\
\hline
\end{tabularx}
\end{scriptsize}
\end{table}

\clearpage
\newpage

\begin{table}
\begin{scriptsize}
\begin{tabularx}{\textwidth}{|l|l|p{10mm}|l|p{10mm}|p{40mm}|l|}
\hline
 Date & $\lambda$ & Phase angle & Duration  & $\sigma$ & Observer & Site \\
      &   [deg]   &   [deg]     & [hours]   &  [mag]   &          &      \\
\hline
\multicolumn{7}{l}{(329) Svea}\\
\hline
 2006 Jul 24.0 & 325.6 & 11.9 & 4.3 & 0.022 & L. Bernasconi & Obs. des Engarouines \\
 2006 Jul 26.0 & 325.2 & 11.2 & 6.2 & 0.028 & L. Bernasconi & Obs. des Engarouines \\
 2006 Jul 29.0 & 324.5 & 10.2 & 5.1 & 0.022 & L. Bernasconi & Obs. des Engarouines \\
 2006 Jul 30.0 & 324.3 &  9.9 & 5.8 & 0.023 & L. Bernasconi & Obs. des Engarouines \\
 2006 Aug 21.0 & 318.9 &  7.3 & 6.4 & 0.009 & R. Poncy & Le Cr{\`e}s\\
 2006 Aug 22.0 & 318.7 &  7.5 & 6.2 & 0.010 & R. Poncy & Le Cr{\`e}s\\
 2006 Aug 27.9 & 317.3 &  9.0 & 6.2 & 0.012 & R. Poncy & Le Cr{\`e}s\\
 &&&&&&\\                            
 2014 Jul 30.0 & 352.9 & 17.3 & 3.2 & 0.004 & A. Marciniak & Borowiec\\
 2014 Aug 03.0 & 352.6 & 16.1 & 4.5 & 0.007 & A. Marciniak & Borowiec\\
 2014 Aug 09.0 & 351.9 & 14.0 & 5.0 & 0.010 & A. Marciniak & Borowiec\\
 2014 Aug 28.0 & 348.2 &  6.2 & 6.5 & 0.007 & A. Marciniak & Borowiec\\
 2014 Sep 04.0 & 346.5 &  3.4 & 6.5 & 0.009 & A. Marciniak & Borowiec\\
 2014 Sep 19.0 & 342.7 &  5.6 & 5.0 & 0.011 & A. Marciniak & Borowiec\\
 2014 Sep 28.8 & 340.6 &  9.8 & 3.0 & 0.017 & A. Marciniak & Borowiec\\
 2014 Oct 03.8 & 339.7 & 11.8 & 5.7 & 0.025 & K. Sobkowiak & Borowiec\\
 2014 Oct 09.8 & 338.9 & 14.0 & 2.4 & 0.006 & J. Horbowicz & Borowiec\\
 2014 Oct 10.2 & 338.9 & 14.2 & 4.7 & 0.007 & F. Pilcher & Organ Mesa Obs.\\
 2014 Nov 26.2 & 341.3 & 22.9 & 3.7 & 0.007 & K. Kami{\'n}ski & Winer Obs.\\
 &&&&&&\\                            
 2015 Nov 25.1 & 110.0 & 17.8 & 6.2 & 0.009  & R. Hirsch & Borowiec\\
 2015 Dec 14.0 & 107.5 & 12.9 & 5.9 & 0.002 & - & Montsec Obs.\\
 2015 Dec 16.0 & 107.1 & 12.4 & 7.4 & 0.010 & - & Montsec Obs.\\
 2015 Dec 17.0 & 107.0 & 12.2 & 8.3 & 0.008 & - & Montsec Obs.\\
 2015 Dec 18.0 & 106.7 & 11.9 & 7.4 & 0.005 & - & Montsec Obs.\\
 2015 Dec 19.0 & 106.5 & 11.7 & 7.5 & 0.009 & - & Montsec Obs.\\
 2015 Dec 22.0 & 105.8 & 11.0 & 7.6 & 0.004 & - & Montsec Obs.\\
 2015 Dec 23.0 & 105.5 & 10.8 & 7.5 & 0.006 & - & Montsec Obs.\\
 2015 Dec 28.0 & 104.3 &  9.9 & 7.4 & 0.025 & - & Montsec Obs.\\
 2015 Dec 29.9 & 103.8 &  9.7 & 3.8 & 0.004 & - & Montsec Obs.\\
 2015 Dec 30.9 & 103.5 &  9.6 & 4.3 & 0.002 & - & Montsec Obs.\\
 2016 Jan 24.9 &  97.4 & 13.0 & 7.4 & 0.006 & - & Montsec Obs.\\
 2016 Feb 01.9 &  96.0 & 15.2 & 7.6 & 0.005 & - & Montsec Obs.\\
 2016 Mar 02.2 &  95.4 & 21.5 & 3.0 & 0.004 & K. Kami{\'n}ski & Winer Obs.\\
&&& 181.7 total &&&\\
\hline
\end{tabularx}
\end{scriptsize}
\end{table}

\clearpage
\newpage

\begin{table}
\begin{scriptsize}
\begin{tabularx}{\textwidth}{|l|l|p{10mm}|l|p{10mm}|p{40mm}|l|}
\hline
 Date & $\lambda$ & Phase angle & Duration  & $\sigma$ & Observer & Site \\
      &   [deg]   &   [deg]     & [hours]   &  [mag]   &          &      \\
\hline
\multicolumn{7}{l}{(478) Tergeste}\\
\hline
2005 Jul 16.0 & 315.2 &  8.3 & 4.7 & 0.006 & L. Bernasconi & Obs. des Engarouines\\
2005 Jul 17.0 & 315.0 &  8.1 & 5.2 & 0.021 & L. Bernasconi & Obs. des Engarouines\\
2005 Aug 06.0 & 311.1 &  5.7 & 5.0 & 0.007 & L. Bernasconi & Obs. des Engarouines\\
2005 Aug 08.0 & 310.7 &  5.6 & 5.8 & 0.010 & L. Bernasconi & Obs. des Engarouines\\
2005 Aug 08.0 & 310.7 &  5.8 & 4.9 & 0.010 & R. Crippa, F. Manzini & Stazione Astro. di Sozzago\\
2005 Aug 09.9 & 310.3 &  6.0 & 2.4 & 0.010 & R. Crippa, F. Manzini & Stazione Astro. di Sozzago\\
2005 Aug 12.0 & 309.8 &  6.3 & 5.0 & 0.010 & L. Bernasconi & Obs. des Engarouines\\
2005 Aug 12.9 & 309.7 &  6.5 & 4.2 & 0.006 & R. Stoss, P. Korlevic, M. Hren,  & OAM-Mallorca\\
              &       &      &  &  & A. Cikota, L. Jerosimic          & \\
2005 Aug 13.0 & 309.6 &  6.5 & 4.2 & 0.006 & R. Crippa, F. Manzini & Stazione Astro. di Sozzago\\
2005 Aug 13.0 & 309.6 &  6.5 & 5.8 & 0.013 & L. Bernasconi & Obs. des Engarouines\\
2005 Aug 15.0 & 309.2 &  6.8 & 3.2 & 0.007 & R. Crippa, F. Manzini & Stazione Astro. di Sozzago\\
2005 Aug 16.0 & 309.0 &  7.0 & 3.2 & 0.007 & R. Crippa, F. Manzini & Stazione Astro. di Sozzago\\
&&&&&&\\                      
2012 Nov 26.2 & 108.5 & 14.7 & 1.5 & 0.006 & M. Murawiecka & Borowiec \\
2013 Feb 15.1 &  95.7 & 16.5 & 2.8 & 0.003 & F. Pilcher & Organ Mesa Obs.\\
2013 Feb 16.1 &  95.6 & 16.7 & 4.1 & 0.003 & F. Pilcher & Organ Mesa Obs.\\
2013 Feb 22.1 &  95.6 & 17.9 & 4.3 & 0.008 & F. Pilcher & Organ Mesa Obs.\\
2013 Feb 24.1 &  95.7 & 18.3 & 4.2 & 0.005 & F. Pilcher & Organ Mesa Obs.\\
2013 Mar 12.1 &  97.0 & 20.3 & 2.7 & 0.003 & F. Pilcher & Organ Mesa Obs.\\
2013 Mar 15.1 &  97.4 & 20.6 & 3.7 & 0.004 & F. Pilcher & Organ Mesa Obs.\\
2013 Mar 17.2 &  97.7 & 20.7 & 3.9 & 0.005 & F. Pilcher & Organ Mesa Obs.\\
2013 Mar 26.2 &  99.3 & 21.1 & 3.8 & 0.006 & F. Pilcher & Organ Mesa Obs.\\
2013 Mar 26.8 &  99.5 & 21.1 & 2.5 & 0.009 & R. Hirsch & Borowiec \\
&&&&&&\\                       
2014 Apr 18.0 & 201.3 &  4.0 & 4.5 & 0.008 & - & Montsec Obs.\\
2014 Apr 19.0 & 201.1 &  4.2 & 4.6 & 0.011 & - & Montsec Obs.\\
2014 Apr 24.0 & 200.1 &  5.5 & 4.2 & 0.012 & - & Montsec Obs.\\
2014 May 15.2 & 196.8 & 12.0 & 4.5 & 0.005 & F. Pilcher & Organ Mesa Obs.\\
2014 May 16.2 & 196.6 & 12.3 & 4.6 & 0.005 & F. Pilcher & Organ Mesa Obs.\\
2014 May 23.9 & 196.0 & 14.2 & 2.0 & 0.005 & - & Montsec Obs.\\
2014 May 27.0 & 195.9 & 14.9 & 2.0 & 0.007 & - & Montsec Obs.\\
&&&&&&\\                         
2015 Jun 18.0 & 279.8 &  5.6 & 4.7 & 0.005 & - & Montsec Obs.\\
2015 Jun 19.0 & 279.6 &  5.4 & 5.5 & 0.006 & - & Montsec Obs.\\
2015 Jun 21.0 & 279.3 &  5.0 & 5.2 & 0.011 & - & Montsec Obs.\\
2015 Jun 27.1 & 278.0 &  4.2 & 6.8 & 0.003 & A. Marciniak & Obs. del Teide\\
2015 Jun 28.9 & 277.6 &  4.1 & 1.5 & 0.003 & A. Marciniak & Obs. del Teide\\
2015 Jul 18.0 & 273.9 &  7.6 & 4.5 & 0.010 & - & Montsec Obs.\\
2015 Jul 26.9 & 272.5 &  9.9 & 1.5 & 0.006 & A. Marciniak & Borowiec\\
2015 Aug 03.7 & 271.6 & 11.9 & 4.2 & 0.003 & M. {\.Z}ejmo & Adiyaman Obs. \\
2015 Aug 08.8 & 271.6 & 13.0 & 2.2 & 0.004 & M. {\.Z}ejmo & Adiyaman Obs. \\
&&&&&&\\                       
2016 Aug 02.0 & 357.6 & 14.1 & 3.4 & 0.008 & K. {\.Z}ukowski & Borowiec\\
2016 Aug 07.9 & 357.1 & 12.9 & 3.0 & 0.008 & A. Marciniak & Borowiec\\
2016 Aug 08.9 & 357.0 & 12.6 & 3.0 & 0.005 & K. {\.Z}ukowski & Borowiec\\
2016 Aug 25.0 & 354.8 &  8.7 & 4.9 & 0.006 & K. {\.Z}ukowski & Borowiec\\
2016 Aug 26.0 & 354.6 &  8.5 & 7.2 & 0.003 & A. Marciniak & Borowiec\\
2016 Aug 28.8 & 354.1 &  7.8 & 2.5 & 0.002 & R. Hirsch & Borowiec\\
2016 Sep 19.9 & 349.6 &  5.9 & 6.5 & 0.008 & R. Hirsch & Borowiec\\
&&& 180.1 total &&&\\
\hline
\end{tabularx}
\end{scriptsize}
\end{table}

\clearpage
\newpage

\begin{table}
\begin{scriptsize}
\begin{tabularx}{\textwidth}{|l|l|p{10mm}|l|p{10mm}|p{40mm}|l|}
\hline
 Date & $\lambda$ & Phase angle & Duration  & $\sigma$ & Observer & Site \\
      &   [deg]   &   [deg]     & [hours]   &  [mag]   &          &      \\
\hline
\multicolumn{7}{l}{(487) Venetia}\\
\hline
 2006 Apr 29.1 & 236.4 &  7.6 & 2.5 & 0.013 & L. Bernasconi & Obs. des Engarouines\\
 2006 May 10.0 & 234.0 &  5.1 & 5.9 & 0.015 & L. Bernasconi & Obs. des Engarouines\\
 2006 May 11.0 & 233.8 &  5.0 & 5.8 & 0.009 & L. Bernasconi & Obs. des Engarouines\\
 &&&&&&\\                      
 2012 Oct 29.0 &  62.3 & 11.7 & 7.5 & 0.012 & M. Bronikowska & Borowiec\\
 2012 Nov 10.2 &  59.8 &  7.5 & 1.0 & 0.009 & W. Og{\l}oza, E. Kosturkiewicz & Suhora\\
 2012 Nov 11.1 &  59.6 &  7.2 & 4.5 & 0.007 & W. Og{\l}oza, E. Kosturkiewicz & Suhora\\
 2012 Dec 28.8 &  51.2 & 17.1 & 7.5 & 0.008 & K. Sobkowiak & Borowiec\\
 2013 Mar 02.8 &  61.7 & 22.9 & 3.2 & 0.006 & R. Hirsch & Borowiec\\
 2013 Mar 03.8 &  62.0 & 22.9 & 2.7 & 0.005 & M. Bronikowska & Borowiec\\
 &&&&&&\\                      
 2014 Feb 05.1 & 174.6 & 13.0 & 7.8 & 0.006 & R. Hirsch & Borowiec\\
 2014 Feb 06.1 & 174.5 & 12.7 & 2.7 & 0.006 & A. Marciniak & Borowiec\\
 2014 Feb 23.1 & 171.5 &  7.2 & 3.8 & 0.008 & K. Sobkowiak & Borowiec\\
 2014 Feb 23.8 & 171.4 &  7.0 & 5.5 & 0.014 & P. Kankiewicz & Kielce\\
 2014 Mar 09.1 & 168.3 &  4.4 & 5.1 & 0.007 & R. Hirsch & Borowiec\\
 2014 Mar 10.1 & 168.1 &  4.5 & 2.1 & 0.016 & J. Horbowicz & Borowiec\\
 2014 Mar 30.0 & 163.8 &  9.6 & 6.6 & 0.004 & W. Og{\l}oza, E. Kosturkiewicz & Suhora\\
 2014 Apr 11.9 & 161.9 & 13.6 & 5.5 & 0.003 & M. Siwak, E. Kosturkiewicz & Suhora\\
 2014 Apr 12.9 & 161.8 & 13.8 & 4.8 & 0.006 & M. Siwak, E. Kosturkiewicz & Suhora\\
 2014 May 21.9 & 162.6 & 20.2 & 3.2 & 0.006 & R. Hirsch & Borowiec\\
 &&&&&&\\                       
 2015 May 08.0 & 263.6 & 12.5 & 3.2 & 0.005 & W. Og{\l}oza & Suhora\\
 2015 May 10.4 & 263.3 & 11.8 & 2.3 & 0.004 & K. Kami{\'n}ski & Winer\\
 2015 May 19.0 & 262.0 &  9.0 & 2.4 & 0.002 & M. {\.Z}ejmo & Adiyaman\\
 2015 May 31.0 & 259.6 &  5.1 & 4.7 & 0.004 &  -  & Montsec\\
 2015 Jun 13.9 & 256.4 &  4.0 & 4.0 & 0.004 & M. {\.Z}ejmo & Adiyaman\\
 2015 Jun 15.3 & 256.1 &  4.3 & 6.2 & 0.003 & F. Pilcher & Organ Mesa Obs.\\
 2015 Jun 17.3 & 255.7 &  4.9 & 3.5 & 0.003 & F. Pilcher & Organ Mesa Obs.\\
 2015 Jun 18.0 & 255.5 &  5.1 & 5.3 & 0.005 &  -  & Montsec\\
&&& 119.3 total &&&\\
\hline
\end{tabularx}
\caption{Observation details: mid-time observing date, ecliptic longitude of the target, 
sun-target-observer phase angle, duration of the observing run, brightness scatter, 
observer, and site name. See Table \ref{sites} for telescope and site details.}
\label{obs}
\end{scriptsize}
\end{table}

\clearpage



\clearpage

\clearpage


\clearpage

\clearpage

\clearpage

\clearpage

\clearpage


\section*{Appendix B}
List of stellar occultation observers.\\
\\
(159) Aemilia (2009-05-02), USA\\
---------------------------\\
S. Meesner (Northfield, Minnesota)\\
S. Conard (Gamber, Maryland)\\
B. Koch (Faribault, Minnesota)\\
A. Scheck (Laurel, Maryland)\\
\\
(329) Svea (2011-12-28), Japan\\
------------------------\\
H. Tomioka (Hitachi city, Ibaraki Prefecture)\\
H. Takashima (Kashiwa, Chiba)\\
K. Kitazato (Musashino, Tokyo)\\
Y. Watanabe (Inabe, Mie)\\
S. Ida (Higashiomi, Shiga)\\
M. Ishida (Moriyama, Shiga)\\
M. Owada (Hamamatsu, Shizuoka)\\
K. Kasazumi (Takatsuki, Osaka)\\
S. Okamoto (Tsuyama, Okayama)\\
N. Tatsumi (Akaiwa, Okayama)\\
Hironaka and Miyamaoto (Hiroshima University Observatory, Hiroshima)\\
\\
(329) Svea (2013-03-07), USA\\
------------------------\\
P. Maley (5 sites, Forida)\\
D. Liles (Florida)\\
A. Cruz (Glen St. Mary, Florida)\\
E. Gray (Macclenny, Florida)\\
J. Brueggemann (Florida)\\
C. McDougal (Tampa, Florida)\\
T. Campbel (3 sites, Florida)\\

\end{document}